\documentclass{ieeeaccess}
\IEEEoverridecommandlockouts
\usepackage{threeparttable}
\usepackage{nomencl}
\makenomenclature
\usepackage{gensymb}
\usepackage{cite}
\usepackage{amsmath,amssymb,amsfonts}
\usepackage{graphicx}
\usepackage{textcomp}
\usepackage{amsthm}
\usepackage{xcolor}
\usepackage{etoolbox}
\renewcommand\nomgroup[1]{%
  \item[\bfseries
  \ifstrequal{#1}{A}{Abbreviations}{%
  \ifstrequal{#1}{V}{Variables}{}}%
]}
\usepackage{enumitem}
\usepackage{graphicx}
\usepackage{amsmath}
\usepackage{multirow}
\usepackage{amssymb}
\usepackage{array}
\newcommand\discussion[1]{\textcolor{red}{#1}}
\renewcommand\discussion[1]{} % comment this to see discussions, uncomment to hide the discussions.

\usepackage{amsthm}

\newtheorem*{theorem*}{Theorem}

\newtheorem*{lemma*}{Lemma}
\usepackage{tabularx}
\usepackage{url}
\usepackage{hyperref}

\newcolumntype{P}[1]{>{\centering\arraybackslash}p{#1}}

\usepackage{enumitem}
\newlist{legal}{enumerate}{10}
\setlist[legal]{label*=\arabic*.}

\newtheorem*{definition*}{Definition}

\newtheorem*{condition*}{Condition}
\newtheorem*{proof*}{Proof}

\usepackage{mathtools}
\newsavebox\myboxA
\newsavebox\myboxB
\newlength\mylenA

\newcommand{\subparagraph}{}

\newcolumntype{b}{>{\hsize=1.25\hsize}X}
\newcolumntype{s}{>{\hsize=.5\hsize}X}

\usepackage{algorithm}% http://ctan.org/pkg/algorithm
\usepackage{algpseudocode}% http://ctan.org/pkg/algorithmicx
\makeatletter
\def\therule{\makebox[\algorithmicindent][l]{\hspace*{.5em}\vrule height .75\baselineskip depth .25\baselineskip}}%

\newtoks\therules% Contains rules
\therules={}% Start with empty token list
\def\appendto#1#2{\expandafter#1\expandafter{\the#1#2}}% Append to token list
\def\gobblefirst#1{% Remove (first) from token list
	#1\expandafter\expandafter\expandafter{\expandafter\@gobble\the#1}}%
\def\LState{\State\unskip\the\therules}% New line-state
\def\pushindent{\appendto\therules\therule}%
\def\popindent{\gobblefirst\therules}%
\def\printindent{\unskip\the\therules}%
\def\printandpush{\printindent\pushindent}%
\def\popandprint{\popindent\printindent}%

\algdef{SE}[WHILE]{While}{EndWhile}[1]
{\printandpush\algorithmicwhile\ #1\ \algorithmicdo}
{\popandprint\algorithmicend\ \algorithmicwhile}%
\algdef{SE}[FOR]{For}{EndFor}[1]
{\printandpush\algorithmicfor\ #1\ \algorithmicdo}
{\popandprint\algorithmicend\ \algorithmicfor}%
\algdef{S}[FOR]{ForAll}[1]
{\printindent\algorithmicforall\ #1\ \algorithmicdo}%
\algdef{SE}[LOOP]{Loop}{EndLoop}
{\printandpush\algorithmicloop}
{\popandprint\algorithmicend\ \algorithmicloop}%
\algdef{SE}[REPEAT]{Repeat}{Until}
{\printandpush\algorithmicrepeat}[1]
{\popandprint\algorithmicuntil\ #1}%
\algdef{SE}[IF]{If}{EndIf}[1]
{\printandpush\algorithmicif\ #1\ \algorithmicthen}
{\popandprint\algorithmicend\ \algorithmicif}%
\algdef{C}[IF]{IF}{ElsIf}[1]
{\popandprint\pushindent\algorithmicelse\ \algorithmicif\ #1\ \algorithmicthen}%
\algdef{Ce}[ELSE]{IF}{Else}{EndIf}
{\popandprint\pushindent\algorithmicelse}%
\algdef{SE}[PROCEDURE]{Procedure}{EndProcedure}[2]
{\printandpush\algorithmicprocedure\ \textproc{#1}\ifthenelse{\equal{#2}{}}{}{(#2)}}%
{\popandprint\algorithmicend\ \algorithmicprocedure}%
\algdef{SE}[FUNCTION]{Function}{EndFunction}[2]
{\printandpush\algorithmicfunction\ \textproc{#1}\ifthenelse{\equal{#2}{}}{}{(#2)}}%
{\popandprint\algorithmicend\ \algorithmicfunction}%
\makeatother

\newcommand*\xoverline[2][0.75]{%
	\sbox{\myboxA}{$\m@th#2$}%
	\setbox\myboxB\null% Phantom box
	\ht\myboxB=\ht\myboxA%
	\dp\myboxB=\dp\myboxA%
	\wd\myboxB=#1\wd\myboxA% Scale phantom
	\sbox\myboxB{$\m@th\overline{\copy\myboxB}$}%  Overlined phantom
	\setlength\mylenA{\the\wd\myboxA}%   calc width diff
	\addtolength\mylenA{-\the\wd\myboxB}%
	\ifdim\wd\myboxB<\wd\myboxA%
	\rlap{\hskip 0.5\mylenA\usebox\myboxB}{\usebox\myboxA}%
	\else
	\hskip -0.5\mylenA\rlap{\usebox\myboxA}{\hskip 0.5\mylenA\usebox\myboxB}%
	\fi}
\makeatother

\usepackage{scalerel}[2014/03/10]
\usepackage{stackengine}

\newlength\myindent
\usepackage{epsfig}
\usepackage{tabularx}
\usepackage{supertabular,booktabs}
\usepackage{capt-of}

\usepackage{longtable}
\usepackage{booktabs, siunitx}

\usepackage{subcaption}

\renewcommand{\algorithmicforall}{\textbf{for each}}

\algrenewcommand\algorithmicrequire{\textbf{Precondition:}}
\algrenewcommand\algorithmicensure{\textbf{Postcondition:}}

\usepackage{etoolbox}
\usepackage{textcomp}
\usepackage[normalem]{ulem}
\usepackage{mathtools}
\usepackage{float}
\usepackage{bm}

\usepackage{caption,setspace}
\def\BibTeX{{\rm B\kern-.05em{\sc i\kern-.025em b}\kern-.08em
    T\kern-.1667em\lower.7ex\hbox{E}\kern-.125emX}}

\begin{document}
	\bstctlcite{IEEEexample:BSTcontrol}

	\history{Date of publication xxxx 00, 0000, date of current version xxxx 00, 0000.}
\doi{10.1109/ACCESS.2017.DOI}

\title{A Comprehensive Review: Impacts of Extreme Temperatures due to Climate Change on Power Grid Infrastructure and Operation}
\author{\uppercase{Kishan Prudhvi Guddanti}\authorrefmark{1}, \IEEEmembership{Member, IEEE}, \uppercase{Alok Kumar Bharati}\authorrefmark{1}, \IEEEmembership{ Member, IEEE}, \uppercase{Sameer Nekkalapu}\authorrefmark{1}, \IEEEmembership{ Member, IEEE}, \uppercase{Joseph McWherter}\authorrefmark{2}, \IEEEmembership{Student Member, IEEE}, \uppercase{Scott Morris}\authorrefmark{1}\IEEEmembership{}}
\address[1]{Pacific Northwest National Laboratory, Richland, WA 99354 USA}
\address[2]{Chapman University, Orange, CA 92866 USA}
\tfootnote{This work was supported by project from DOE-CESER}

\corresp{Corresponding author: Alok Kumar Bharati (e-mail: ak.bharati@pnnl.gov).}

	\titlepgskip=-50pt
	\begin{abstract}
	The power grid is experiencing a multi-fold transformation while the global climate evolves with record-breaking extreme temperatures during heat domes, polar vortices, and severe ice. Over the decades, these extreme temperature events have increased in frequency, duration, and intensity. The power grid infrastructure is geographically spread over thousands of square miles with millions of small and large components, and the impact of extreme temperature operations on the grid infrastructure needs to be researched further. This paper reviews academic literature, standards, industry articles, and federal reports to identify the impacts of heat domes, polar vortices, and icing on all the T\&D grid equipment, including substations (assets owned and operated by the utilities and independent system operators). This paper classifies the equipment into primary and auxiliary equipment and determines its vulnerability to extreme temperatures for a deeper analysis of a more critical and vulnerable set of grid equipment. For each equipment under consideration, its fundamental role in the system, the impact of extreme temperatures on its operation, available monitoring, and mitigation of these impacts are discussed. The paper develops insights on standards readiness and identifies gaps concerning extreme temperature definitions. The paper also develops summary tables to identify the critical failure modes for each type of equipment, failure influence diagrams, and cascading influence diagrams to highlight and aid in translating the equipment vulnerability information into power grid contingency definitions that need to be considered in grid planning and operations.	
	\end{abstract}

% \begin{keywords}
% Transmission planning, large interconnected grids, dynamic simulations, machine learning, dynamic security assessment, high renewable penetration
% \end{keywords}

\titlepgskip=-50pt

% \IEEEpeerreviewmaketitle
\maketitle
% % below .tex has nomenclature or abbreviations example.
% \input{input_files/tiltles}

% \vspace{-2em}
\section{Introduction}
\label{sec:introduction}
Electric power grids, consisting of transmission and distribution (T\&D) systems spread over thousands of square miles, are among the largest infrastructures on Earth \cite{Glover}. The transmission system is largely meshed, and the distribution system is generally radial. Each system has equipment to carry power, as well as auxiliary equipment and components for measurement, communication, and control. Together, the grid contains hundreds of small and large components and systems that work in a coordinated manner to ensure safe and reliable grid operation \cite{fang2011smart}. These components range from large outdoor high-voltage, high-power transformers, to small outdoor components like vibration dampers, to large and small indoor equipment \cite{mei2011power}. Power grid equipment is typically designed to last for decades, because more frequent maintenance and repairs would be very expensive \cite{alvarez2022power}. However, equipment lifespans are estimated on design basis assumptions for historically stable operating conditions. There has been a significant shift in the operating conditions of the T\&D grid in recent decades—--in particular, higher incidence of extreme weather related to climate change \cite{beard2009key}. Therefore, it is essential to understand the how operating conditions for T\&D equipment may differ from when this equipment was designed or commissioned and assess its vulnerability to extreme operating conditions. 

\begin{table}[H]
\centering
\begin{tabular}{ |p{0.46\textwidth}| } 
\hline
This paper is under review with IEEE Access. If accepted, the link to final paper will be provided here. \\
\hline
\end{tabular}
\end{table}

\begin{figure}
    \centering
    \includegraphics[width=\linewidth]{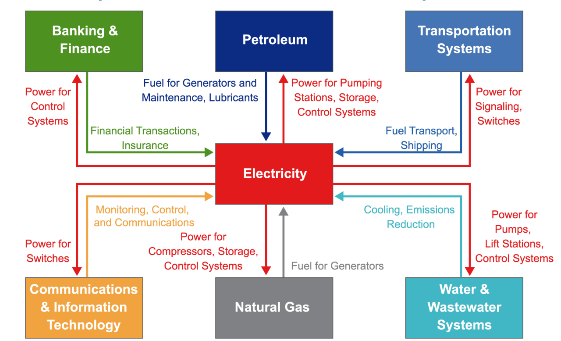}
    \caption{Interdependence of critical infrastructure on electricity \cite{doe2015quadrennial}}
    \label{fig:interdependencies}
\end{figure}

\begin{figure*}
    \centering
    \includegraphics[width=\linewidth]{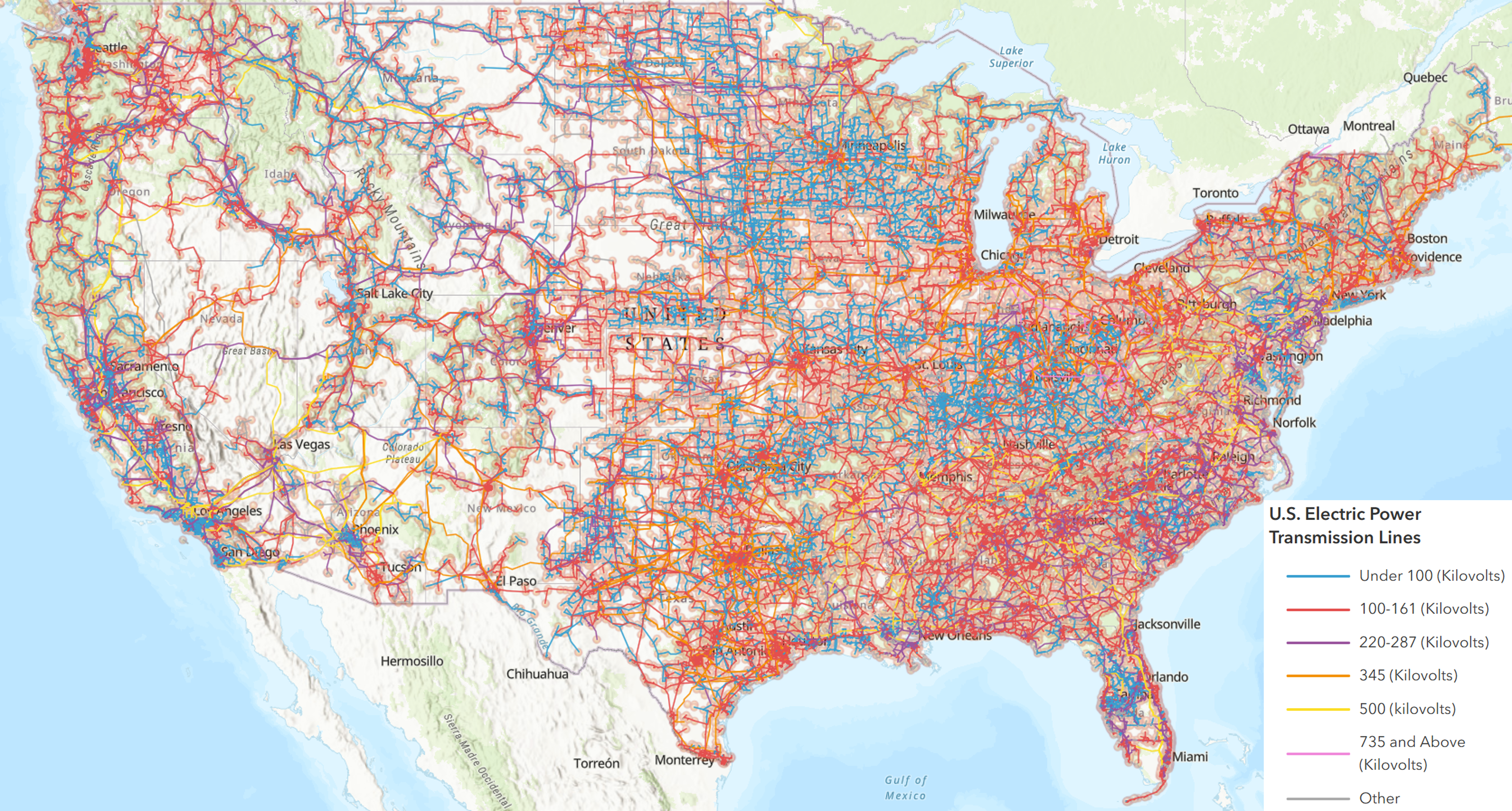}
    \caption{Map of high-voltage substations (light orange bubbles at terminals of the transmission lines in the map) and transmission lines in U.S. \cite{hifld_1}}
    \label{fig:lines_us}
\end{figure*}

Importantly, the electric power grid is a central critical infrastructure; it is essential for all other critical infrastructures to function (Fig.~\ref{fig:interdependencies}). Fig.~\ref{fig:lines_us} shows the density of high-voltage substations and transmission lines in the U.S.; most T\&D equipment is above ground and exposed to weather. Two weather events of interest are heat domes and polar vortices. These extreme hot and cold events are projected to threaten power grid equipment more and more as average global temperatures rise \cite{us2017climate, chen2023projected, zhang2023increased, fischer2015anthropogenic, climategov_arctic}. They are key areas of focus when considering future impacts to the grid, which in turn have compounded effects to other critical infrastructures.

\subsection{TEMPERATURE TRENDS AFFECTING POWER GRID EQUIPMENT}
\subsubsection{EXTREME HEAT}
\begin{figure}
    \centering
    \includegraphics[width=\linewidth]{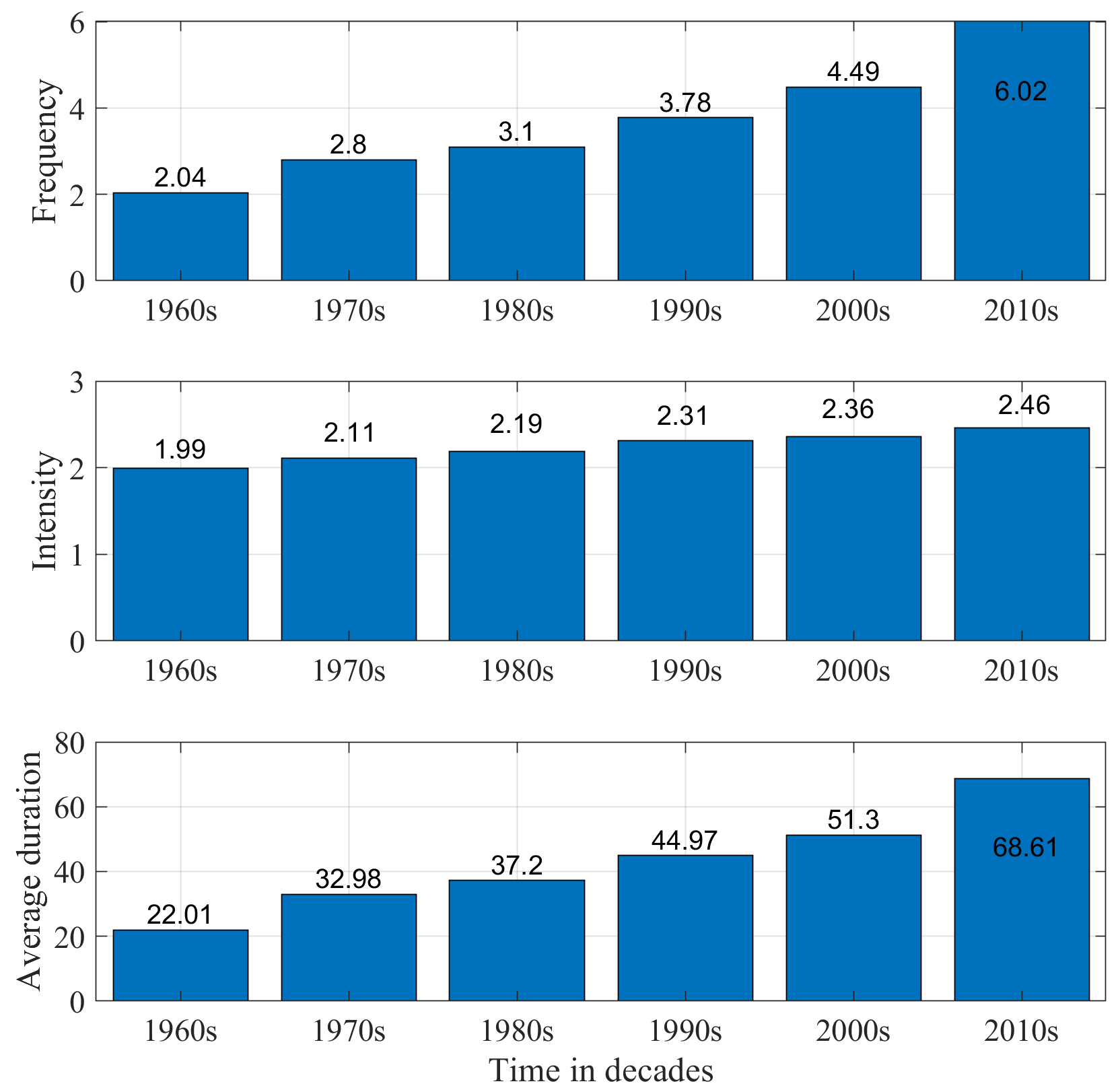}
    \caption{Heat wave characteristics in the U.S. over the past six decades \cite{heat_wave_stats_1}. Frequency is the average number of heat wave occurrences. Intensity is the average temperature (in $\degree$F) above local threshold during heat waves. Average duration is the average length of annual heat wave season in days.}
    \label{fig:heatwave_charac}
\end{figure}

Average temperatures have increased in recent decades across the globe, leading to an increased number of heat waves, a longer average duration per heat wave, and a longer heat wave season (Fig.~\ref{fig:heatwave_charac}). Heat domes contribute strongly to the most extreme heat waves \cite{zhang2023increased}.

% \begin{figure}
%     \begin{subfigure}{0.5\textwidth}
%         \centerline{\includegraphics[width=\linewidth]{images2/charac_1.png}}
%         \caption{{Heat wave occurrences and intensity in the United States by decade, 1961--2019.}}
%         % \vspace{3em}
%         \label{fig:heatwave_charac_1}
%     \end{subfigure}
%     % \vspace{5em}
%     \hspace{1em}
%     \begin{subfigure}{0.5\textwidth}
%         \centerline{\includegraphics[width=\linewidth]{images2/charac_2.png}}
%         \caption{{Heat wave season duration in the United States by decade, 1961--2019.}}% The time taken by the proposed index is almost negligible as it is a non-iterative index.}
%         \label{fig:heatwave_charac_2}
%     \end{subfigure}
%     \caption{{Heat wave characteristics in the U.S. over the past six decades \cite{heat_wave_stats_1}}}
%     \label{fig:heatwave_charac}
% \end{figure}

\begin{figure}
    \centering
    \includegraphics[width=\linewidth]{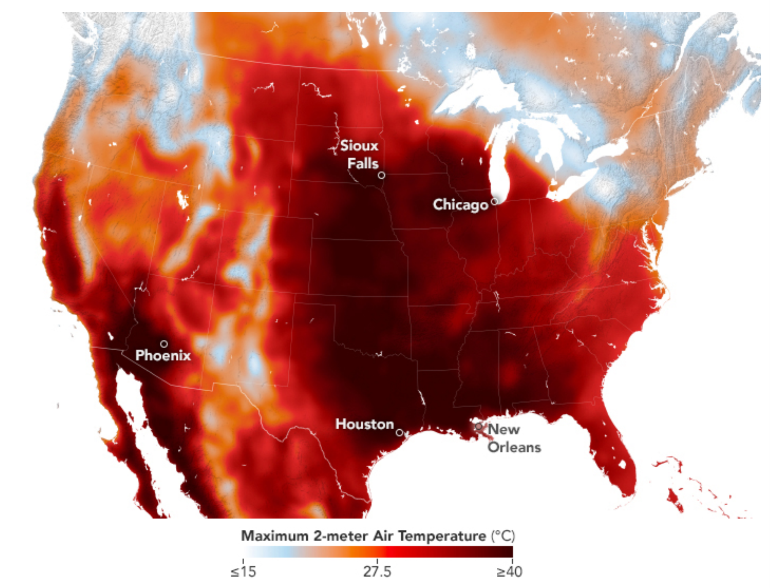}
    \caption{Temperature map of a heat dome over the U.S. on August 23, 2023 \cite{hifld_2}}
    \label{fig:dome_us}
\end{figure}

Heat domes are weather events in which atmospheric currents trap warm air against the Earth's surface. These conditions prevent cloud formation, allowing sunlight to further heat the air near the ground. Heat domes cause extremely high temperatures that can last for days \cite{heat_blast_1} and cover multiple states (Fig.~\ref{fig:dome_us}), significantly affecting operating conditions for T\&D equipment and simultaneously increasing cooling-related electric demand. These factors make the power grid extremely vulnerable during such conditions. 

\begin{figure}
    \centering
    \includegraphics[width=\linewidth]{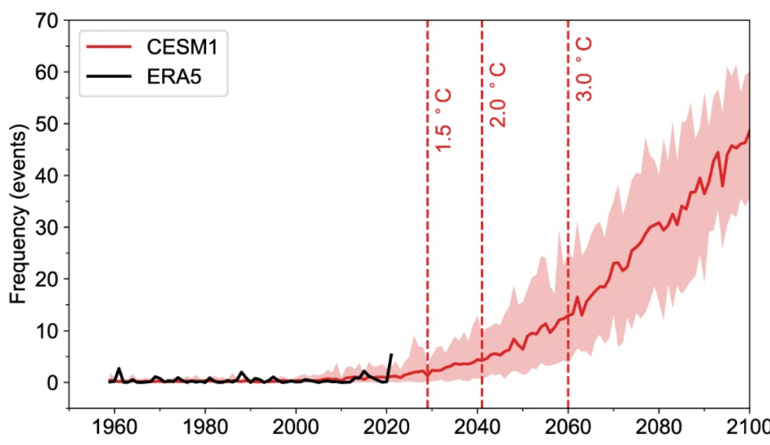}
    \caption{Simulated heat wave frequency over  Northwest America \cite{zhang2023increased}}
    \label{fig:heatwave_frequency}
\end{figure}

The most recent extreme heat dome was in June 2021 in the United States' Pacific Northwest region \cite{white2023unprecedented}. Historical and projected data suggest that high temperatures caused by heat dome-like events are increasing in intensity at a higher rate than the average global temperatures \cite{zhang2023increased}. One projection anticipates that the probability of extreme high-temperature events will grow at an increasing, non-linear rate as the global average temperature rises (Fig.~\ref{fig:heatwave_frequency}) \cite{fischer2015anthropogenic}.


\subsubsection{EXTREME COLD}
\begin{figure}
    \centering
    \includegraphics[width=\linewidth]{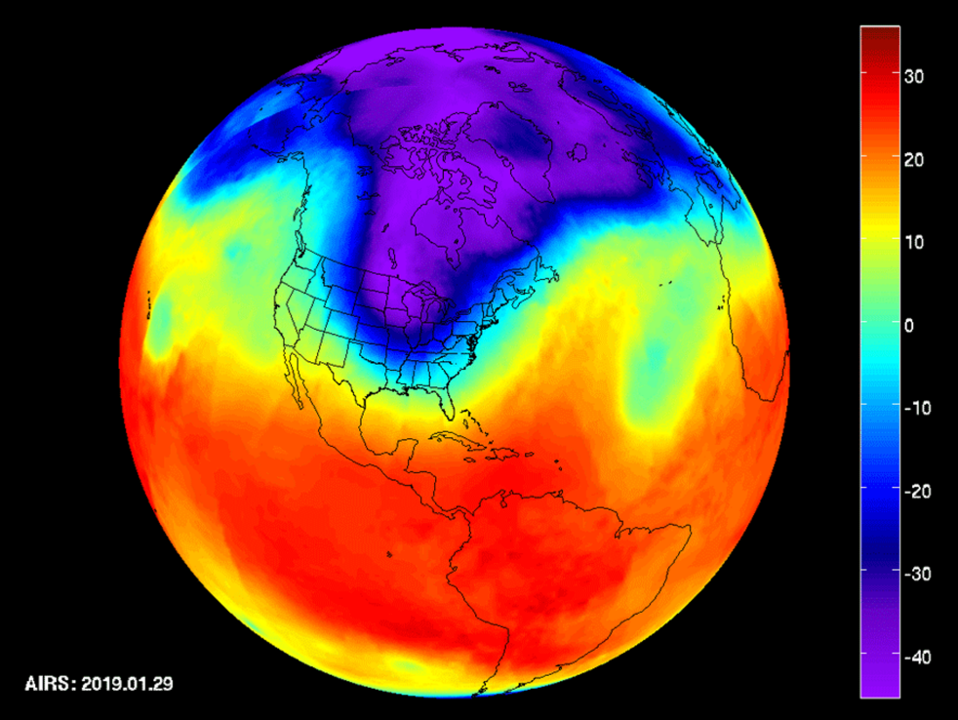}
    \caption{Polar vortex temperature heat map \cite{polar_vortex_1}.}
    \label{fig:polar_vertex1}
\end{figure}
\begin{figure}
    \centering
    \includegraphics[width=\linewidth]{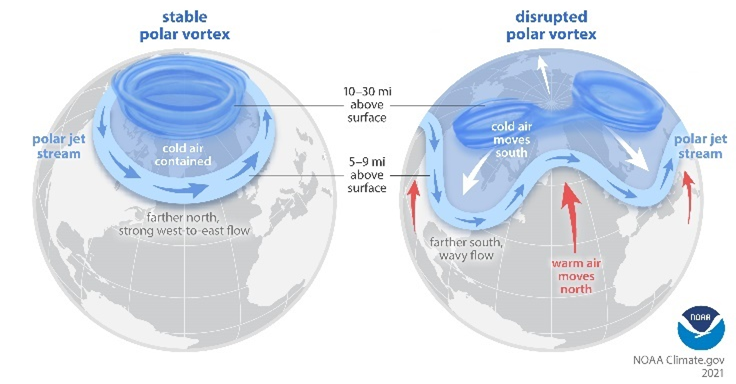}
    \caption{Stable and unstable polar vortices \cite{climategov_arctic}.}
    \label{fig:polar_vertex}
\end{figure}

The overall increase in global temperatures is causing extreme cold events as well as extreme heat events. Warmer temperatures disrupt polar vortices and cause them to reach as far as the southern states in the U.S. \cite{climategov_arctic}. The polar vortex is a region of winds that circulate from west to east in the stratosphere above the Arctic. When the vortex weakens, it can move south from the pole as warmer air moves north (Fig.~\ref{fig:polar_vertex1}) \cite{polar_vortex_1, climategov_arctic}. This pattern can send cold Arctic air into the U.S. as far south as Florida and Texas \cite{polar_vortex_1, polar_vortex_2} and result in freezing temperatures that last for days \cite{climategov_arctic}. Warmer temperatures across the Arctic cause more polar vortex disruptions, bringing these events to the U.S. with greater frequency and intensity \cite{climategov_arctic}. Fig.~\ref{fig:polar_vertex} shows the critical differences between a stable and disrupted polar vortex.

The extreme cold conditions caused by disrupted polar vortices can be accompanied by snow and ice storms in regions where these weather events are not regularly anticipated, challenging electric grid reliability even as heating-related electric demand reaches potentially unforeseen levels. The exposure of T\&D equipment to colder air temperatures is also a cause for concern.

The paper is focuses on the impact of the equipment from a electrical function and does not explicitly focus on structural integrity that can arise due to corrosion in heat domes (it is a combination of temperature and humidity). While structural integrity related failures or degradation can impact the electrical function, this paper is a review of findings in the open literature on the electrical functionality that gets impacted during extreme heat and extreme cold related events.

\subsection{KEY CONTRIBUTIONS}
The overarching contribution of this review paper is that it collates and combines a vast literature into a concise  yet comprehensive document that enables the reader to understand and reflect on the key impacts due to extreme temperature operation of the power grid. The key contributions of the paper are as follows:

\begin{enumerate}
    \item The paper provides a detailed review of acute failure modes, monitoring techniques, and mitigation approaches for all major and minor power grid equipment under extreme temperature operation. The power grid equipment spans across the transmission, distribution, and substation assets.
    \item This paper consolidates the failure modes in a comprehensive manner for various power grid assets due to extreme heat and extreme cold scenarios.
    \item The paper also consolidates the information from equipment standards to reflect the readiness of power grid standards to handle the extreme temperatures in the recent years. Furthermore, this paper also identifies the potential operation of power grid assets in non-compliance with standard specifications due to worsening climatic conditions such as heat wave in cold region, ice storm in a hot region, etc.
    \item The paper provides two kinds of failure maps: influence maps that show the inter-dependency between different equipment failures that are caused due to extreme temperatures; and cascading failure maps that provides the details of how failure of a single critical equipment can cascade into failure of other system equipment.
\end{enumerate}

\subsection{ORGANIZATION OF THE PAPER} The paper is organized as follows. Section~\ref{sec:events_here} provides illustrative examples of extreme hot and cold weather events in the U.S. that caused failures of T\&D equipment. Section~\ref{sec:scoping_study} describes the methods of this study for choosing which classes of T\&D equipment to evaluate and presents the results of that assessment. Sections~\ref{sec:TS}, \ref{sec:DS}, and \ref{sec:SS} discuss transmission, distribution, and substation equipment, respectively. For each class of equipment within those categories, these sections describe: 1)~Impact of Extreme Temperature Operations, 2)~methods to measure impacts of extreme weather, and 3)~ methods to mitigate these impacts. Note that in cases where equipment fits into more than one category, this paper places it in the section where the impact of failure is most prevalent. For example, fuses are in the distribution section; transformers are in the substation section; insulators and bushings are in the transmission section; etc. Section~\ref{sec:ST} summarizes standards related to extreme temperatures for different power grid equipment. Section~\ref{sec:discussion} presents a consolidated discussion of equipment failure modes during extreme temperatures, how current and future  temperatures may violate standards for T\&D equipment, and how the interdependency of grid equipment may lead to cascading failures during extreme temperatures. Section~\ref{sec:conclusion} presents a brief background, conclusion, and future work.

% including transmission, distribution, and substation equipment. A dedicated section has also been provided to summarize the standards for the different T \& D grid equipment from the perspective of how applicable the current recommended standards are for extreme temperatures. For the transmission, distribution, and substation equipment sections, some equipment types go across the ratings and subsystems, such as transformers, conductors, circuit breakers, fuses, and insulators. For these equipment types there is a single subsection and is placed in the section where the impact is most prevalent or the section which has maximum types of these equipment. For example, the fuses subsection is in the distribution system; the transformer are in the substation section; insulators and bushings are in the transmission section, etc.

% The discussion section summarizes the different subsections and provides insights into the direct and indirect failure impacts of extreme temperatures. This section also discusses the failure influence diagrams and the cascading impact of these failures.

% The critical aspects of the paper are summarized in the conclusion section.

\section{CASE STUDIES: TEMPERATURE-RELATED POWER OUTAGES} \label{sec:events_here}
\subsection{Hot temperature-related events}
Heat waves have caused many power outages in the U.S. Some of the most prominent heat-related outages, starting from 1996, are overviewed below. 

\subsubsection{August 10, 1996, West Coast} Hot temperatures caused two transmission lines in the Pacific Northwest with light loads to sag and come into contact with trees. A third transmission line with a heavier load also sagged into a tree due to heat expansion in the conductor. The third fault caused other transmission lines to overload and trip offline \cite{muir2004final}. In addition, the 13 turbines of the McNary Dam went offline because of incorrect relay tripping \cite{event_1_kg}. These events caused power outages throughout the West Coast down to Southern California \cite{event_1_kg}.

\subsubsection{June 10--11, 1999, Ohio, Michigan, Indiana} High temperatures that lingered for multiple days in Ohio, Michigan, and Indiana resulted in heavy transmission loading, exacerbated by equipment outages around Northwest Ohio. American Electric Power (AEP) transferred 4,270 MW of power to neighboring utilities when low voltage conditions warranted transmission loading relief. 175 MW from AEP to First Energy and 308 MW from Virginia Power to ComEd had to be curtailed. A 345 kV circuit and a 345/138 kV transformer went out because of a trip from a fault and the misoperation of a breaker.

On June 11, a transmission loading relief was called for the Dumont 765/345 kV transformer due to a contingency loss of another transformer. This caused around an additional 800 MW of power to be shed \cite{event_15}.

\subsubsection{July 1999, Mid-Atlantic} The transmission system in the Mid-Atlantic region was under high load due to sustained high temperatures and an abnormally high-temperature humidity index. Even with management methods such as voluntarily dropping customer load and reducing the voltage by 5\%, the voltage dropped further. The Pennsylvania-New Jersey-Maryland Interconnection cut 1200 MW of transactions to reduce 200 MW of flow through the region. Sub-regions underwent forced outages to keep electricity distribution reliable; the utility connectivity reduced its load by about 140 MW. Multiple transformer failures at a substation in New Jersey caused the loss of 600 MW of power \cite{event_15}.

\subsection{Cold temperature-related events} Equipment failures and power outages in winter months have frequently been caused by polar vortices' severe cold and icing conditions. Some of the most significant disturbances and their causes are reviewed below.

\subsubsection{December 4--5, 2002, North Carolina} An ice storm in North Carolina coated the state in 0.75--1 inch of ice. This ice built up on power lines, power-line-carrying equipment, and trees. Many trees still had their leaves, because the storm occurred early in the winter. This extra surface area allowed more ice to accumulate on the trees, which caused a great number of them to fall onto power lines. North Carolina suffered around 362,000 outages, and its three principal utilities experienced many equipment failures. Across these utilities, 4,523 poles, 5,587 cross-arms, and 40,760 insulators were damaged. Additionally, 4,503 transformers and 116,048 fuses were replaced \cite{event_40}.

\subsubsection{February 1--5, 2011, Southwest} The power outages during this cold weather event were primarily due to generation failure and a low natural gas supply. However, two transmission failures were reported, and they caused a loss of 89 MW of power. Additionally, cold grease in a circuit breaker slowed its breaking, which caused six generators to trip offline \cite{event_41}.

\subsubsection{October 29--30, 2011, Northeast} A snowstorm in the Northeast caused significant power outages that lasted up to a week. In total, 74 transmission lines had outages, six of which were caused by direct snow and ice accumulation on the equipment. Transmission failures caused forty-four transmission substations to go offline as well, and 130,000 customers across six states lost power. Trees that fell because of snow and ice build-up also brought down or shortened many distribution lines. Many more distribution lines were damaged than transmission lines because less clearance was kept between lower-voltage lines and trees.

Additionally, a 115 kV power line was taken offline for five hours because a relay was tripped incorrectly, although no customers were affected. Another 115 kV line was taken out of service when a circuit breaker failed to operate. The circuit breaker was stuck and could not isolate a faulted line, causing 4,900 consumers to lose power. An additional 3,800 consumers lost power because of unknown failures that could result from trees touching lines or icing across insulators \cite{event_42}.

\subsubsection{February 11--20 2021, Texas} In 2021, an unusually long and cold polar vortex hit Texas ("the Great Texas Freeze"). The cold temperatures caused icing to form on wind turbine blades and in their gearboxes, and there were issues with natural gas supply to electrical generation plants. While energy demand increased as customers turned their heaters up, large generators started to go offline unexpectedly. Weather-related outages increased from 15,000~ MW to 30,000~MW in 24 hours, and transmission and substation outages caused 1900~MW of solar and wind generation to go offline. The frequency began to drop at the end of February 14 because the load was reaching the maximum capacity, and on February 15, generators began shedding load. At one point, the frequency dropped to a near-critical point but recovered before a complete blackout occurred \cite{event_43}. Overall, 111 people lost their lives, 4.5 million homes lost power, and \$130 billion of losses and property damage occurred as a result of the event \cite{event_44}.

While these events do not necessarily highlight only T\&D equipment failures, they involves lot of impacts on the power grid in general, however, this paper is an attempt to review the impacts of extreme heat and extreme cold related events on the operation of the various T\&D grid equipment.

\section{SCOPING STUDY}
\label{sec:scoping_study}
The power grid comprises many kinds of equipment and systems, including legacy devices as well as new upgrades and expansions. To determine which equipment is to be evaluated, this study analyzes equipment based on two criteria: its importance in grid operations and its vulnerability to extreme temperatures. The equipment that are scoped for a deeper literature review are identified.

While the scoping study mainly looked at the electrical behavior, there are other known vulnerabilities that directly impact the structural integrity of various equipment that can also lead to significant damage to the T\&D grid infrastructure. However a followup study and its findings are planned to be reported in a future publication. The scope would be too large to address in a single publication.

Tab.~\ref{tab:my_table} overviews different classes of transmission, distribution, and substation equipment and notes which types are in scope for this paper. The column ``In Scope (Yes/No)'' specifies if an equipment is considered in this paper. The decision ``Yes'' is selected in two scenarios. These scenarios are as follows: scenario 1 - either of the columns ``Importance During Grid Events'' or ``Vulnerability to Extreme Temperature'' is a ``high'', scenario 2 - when both of the columns ``Importance During Grid Events'' and ``Vulnerability to Extreme Temperature'' are moderate. In any other combination of scenarios, the scope selected is a ``No''. For example, insulators (high importance) are in scope, but vibration dampers (low importance, minimal temperature vulnerability) are not.

% \onecolumn

\begin{table*}
\centering
\caption{Scoping of transmission, distribution, and substation equipment based on importance and vulnerability to extreme temperatures.}
\label{tab:my_table}
\begin{tabular}{|>{\centering\arraybackslash}p{0.04\linewidth}|>{\centering\arraybackslash}p{0.17\linewidth}|>{\centering\arraybackslash}p{0.23\linewidth}|>{\centering\arraybackslash}p{0.12\linewidth}|>{\centering\arraybackslash}p{0.09\linewidth}|>{\centering\arraybackslash}p{0.1\linewidth}|>{\centering\arraybackslash}p{0.07\linewidth}|} \hline        
\textbf{S.No} & \textbf{Equipment} & \textbf{Functionality} & \textbf{Category} & \textbf{Importance During Grid Events} & \textbf{Vulnerability to Extreme Temperature} & \textbf{In Scope (Yes/No)}\\ \hline 
1 & vibration dampers & prevent mechanical vibrations & performance & low & minimal & no \\ \hline 
2 & insulators & insulate conductors from support structures & protection & high & minimal & yes \\ \hline 
3 & arcing horns & flashover protection & protection & moderate & null & no \\ \hline 
4 & optical ground wire & lightning protection and telecommunication & communication & moderate & null & no \\ \hline 
5 & conductor & power transmission & operation & high & high & yes \\ \hline 
6 & corona ring & avoid corona discharge & performance & moderate & moderate & yes \\ \hline 
7 & jumper wire & connector between conductors & performance & high & null & yes \\ \hline 
8 & spacers & spacing between conductors & protection & moderate & null & no \\ \hline 
9 & transformers & measurement and conversion of voltage levels & operation & high & high & yes \\ \hline 
10 & riser & connect conduit from ground to overhead lines via potheads & operation & high & null & yes \\ \hline 
11 & circuit breakers (air, oil, SF$_6$, vacuum) & protection & protection and control & high & high & yes \\ \hline 
12 & batteries & energy storage & storage & high & high & yes \\ \hline 
13 & bus support insulators/bushings & isolate bus bar switches and support structures & protection & high & minimal & yes \\ \hline 
14 & capacitor bank & control voltage level & regulation & high & high & yes \\ \hline 
15 & circuit switchers & protection & protection & moderate & moderate & yes \\ \hline 
16 & voltage regulator & voltage regulation & regulation & high & high & yes \\ \hline 
17 & distribution busbars & steel structure array of switches used to route power out of a substation & operation & moderate & moderate & yes \\ \hline 
18 & frequency changers & changes power of an alternating current system from one frequency to one or more different frequencies & regulation & high & high & yes \\ \hline 
19 & grounding resistors & limits transient currents & protection & high & high & yes \\ \hline 
20 & fuses & protection & protection & high & high & yes \\ \hline 
21 & lightning arresters & protection & protection & high & high & yes \\ \hline 
22 & relays & Protection & protection and control & high & high & yes \\ \hline 
23 & shunt reactors & control voltage level & regulation & high & high & yes \\ \hline 
24 & suspension insulators & protection & protection & high & moderate & yes \\ \hline 
25 & distribution transformers & conversion & operation & high & high & yes \\ \hline 
26 & underground cables & power transmission & operation & high & high & yes \\ \hline 
27 & Overhead line insulators (distribution and medium voltages) & protection & protection & high & high & yes \\ \hline 
28 & power line carrier equipment & communication & communication & high & high & yes \\ \hline 
29 & CTs & measurement and control & measurement & high & high & yes \\ \hline 
\end{tabular}

\end{table*}

% \twocolumn

\begin{table*}
\centering
\ContinuedFloat
\caption{Scoping of transmission, distribution, and substation equipment based on criticality and vulnerability to extreme temperatures.}
\label{tab:my_table}
\begin{tabular}{|>{\centering\arraybackslash}p{0.04\linewidth}|>{\centering\arraybackslash}p{0.17\linewidth}|>{\centering\arraybackslash}p{0.23\linewidth}|>{\centering\arraybackslash}p{0.12\linewidth}|>{\centering\arraybackslash}p{0.09\linewidth}|>{\centering\arraybackslash}p{0.1\linewidth}|>{\centering\arraybackslash}p{0.07\linewidth}|} \hline       
\textbf{S.No} & \textbf{Equipment} & \textbf{Functionality} & \textbf{Category} & \textbf{Importance during grid events}  & \textbf{Vulnerability to extreme temperature} & \textbf{Scoped-in (Yes/No)} \\ \hline 
30 & PTs & measurement and control & measurement & high & high & yes \\ \hline 
31 & tap changers & conversion & regulation & moderate & high & yes \\ \hline 
32 & static VAR compensators and VAR devices & voltage control & regulation & moderate & high & yes \\ \hline 
33 & flexible AC transmission devices (STATCOM, SSSC, GUPFC, IPFC) & control and regulation & regulation & high & high & yes \\ \hline 
34 & switchgear panels & protection & protection & high & high & yes \\ \hline 
35 & fuses & protection & protection & high & high & yes \\ \hline 
36 & high-voltage direct current converters & conversion and power transmission & operation & high & high & yes \\ \hline 
37 & harmonic filters & power quality & operation & high & high & yes \\ \hline 
38 & disconnect switches and isolators & protection & control & high & high & yes \\ \hline 
% 39 & Capacitor Bank & Control voltage level & Regulation & Moderate & High & Yes \\ \hline 
% 40 & batteries & Energy storage & Storage & High & High & Yes \\ \hline 
39 & dynamic line rating system & calculate line capacity & measurement & moderate & minimal & no \\ \hline 

\end{tabular}

\end{table*}

\section{TRANSMISSION GRID EQUIPMENT IMPACTS}
\label{sec:TS}
\subsection{Insulators and Bushings}
\begin{figure}
    \centering
    \includegraphics[width=\linewidth]{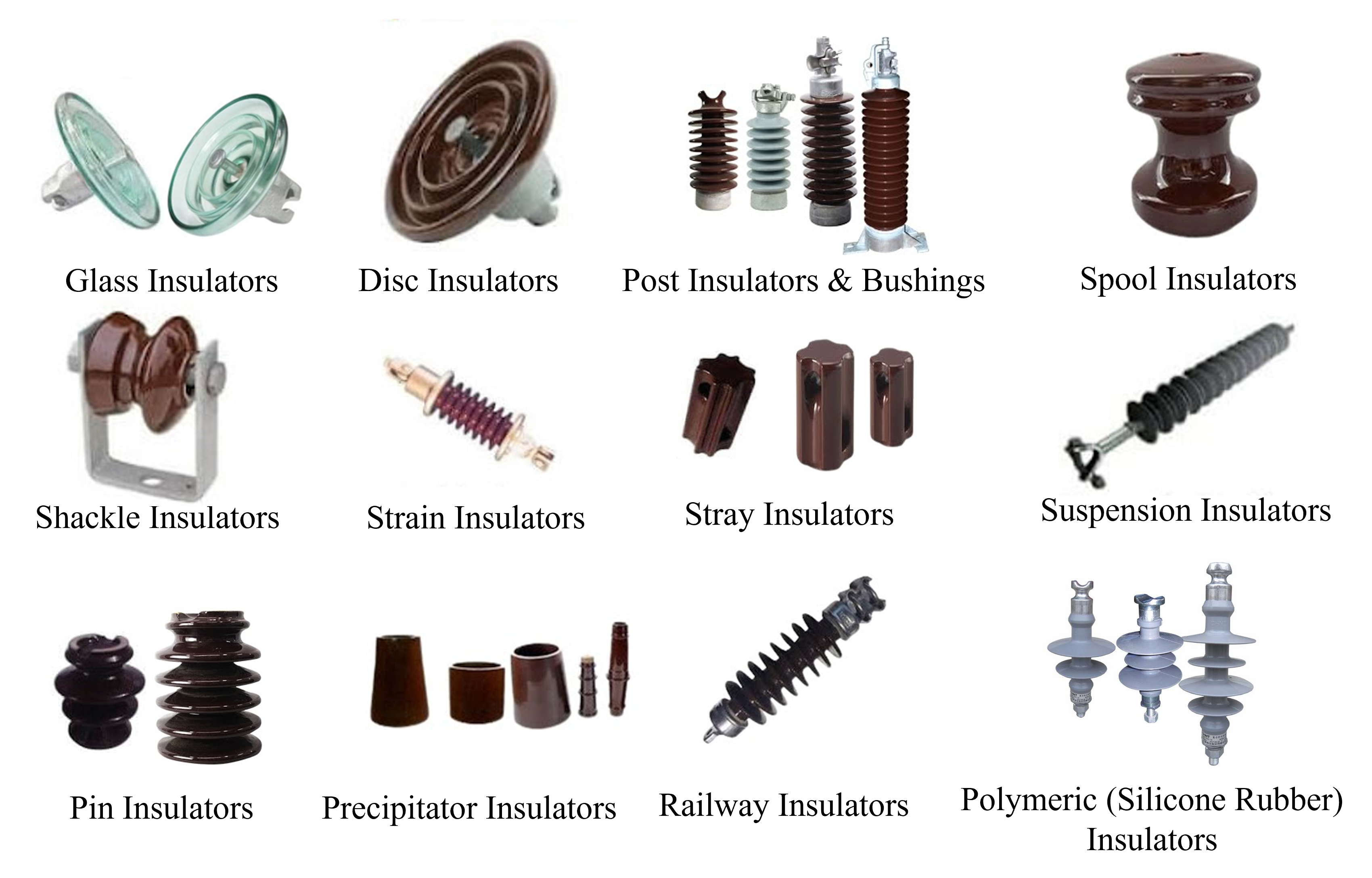}
    \caption{Types of insulators in T\&D systems}
    \label{fig:insulator_types}
\end{figure}

% \textcolor{red}{Alok}
\label{subsec: insulators_ts}
Insulators prevent the flow of electric power from conductors to supporting structures. When insulators fail, large currents can flow to ground, activating protection systems (circuit breakers and fuses). Insulators are classified into several types depending on location and function, as shown in Fig.~\ref{fig:insulator_types}. However, the materials and their performance with respect to environmental conditions is expected to be similar \cite{insulator_1,insulator_2}.

Most low-voltage insulators (typically in distribution systems) are primarily made of polymers or ceramics like glass and porcelain \cite{insulator_2}, but these insulating materials are joined with other components. Overhead line insulators may include rubber, cement, iron couplings, and other materials \cite{insulator_3,insulator_4, insulator_5}. Extreme heat can compromise the joints between materials. Literature reports extreme heat under conditions of wildfires, arc flash, or flashovers. Ambient conditions under heat domes and cycling of extreme temperatures can also trigger failures of insulator structures \cite{insulator_5}.
Insulators used on busbars inside of switchgear and around auxiliary equipment are more prone to failures due to extreme heat than insulators on overhead lines. These insulators include standoff insulators, epoxy powder coating, heat-shrink tubes, and insulating films \cite{insulator_6}. Extreme heat coupled with high load currents can cause temperatures to rise beyond safe operating ranges, resulting in insulator failures \cite{insulator_7}. The thermal monitoring systems inside switchgear help to prevent these high-temperature-related failures, as described in subsection \ref{subsec:switchgear_ss}. 

 \textbf{Impact of Extreme Temperature Operations:}
While high temperatures can affect the mechanical components of insulators, which can lead to fatigue and failure, extreme cold conditions like polar vortices and winter storms can cause severe icing and snow deposition that can compromise their electrical function \cite{insulator_8}. Icing and snow deposition on outdoor insulators can lead to increased risk of leakage, arcing, and flashovers that can cause temporary and permanent damage to the insulators \cite{insulator_9}.  

\begin{figure}%[H]
    \centering
    \includegraphics[width=\linewidth]{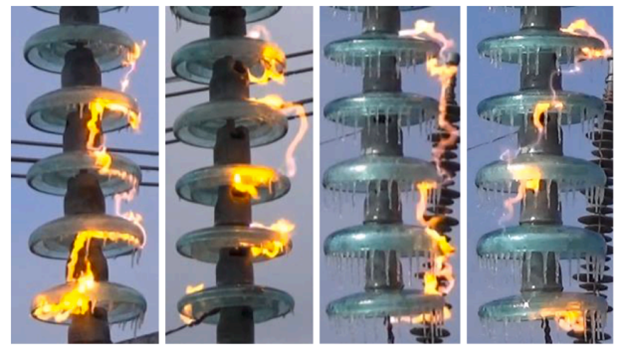}
    \caption{Flashover on ice deposited insulators from \cite{insulator_14}.}
    \label{fig:ice_flashover}
\end{figure}

Porcelain/Ceramic Insulators: Extreme heat can cause issues with the parts used for coupling to fixtures or conductors. These insulators can also have issues with degraded performance, flashovers, or arcing due to snow or ice accumulation in extreme cold conditions during polar vortices and winter storms. In fig.~\ref{fig:ice_flashover}, ice has covered the entire surface of the insulator and created an easy path for current, leading to flashover \cite{insulator_8,insulator_9,insulator_10}.

Polymeric/Silicone Rubber (SR) Insulators: Polymeric insulating materials have better dielectric properties than ceramics, so polymeric insulators are more compact. However, unlike ceramic insulators, polymeric insulators are highly vulnerable to damage from flashovers. This results in a higher risk from extreme cold and ice storms, because ice deposition that triggers a flashover can cause permanent failure of the insulator.

\begin{figure}%[H]
    \centering
    \includegraphics[width=\linewidth]{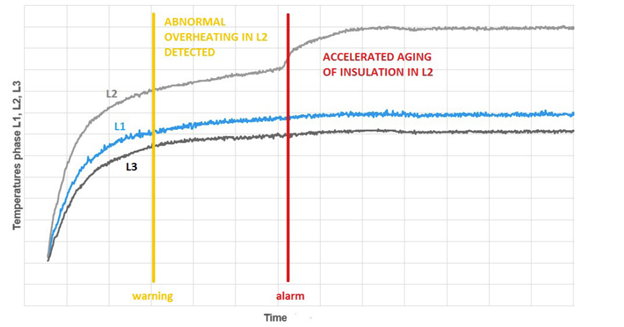}
    \caption{Monitoring cable insulation temperature from \cite{insulator_12}}
    \label{fig:cable_insulation}
\end{figure}

Cable Insulation: Cables and conductors in panels often have polymer-based flexible insulation that can fail in extreme heat (e.g., during heat domes). Figure~\ref{fig:cable_insulation} shows how increased cable temperature can lead to insulation breakdown, in turn causing temperature to rise further. Cable insulation failure can result in arc flash and equipment damage.

\textbf{Existing Methods to Measure the Impacts:}
Visual inspection is the main method of monitoring insulator failure. Large failures can result in a significant leakage current that can be detected by some relays, depending on the sensing thresholds.
There is discussion in the literature on power system insulator condition monitoring  \cite{insulator_13} that could prove to be useful for for both monitoring and preventive maintenance.

\textbf{Existing Methods to Mitigate the Impacts:}
Effective derating to consider the life and expected operating temperatures is a standard method to prevent failures during extreme heat. For managing extreme cold temperatures and severe icing, surface treatments like hydrophobic coatings that are typically used for switchgear panels and other power equipment could be explored for insulator and bushings \cite{insulator_10}. While there are no specific methods discussed in the literature, additional planning and maintenance practices can be adopted to account for the changing climate, which could include further derating insulators to ensure performance during extreme climate conditions. Reference \cite{insulator_11} describes one such method that could prove to be effective in predicting insulator failures under extreme heat and cold or severe ice. 

\subsection{Conductor}
\label{subsec: conductors_ts}
There are various types of power lines in the T\&D system, but they are essentially categorized as either overhead or underground power lines. These two types of lines are present in both transmission and distribution power grids.

Overhead lines are constructed by setting up conductors between poles. These poles can be built in any type of terrain, and overhead lines are easier to repair than underground lines \cite{LE_C}. However, overhead lines are vulnerable to extreme weather conditions because of their direct exposure to weather.

Underground lines are constructed by laying the conductors below the surface, protected by conduits. Underground lines are less susceptible to weather, but they are three to five times more costly than overhead lines \cite{NPR_C}.

Power lines are made of different conducting materials such as copper and aluminum reinforced with steel cores. The conductors come in different configurations such as single core or multicore. The material and structural configuration depends on the expected powerflow, losses, weather impact, etc. Copper is highly conductive but expensive; while aluminum is $60$\% of copper's conductivity, it is much more economical \cite{conductor_type_SN}. Most power lines today are aluminum wrapped around a steel core 
\cite{power_possibility_NREL_SN}.  Note that an increase in temperature decreases the conductivity of a conductor irrespective of the material, but reduction of conductivity varies for different materials. There are jumper cables used often to link ends of conductors that have similar properties as conductors but the mechanical joints for the jumper cables often need extra care to ensure their fitment and installation is accurate. 

\begin{figure}
    \centering
    \includegraphics[width=\linewidth]{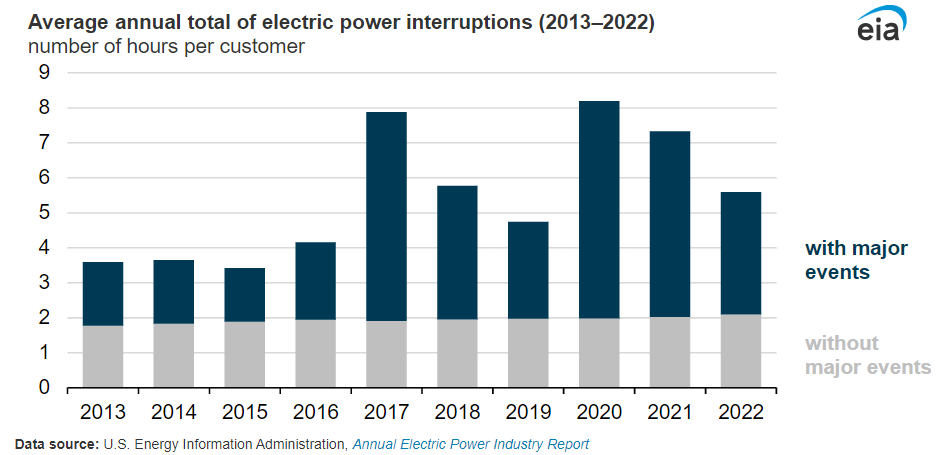}
    \caption{Average annual total power interruptions per customer from 2013 to 2022 from \cite{Electricity_Customers_SN}.}
    \label{fig:total_stats}
\end{figure}

\begin{figure}
    \centering
    \includegraphics[width=\linewidth]{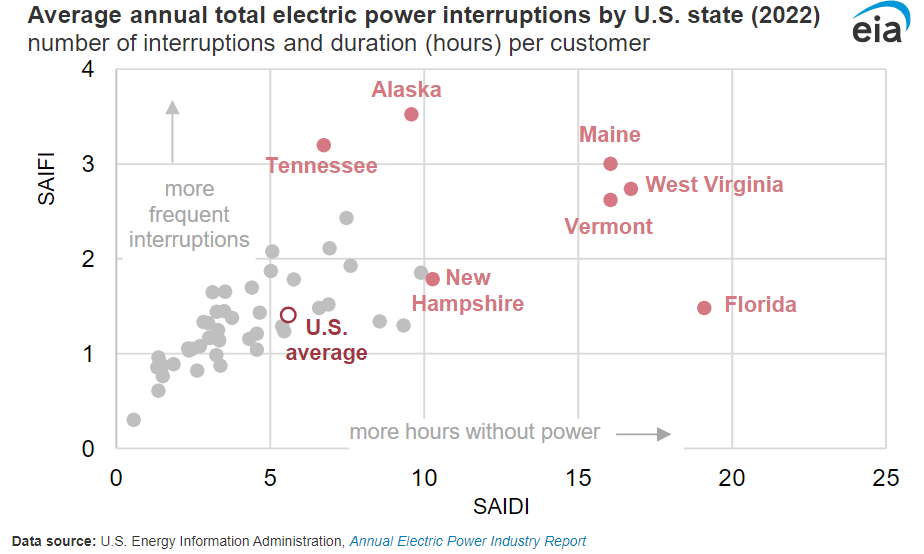}
    \caption{Average annual power interruption by U.S. states in 2022 from \cite{Electricity_Customers_SN}.}
    \label{fig:saifi}
\end{figure}

\textbf{Impact of Extreme Temperature Operations}: Recent temperature extremes have made it more difficult to ensure resilience for power lines.  In 2021, temperatures were so high in the Pacific Northwest that some power lines were reported to have melted \cite{NPR_C}.  Even underground cables can experience overheating when the loading in the power lines is underestimated, an effect which can be mitigated with forced cooling systems \cite{brown1978forced} but only at significant cost. Extreme cold has also caused disruptions in the winter \cite{time_c}. Fig.~\ref{fig:total_stats} reports that the average annual total interruptions increase by 2 hours after excluding major events, and it is reported that the primary reasons for major interruptions are due to weather, followed by vegetation and maintenance issues \cite{Electricity_Customers_SN}. Fig.~\ref{fig:saifi} shows the U.S. states that have more frequent and more longer power interruptions. The U.S. has had 18 weather-related disasters that exceeded $1$ billion dollars in damages in 2022, including the winter storm Elliott, which left Texas, Florida, Arkansas, Tennessee, West Virginia, and Wisconsin without power and caused four feet of snowfall in New York \cite{Electricity_Customers_SN}.

In extreme cold conditions, icing increases the weight of power lines and tree limbs. Trees can lean onto distribution lines, causing outages, and $1/2''$ of ice accumulation on power lines can add 500 pounds of extra weight \cite{entergy_c}. Freezing rain in combination with steady winds also results in a skipping-rope-like movement of power lines (galloping) that causes outages. Icing not only affects overhead lines but also (much more rarely) underground lines, because ice follows tree roots and builds up \cite{entergy_c}.

In addition these, extreme heal and cold operations make the jumper cables and their joints extremely vulnerable to become the weak links causing overall failure of the conductors. these can also become weak over the life of the conductor and the impact  severity is increased on aged infrastructure operating under extreme temperatures. 

\textbf{Existing Methods to Measure the Impacts}: Advanced weather monitoring tools are popular to measure or monitor the impacts of extreme weather on transmission lines \cite{advance_monitor_conductors}. Another solution is to directly monitor the transmission line temperature and its impact on dynamic line ratings \cite{lunalios_c}.  In addition to conductor temperature and line ratings, this real-time measurement technology can also monitor the conductor clearance, conductor sag, and line current \cite{lindsey_c}. Refs. \cite{wei2022temperature, lin2015field} also show how this temperature monitoring can be achieved based on Raman distributed optical fiber sensing technology.

\textbf{Existing Methods to Mitigate the Impacts}: There are several approaches to protecting power lines against extreme temperatures. One is more informed planning; risks from climate change and historical climate risk are calculated and then used to identify the changes in power line ratings and other properties such as sag, clearance, etc. \cite{hu2013impact, ESCI_c}. Another solution is to adopt underground cables, which are less vulnerable to extreme temperatures than overhead lines. However, it is not practical to have very high voltage cables underground because they require additional cooling solutions, making them very costly. Despite of these challenges, underground cables still significantly help to prevent outages due to extreme temperature at the distribution level. Underground cables are also used at transmission level, but it is primarily done to navigate terrain and not common. Other solutions include rating the assets for different temperatures, continuous temperature monitoring, and guarding against fires close to the power lines \cite{NG_c}.

\subsection{Corona Ring}
Corona rings are made of metal and used in high-voltage transmission equipment to distribute the electric field gradient and lower its value below the corona threshold (i.e., the threshold above which electric current on a surface ionizes the surrounding air). 

\textbf{Impact of Extreme Temperature Operations:}
Temperature affects corona rings in various ways. The corona onset voltage is the voltage at which corona discharge begins. Ref. cite{zeng2023negative, huang2021research} presents the dependency of corona onset voltage and temperature; as temperature increases, the corona onset voltage decreases. This is because rise in temperature increases the effective ionization coefficient. Ref. \cite{yan2016experimental} discusses the impact of high temperatures on additional corona current due to factors such as gas molecule ionization, free electron detachment, and thermionic electron production. Increase in corona discharge has a number of negative impacts on the grid such as power loss, equipment damage, electromagnetic interference, insulation degradation, etc. Ref. \cite{corona_ring_1} also shows that not only do high temperatures degrade corona rings, but low temperatures cause them to operate below optimal efficiency. For this reason, corona rings are typically operated at ambient air or oil at room temperature (around $20\degree$C) even though they can be operated at a wide range of temperatures.

\textbf{Existing Methods to Measure the Impacts:} The temperature of corona rings is primarily measured using infrared thermography to detect the heat emitted from the ring as a non-contact method \cite{ma2022experimental}.

\textbf{Existing Methods to Mitigate the Impacts:} Mitigation is a challenge due to outdoor nature and placement on the grid. The mitigation could come from design of corona rings to handle extreme weather conditions.
% \textcolor{red}{Kishan}

\section{DISTRIBUTION GRID EQUIPMENT IMPACTS}
\label{sec:DS}
% \subsection{Distribution Transformer}

% \subsection{Underground Cable}

%\subsection{Overhead Line Insulator}
%\textcolor{red}{Alok}
%\label{subsec:insulators_ds}
%\input{tex_files/insulators_ds}

% \subsection{Power Line Carrier Equipment}
% \textcolor{red}{Alok}

\subsection{Fuses}
\label{subsec:fuses_ds}
% \textcolor{red}{Alok}

Fuses are important protection components in distribution systems that carry electric power and are often classified as primary grid equipment. Reference \cite{fuse_1} provides some fundamental information on parts and operation of fuses. The fuses operate to isolate faulty parts of the system very quickly when the fault currents are very large. Fuses are designed such that when the large amount of fault current flows through them, the increase in the $I^2 R$ losses cause excessive heating of the fuse element and in turn disconnects the circuit by blowing up. Fuse operation is heavily influenced by temperatures and extreme ambient temperatures can alter the normal operation of a fuse \cite{fuse_2,fuse_3, fuse_4}.

\begin{figure}
    \centering
    \includegraphics[width=\linewidth]{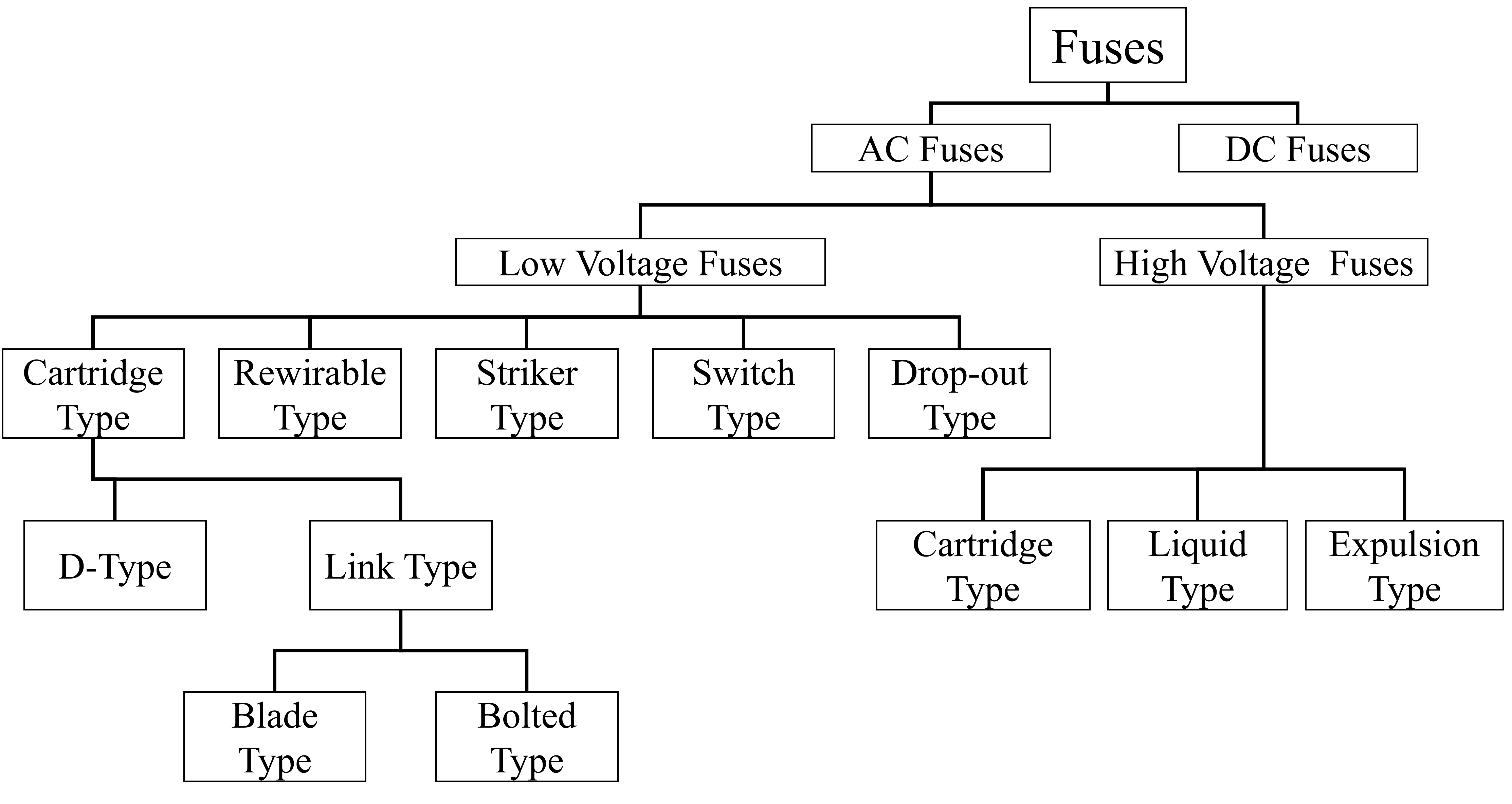}
    \caption{Types of fuses in T\&D systems from \cite{fuse_9}}
    \label{fig:fuse_types}
\end{figure}

Typical lifespan of a fuse is 20--30 years \cite{fuse_8}; performance degradation occurs over that lifespan. The key failure mode of the fuse is exposure to hard environmental conditions \cite{fuse_10}. 
There are different types of fuses in power systems, as shown in Fig. \ref{fig:fuse_types}.

\begin{figure}
    \centering
    \includegraphics[width=\linewidth]{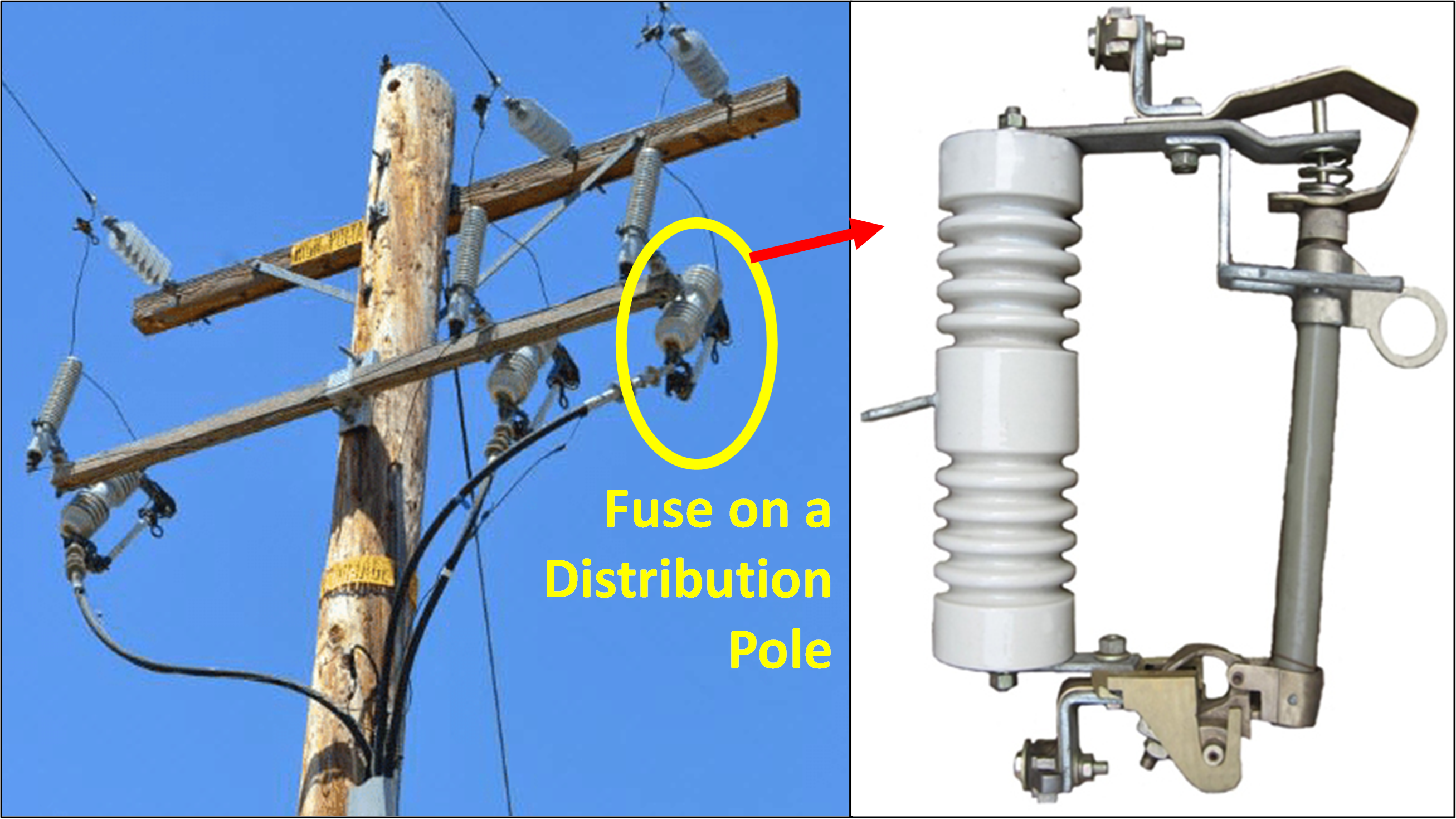}
    \caption{Typical fuse on a pole exposed to climate changes from \cite{fuse_11}}
    \label{fig:fuse_pole}
\end{figure}

\textbf{Impact of Extreme Temperature Operations}:
Operating a fuse in extreme heat (as in heat domes) or extreme cold conditions (polar vortices or icing) can lead to faster temperature cycling, causing the expansion and contraction of the fuse element that is the most important part of the fuse. This repeated expansion and contraction can result is failure of the fuse \cite{fuse_4,fuse_5}. The other parts of the fuse that are sensitive to extreme temperatures are the contact elements and terminals. The parts inside the fuse also tend to degrade and are the other leading failure modes in fuses.
Another common component in a fuse construction is the fuse enclosure, which is exposed to ambient operating conditions including heat domes, polar vortices, and icing. This enclosure has a limited function (simply to contain the fuse internals) but can transfer the ambient temperature to the fuse components. Extreme ambient conditions can impact the overall functioning because the enclosure cannot prevent ambient temperatures from affecting the internal fuse temperatures. Icing can cause mis-operation due to the malfunction of fuse cutout.
Heat domes and polar vortices are often short-duration events (a few consecutive days). During these short durations, derating the power system is important, but in the process, various controls and protection components like fuses and circuit breakers will need additional coordination of settings to prevent unwanted tripping. Such tripping could lead to isolation of parts of the system that were not planned or intended.
Fig.~\ref{fig:fuse_pole} shows a typical fuse on a distribution pole exposed to weather. %\textcolor{red}{From Alok (needs more editing): the risk of open fuses on unused circuits/ paths is a risk that we want to identify. the basic thing is, in distribution circuits, or in large substations, there will be a N-1 path for reroute power in case of failure of equipment.. but on the back-up route, if a fuse is already blown due to climate, that would not be known before}

The failure of a fuse mainly results in an open circuit and its impact on current-carrying circuits is loss of power to downstream equipment or loads. The failure of a fuse on a back-up circuit means the backup may not be functional, causing serious reliability issues.

\textbf{Existing Methods to Measure the Impacts}: Unlike other component failures, the impacts of a blown fuse are not damaging to the system components. Therefore, the key methods to identify if a fuse has failed is usually visual inspection. There are some available products for low-voltage fuses, and examples are shared in references \cite{fuse_6} and \cite{fuse_7}. A similar method could be used for medium- and high-voltage fuses.

\textbf{Existing Methods to Mitigate the Impacts}: During heat domes, the temperature rise and the heat in the ambient conditions affects the performance of the fuse and the fuse element. Therefore derating the fuse in these extremely hot conditions is a standard practice \cite{fuse_5}. There is, however, a need to coordinate the fuse derating with the rest of the protection system. There is no literature that directly reports how the coordination is adopted in real world.

\subsection{Capacitor Bank}
\label{subsec:cap_bank_ss}
% \textcolor{red}{done}

Capacitor banks help with providing reactive power support to the grid to increase the maximum power transfer capability of the system. Generally, capacitor banks are a group of capacitor units connected in series and parallel circuits to achieve desired voltage and VAR rating, respectively. When a capacitor unit fails, it causes unbalance and results in tripping of capacitor bank. Without a protection trip, the capacitor units in the bank would experience a cascading failure \cite{schaefer2014minimizing}.

Excess heat accumulation is one of the common causes of a capacitor bank failure. Therefore, the internal temperature is used as the physical quantity to identify failure of a capacitor bank. This internal temperature is function of both ambient temperature and the radiant heat from the nearby equipment \cite{natarajan2018power}. Thus, ambient temperature not only helps with monitoring of capacitor bank failure but also effects the life of capacitor bank \cite{natarajan2018power, al2018impact}.

% Capacitor banks are critical substation assets that play a vital role in providing reactive power support, thereby increasing the power system capacity. High-voltage capacitor banks are constructed as single-wye, double-wye, or H-bridge configurations and can be grounded or ungrounded. Capacitor banks consist of a number of single-phase capacitor units connected in series and parallel to achieve the desired voltage and VAR rating. The capacitor units can be externally or internally fused, fuseless, or unfused. When the unbalance resulting from unit or element failures becomes too high, the capacitor bank needs to be taken out of service by the protection system before the resulting unit overvoltages lead to a cascading failure and the faulty units must be replaced \cite{schaefer2014minimizing}.

% Excessive heat buildup is one of the common causes of capacitor failure. Generally thermal imagers are used to monitor the internal temperature and when the internal temperature increases, it is an indication that capacitor may fail. This excessive heat can be a combination of both ambient temperature and radiated heat from adjacent equipment \cite{natarajan2018power}.

% Ambient temperature also impacts the life of capacitor \cite{natarajan2018power}.

% \begin{figure}[H]
%     \centering
%     \includegraphics[width=\linewidth]{images2/p5.png}
%     \caption{temp}
%     \label{fig:enter-label}
% \end{figure}

Another indirect reason for capacitor bank failures is because of load growth on an extremely hot day. The temperature rise of a capacitor is a function of the total internal watts dissipated by the windings. Initially, the watts are due to fundamental current (squared) and equivalent series resistance of the windings (ESR). Thus, the geometry of the capacitor windings plays a major role. 

%Additional heating will be produced when harmonic currents flow into the capacitor and when capacitor terminal voltage increases (due to system voltage fluctuations, reactor boosting effects, or due to higher current flow). Voltage transients sustained over-voltage, transient currents, sustained over current, and ambient air temperature all may affect the performance and life expectancy of power factor capacitors.

\begin{figure}
    \centering
    \includegraphics[width=\linewidth]{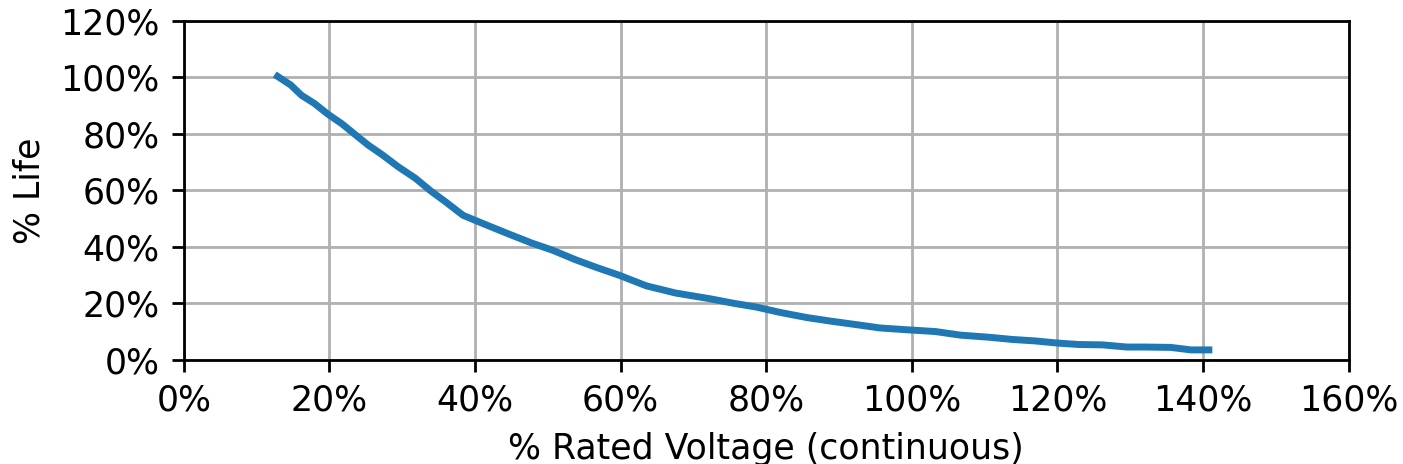}
    \caption{Impact of overvoltage (function of loading conditions) on the life expectancy of capacitor from \cite{Capacitor_Problems_SN}} 
    %\href{https://electrical-engineering-portal.com/capacitor-bank-catastrophic-explosion}{link}.}
    \label{fig:cap_voltage}
\end{figure}

\begin{figure}
    \centering
    \includegraphics[width=\linewidth]{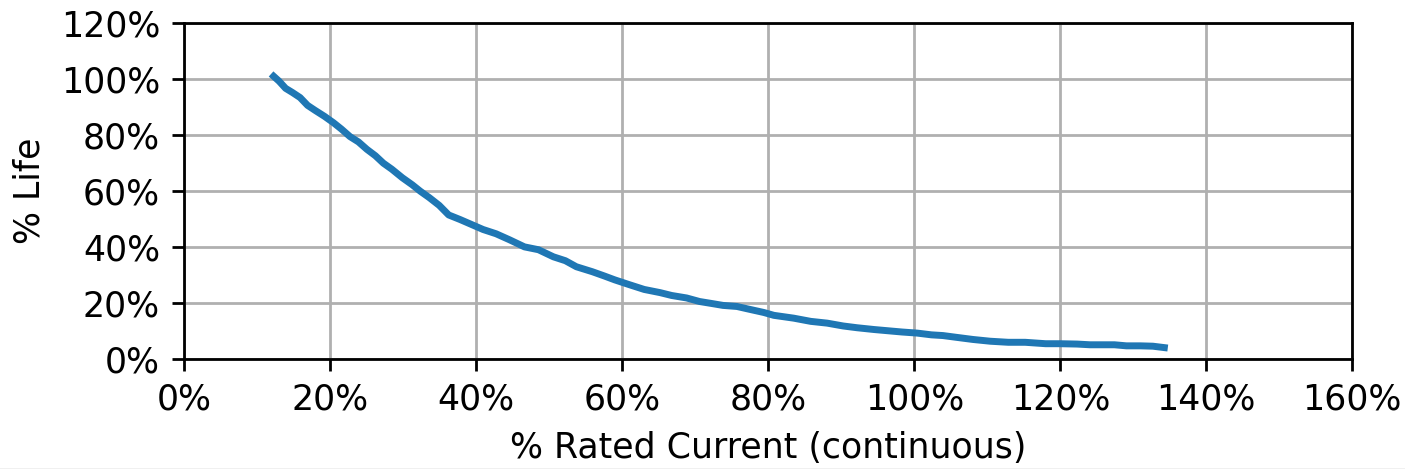}
    \caption{Impact of over current (function of loading conditions) on the life expectancy of capacitor from \cite{Capacitor_Problems_SN}} %\href{https://electrical-engineering-portal.com/capacitor-bank-catastrophic-explosion}{link}.}
    \label{fig:cap_current}
\end{figure}

\begin{figure}
    \centering
    \includegraphics[width=\linewidth]{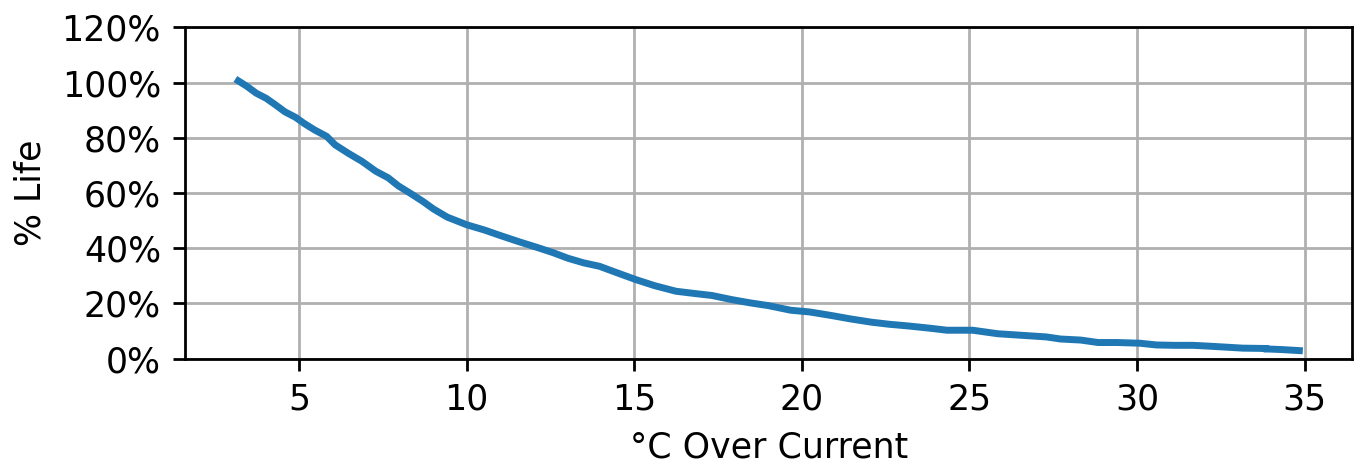}
    \caption{Impact of operating temperature (function of internal and ambient temperatures) on the life expectancy of capacitor from \cite{Capacitor_Problems_SN}} %\href{https://electrical-engineering-portal.com/capacitor-bank-catastrophic-explosion}{link}.}
    \label{fig:etemp_life_cap}
\end{figure}

\textbf{Impact of Extreme Temperatures}
It is reported that operation of the capacitor above its continuous operating temperature rating resulted in a capacitor failure and injuries \cite{ecm_bat1}.

\textbf{Existing Methods to Measure the Impacts}: 
Heat impacts the electronic components and likewise affects the capacitor bank failures as well. As shown in Fig.~\ref{fig:cap_voltage} and Fig.~\ref{fig:cap_current}, the internal temperature of the capacitor bank is not only impacted by overvoltage (lightly loaded -- hot/cold days depending climate zone) and over current (heavily loaded -- hot/cold days depending climate zone) but also by the geometric dimensions of the capacitor winding \cite{Capacitor_Problems_SN}\cite{natarajan2002computer_SN}. % \href{https://electrical-engineering-portal.com/capacitor-bank-catastrophic-explosion}{link} \href{https://www.google.com/search?q=power+system+capacitors+by+ramasamy+natarajan&rlz=1C1GCEJ_enUS994US994&oq=power+system+capacitors+by+rama&gs_lcrp=EgZjaHJvbWUqCAgBEAAYFhgeMgYIABBFGDkyCAgBEAAYFhgeMg0IAhAAGIYDGIAEGIoFMg0IAxAAGIYDGIAEGIoFMg0IBBAAGIYDGIAEGIoFMg0IBRAAGIYDGIAEGIoFMgoIBhAAGIAEGKIEMgoIBxAAGIAEGKIEMgoICBAAGIAEGKIE0gEINzAzNGowajeoAgCwAgA&sourceid=chrome&ie=UTF-8}{link}. 
The overall operating temperature is a function of internal and ambient temperature. Thus depending on the internal temperature, the limits on ambient temperature are strictly imposed. Note that the ambient temperature referred here is not the room's air but the temperature within the capacitor bank's enclosure, i.e., the air that is surrounding the capacitor bank. Generally, this can be 20\degree C above the ambient room's air temperature. Therefore, monitoring the enclosure temperature during extreme heat helps to identify impacts on the capacitor bank. From Fig.~\ref{fig:etemp_life_cap}, for every 7\degree C above the rated operating temperature, the capacitor life is reduced by 50\% \cite{Capacitor_Problems_SN}.
%\href{https://electrical-engineering-portal.com/capacitor-bank-catastrophic-explosion}{link}.

% Heat is known to be a killer of electronic components. This is also a major cause of capacitor failures. Over voltage and over current both increase the internal temperature of a capacitor and lead to reduced life. Surprising to many, the geometry (diameter vs. length) of the capacitor windings has very much to do with the internal temperature rise of a capacitor. Based on its own internal heat generation, a capacitor may face strict limitations in the ambient temperature within which it may be applied. Ambient temperature in the case of capacitors does not refer to the room air, but rather to the temperature within the capacitor enclosure. That is the air immediately surrounding the capacitors. In many cases, this can be 20\degree C above the ambient room air temperature.

% Additionally, placement of a capacitor into an environment that exceeds its continuous temperature capability can shorten its life dramatically. Continuous operation of a capacitor in an ambient temperature that is 7°K above its rated temperature will reduce its life at least by 50\%. For each seven degrees Celcius (7\degree C) above its rated operating temperature, the capacitor life is reduced 50\%. Typical capacitor life reduction due to elevated ambient temperature is illustrated in the Figure.

Apart from temperature monitoring, exposure to sharp temperature variations, and transient overvoltages, aging and manufacturing defects can cause internal failures of capacitor elements. A new method using indicating quantity superimposed reactance (SR) is presented in \cite{jouybari2018new}.

\textbf{Existing Methods to Mitigate the Impacts}: Capacitors that are often operated under uncertain loading conditions can be placed in a climate-controlled environment to help mitigate the risk of failure. A study of capacitor internal heat generation, reported in IEEE Transactions on Industry Applications \cite{el2002thermal}, revealed that for a given capacitance and winding area, short windings of larger diameter generated far less internal temperature rise than tall windings. The article compares four capacitors with equal $\mu$F, voltage rating, film thickness and thickness of metallization. The only difference is the height of the four capacitors. The conclusion was that the optimized winding construction for capacitors is a short coil, which results in a relatively larger diameter. Although it may be less expensive to produce capacitors with tall windings, they do not handle heat as well as short windings.

Further, it is recommended to use a reactor in series with capacitors whenever harmonics are present on the electrical system to which capacitors will be applied. Use a 7\% detuning reactor (tunes capacitor at $3.78^{\textnormal{th}}$ harmonic) to protect power factor capacitors \cite{pqmc_bat}. This reduces heat by reducing over current in the line.

\subsection{Batteries}
\label{subsec:batteries_ds}
% \textcolor{red}{done}

Batteries having varying constructions, anode and cathode materials, and electrolyte compositions. This study does not focus on the electrolyte material but rather discusses how temperature affects battery efficiency and how the impact varies between different types of batteries. According to EIA, in 2022, global investments in battery energy storage exceeded 20 billion US dollars, and 65\% of it is for grid-scale deployment. Battery energy storage investments are projected to exceed 35 billion US dollars in 2023 \cite{iea_bat}.

% “Global investment in battery energy storage exceeded USD 20 billion in 2022, predominantly in grid-scale deployment, which represented more than 65\% of total spending in 2022. After solid growth in 2022, battery energy storage investment is expected to hit another record high and exceed USD 35 billion in 2023, based on the existing pipeline of projects and new capacity targets set by governments” \cite{iea_bat}.

Based on cost and energy density considerations, lithium--iron phosphate batteries, a subset of lithium--ion batteries, are still the preferred choice for grid-scale storage. For applications with limited space and home energy storage, energy-dense chemistries for lithium--ion batteries such as nickel cobalt aluminium (NCA) and nickel manganese cobalt (NMC) are popular. However, flow batteries with low performance degradation are also under development for technology breakthrough energy storage needs with limited space. In 2022, the world's largest vanadium redox flow battery was commissioned in China with a capacity of 100 MW and energy volume of 400 MWh \cite{iea_bat}. At the end of 2022, the entire U.S. grid had more than 1.1 million MW of capacity \cite{eenews_bat}.

% More energy-dense chemistries for lithium-ion batteries, such as nickel cobalt aluminium (NCA) and nickel manganese cobalt (NMC), are popular for home energy storage and other applications where space is limited. 

% Besides lithium-ion batteries, flow batteries could emerge as a breakthrough technology for stationary storage as they do not show performance degradation for 25-30 years and are capable of being sized according to energy storage needs with limited investment. In July 2022 the world’s largest vanadium redox flow battery was commissioned in China, with a capacity of 100 MW and a storage volume of 400 MWh \cite{iea_bat}.

% “At the beginning of August, the U.S. had about 237,000 megawatts of utility-scale solar, wind and battery storage online, up 12 percent from the same time last year, according to the American Clean Power Association. Of that, 10,000 megawatts were added in the first half of 2023. At the end of 2022, the entire U.S. grid had more than 1.1 million megawatts of capacity.” \cite{eenews_bat}.

\textbf{Impact of Extreme Temperature Operations
 }: Storage solutions for grid are either placed in a climate-controlled indoor setup or outdoors with an appropriate cooling setup \cite{pvmag_bat}. High temperatures cause more depletion of energy in a battery. Thus, extreme temperature in combination with any possible point of failure in cooling system can result in battery performance that may not be sufficient to support grid functions. Ref. \cite{Tesla_Fire_SN} %\href{https://www.energy-storage.news/investigation-confirms-cause-of-fire-at-teslas-victorian-big-battery-in-australia/#:~:text=High%20wind%20conditions%20also%20contributed,speeds%20of%20up%20to%2035mph.}{link} 
reports the failure of grid-scale battery packs in Australia due to high temperature and failure of cooling systems.

\begin{figure}
    \centering
    \includegraphics[width=\linewidth]{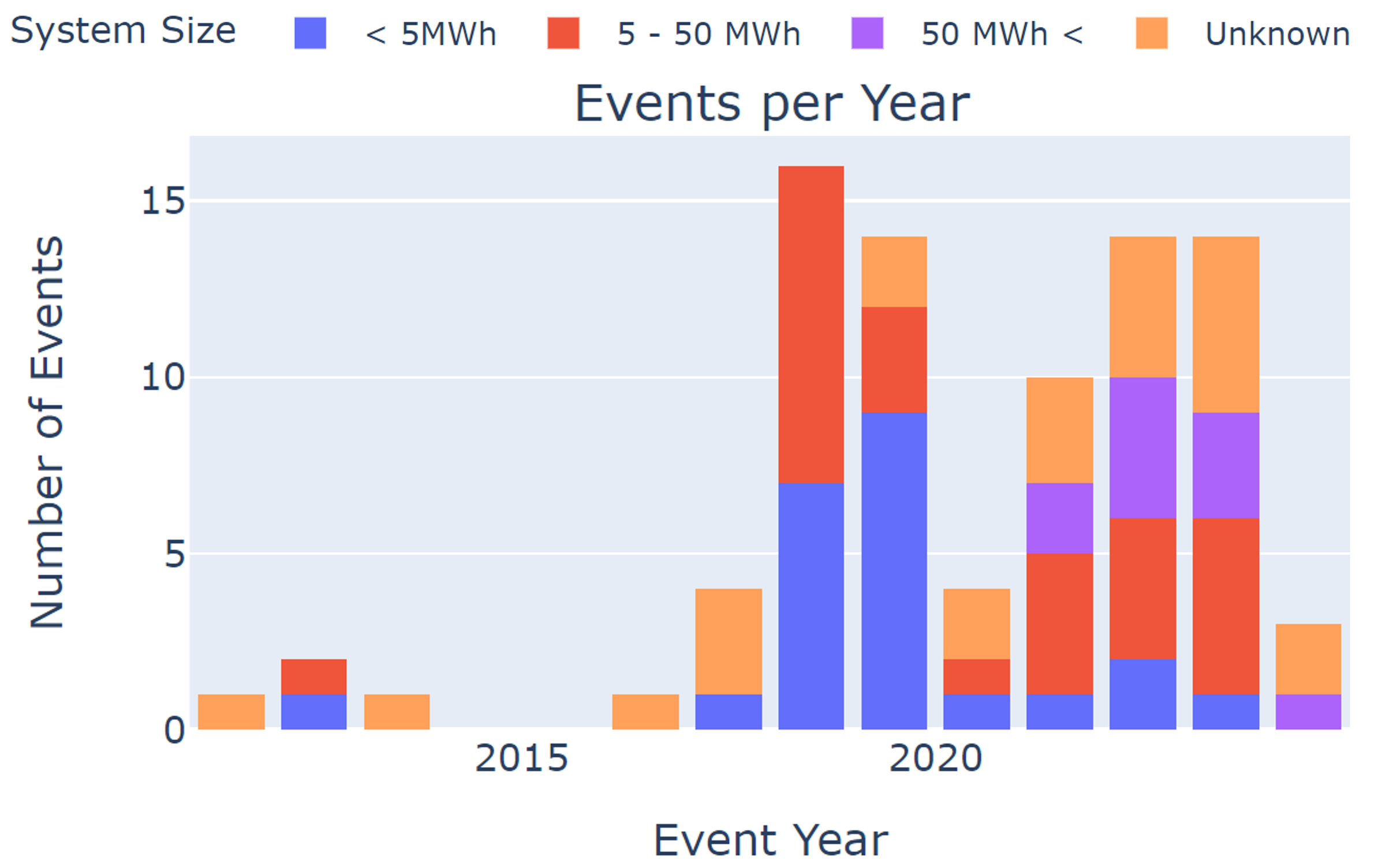}
    \caption{BESS events for different years from \cite{epri_wiki}.}
    \label{fig:bess1}
\end{figure}

\begin{figure}
    \centering
    \includegraphics[width=\linewidth]{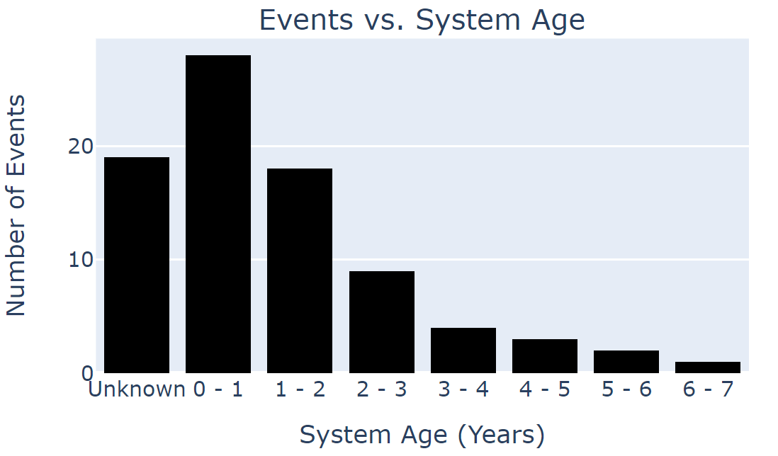}
    \caption{BESS events versus BESS age in years from \cite{epri_wiki}.}
    \label{fig:bess2}
\end{figure}

Another common cause for storage system failure is an indirect one: wildfire-caused transmission faults \cite{nerc_bat1}. Fires are the leading cause of BESS failures \cite{epri_wiki}. Fig.~\ref{fig:bess1} shows the increase in BESS-related events increase in recent years. This is not surprising, because BESS deployment has grown tremendously in recent years. From Fig.~\ref{fig:bess2}, we can observe that the BESS events are often observed when their age between 0 and 3 years. Another useful statistic is that as the system age increases, the number of events reduces \cite{epri_wiki}. More than 50\% of BESS failures occur within the initial two years of operation. 5--50 MWh BESS account for over half of total failure events globally \cite{tdworld_bat}. 48\% of failures have been linked to solar-plus-storage projects \cite{tdworld_bat}. In addition, some locations subjected to rapid temperature variations such as in the mountains can experience dewing leading to damage within the BESS located outdoors if not well controlled \cite{nfpa_bat1}.

% \begin{figure}[H]
%     \centering
%     \includegraphics[width=\linewidth]{images2/p4.png}
%     \caption{temp}
%     \label{fig:enter-label}
% \end{figure}

\textbf{Existing Methods to Measure the Impacts}: The existing method to measure the impact of extreme temperatures on BESS are through battery management systems (BMS). BMS monitors the temperature of different cells in a battery bank, coolant intake and output temperature, state of the battery, etc. Ref. \cite{bhatt2019experimental} presents how different temperatures affect the performance parameters of a battery with a testbed. Sub-zero temperatures are shown to cause capacity loss \cite{vidal2019xev}.

\textbf{Existing Methods to Mitigate the Impacts}: Some ways to mitigate the impacts are as follows \cite{tdworld_bat}. 1) Ensuring sufficient spacing between battery modules is critical as radiant heat impact from adjacent equipment and their failure can cause cascading failures. 2) Carefully selecting the right type of battery for BESS depending on the local climate. 3) Creating spacing standards for BESS units. and 4) Involving OEMs throughout the entire BESS project life cycle.

% \begin{enumerate}
%     \item Ensuring sufficient spacing between battery modules.
%     \item Carefully selecting the right type of battery for BESS depending on the local climate.
%     \item Creating spacing standards for BESS units.
%     \item  Involving OEMs throughout the entire BESS project lifecycle.
% \end{enumerate}

Furthermore, the other solution to mitigate the impacts is through a reliable BMS and redundant climate control options to maintain the temperatures of the BESS within permissible limits. However, redundant systems are more expensive to acquire and maintain. There has been work on early prediction of battery life \cite{zhang2021early}. Ref. \cite{laidig1994technology} describes a method employed by a commercially available battery monitoring system which, through impedance and other measurements, provides the user with a viable and predictive approach to improving overall system reliability. \cite{chunhua2012fault} presents a prediction methodology to identify faults by using redundant sensors.

\section{SUBSTATION GRID EQUIPMENT IMPACTS}
\label{sec:SS}
\subsection{Transformer}
\label{subsec:xfrmr_ss}
% \textcolor{red}{done}

Transformers are critical infrastructure for transmission and distribution circuits of power grids. Transformers used in the power domain can be categorized as 1)~power transformers, 2)~measurement transformers (current and potential transformers), and 3)~distribution transformers.
\begin{enumerate}
    \item \textbf{Power Transformers}: Power transformers facilitate energy transfer between the high-voltage transmission system and the primary distribution system. These transformers are generally large and are located in substations. Based on operational voltage level, these transformers can be segregated into three types, namely 1)~small power transformer (up to 7 MVA), 2)~medium power transformer (50-–100 MVA), and 3)~large power transformer (more than 100 MVA).
    \item \textbf{Measurement transformers}: Measurement transformers, such as current and potential transformers, are used to measure the current and voltage of the power line.
    \item \textbf{Distribution transformers}: Distribution transformers facilitate energy transfer at very low voltage when compared to power transformers. More specifically, 33~kV for industrial applications and 440--220~V for residential applications.
\end{enumerate}

\textbf{Impact of Extreme Temperature Operations}: The impact of extreme temperatures on transformers can be understood by studying the direct and indirect impacts of the ambient temperature. Direct impact involves the ambient temperature’s impact on transformer components. Whereas indirect impact involves the ambient temperature’s impact on electrical loading of the transformer, which in turn causes overheating. For example, the winding and oil inside the transformer gets hotter as the loading and losses increase. Through conduction, the heat that arises in the winding impacts the transformer oil that is used as a coolant. Excessive heat will damage the insulation, resulting in transformer breakdown.

Power transformers' operation is susceptible to temperature changes in two main ways. First, changes in ambient temperature can cause the temperature of coolant oil to change, resulting in transformer aging and sometimes even breakdown \cite{munir2019analysis}. Second, the direct impact and indirect impact of ambient temperature can have a combined effect on the transformer \cite{munir2019analysis}. Heat can cause anomalies in some parts of the transformer which, can increase agiing of transformer \cite{Warman2019Remaining}. According to the IEC standard, power transformers are used at an ambient temperature of 20\degree C \cite{is1972} (30\degree C IEEE standard \cite{transformers_mineral_oil}). However, environmental temperature differences in design and real conditions make a transformer very risky to operate at its maximum capacity \cite{sen2001overloading}. Ref. \cite{okundamiya2021impacts} shows that ambient temperature changes also affect breakdown voltage of the transformer. For example, the allowable temperature rise of a coil is 65\degree C. Based on Class A insulation for winding, the maximum operating temperature is 105\degree C. When the ambient temperature is 40\degree C, the allowable temperature rises to 65\degree C (105\degree C - 40\degree C = 65\degree C). As the temperature of transformer is generally 10\degree C lower than that of winding, the allowable temperature rise of transformer oil is 55\degree C. To prevent oil aging, the temperature rise of upper oil surface shall not exceed 45\degree C. No matter how the ambient air changes, the safe operation of the transformer within the specified service life can be guaranteed only if the temperature rise does not exceed the allowable value of 45\degree C.

Instrument/measurement transformers play a vital role for measurement and protection in power systems. The accuracy of instrument transformers is impacted by the ambient temperature \cite{mingotti2018effect, gedatasheet}. For example, an inductive current transformer does not fulfill the accuracy class requirements when the temperature is lower the conventional ambient temperature of 23\degree C but higher than the minimum working temperature specified by the standards \cite{transformers2009part}. Industry also reports the derating nature of the current transformers due to increase in ambient temperature \cite{peakdemand_xfrmr}.

Data from Chicago, Illinois was used to study the effect of ambient temperature on distribution transformers, and it was shown that the overall temperature of the transformer is affected by ambient temperature and load level \cite{almohaimeed2017steady}. Furthermore, the ratings of the distribution transformer need to be designed based on the highest temperature recorded and altitude of the site \cite{SE_xfrmr_ambient_Temp}. Ref. \cite{fox2019heat} also shows how the heat generation within the padmount transformer leads to failures.

\textbf{Existing Methods to Measure the Impacts}: Techniques exist to quantify the aging of transformers and predict insulation breakdown. Ref. \cite{gouda2012predicting} presents a methodology to estimate the aging of transformers by predicting transformer temperature rise. Industry also focuses on monitoring the temperature and moisture of the winding insulation to track the aging rate using dynamic rate tracking \cite{dynamicratings_company}. For every 6\degree C increase in winding temperature above the rated maximum of 110\degree C, the transformer aging rate roughly doubles \cite{dynamicratings_company}. There are also thermal imaging solutions that are used to monitor the temperature of the transformer, which is further used to calculate the internal winding and oil temperature to determine the aging impact or potential failure \cite{mariprasath2018real, zarco2022conducting, fanchiang2022application}. Refs. \cite{ko2020prediction, bai2013analyzing} developed a prediction model that uses weather data to correlate the distribution transformer failure. Specifically, \cite{bai2013analyzing} provides three metrics that can be used with the help of a closed form expression obtained in the work to study the impacts on transformer life expectancy for any region. %also provides different types of transformer failures and different machine learning techniques such as ANN (artificial neural networks), SVM (support vector machine), NB (naïve Bayes), and DT (decision tree) to predict the transformer failures.

\textbf{Existing Methods to Mitigate the Impacts}: The various methods to prevent overheating of a transformer includes load sizing that is supported by the transformer, auxiliary cooling systems, adequate ventilation (placement proximity of transformer and other equipment), insulation inspection, oil quality checks, and regular maintenance. Out of the various solutions, auxiliary cooling systems and adequate ventilation play a crucial role in preventing extreme temperature impacts on the transformers. Please note that the auxiliary cooling systems are additional systems that are not part of the transformer’s design, i.e., exclusive of cooling methods like AN, AF, ONAN, ONAF, etc. \cite{bower_topic}.

There have been mobile transformers that are used to mitigate the power grid impacts because of transformer failure in extreme temperatures \cite{pe_international_xfrmr}; this enabled to reduce the down time significantly. Another solution is to provide mobile auxiliary cooling services; for example, one substation in the eastern United States was facing 50,000 gallons of overheated transformer oil due to an overloaded transformer where the typical 208V/240V cooling application is not sufficient. In this case, portable trailer-mounted motor cooling was used to cool the transformer \cite{united_rentals}. Another well-known substation solution is thermal imagery techniques to monitor the temperatures of the transformer and its adjacent equipment, which can also radiate heat increasing transformer's temperature \cite{mariprasath2018real, zarco2022conducting, fanchiang2022application}.

% \textbf{TODO}: find more real-world reports from industry reports and news (exclusive search needed). “adequate ventilation” part needs to coincide with the thermal image idea Alok has where multiple equipment impact each other due to close proximity.

\subsection{Circuit Breaker}
\label{subsec:circuit_breakers_ss}
% \textcolor{red}{Alok}

Circuit breakers are essential grid equipment whose role is to protect the system and isolate disturbances or faults in the power grid. Circuit breakers also carry power and have internal device heat during heavy loading conditions on hot days. The insulating medium in circuit breakers changes with the system voltage and has a role to play in reliable operation. High-voltage circuit breakers usually have an insulating medium: SF$_6$ gas, nitrogen, or a mix of these gases. Vacuum circuit breakers are common high- and medium-voltage circuit breakers. Medium- and low-voltage circuit breakers are usually air-insulated circuit breakers. While reference \cite{powercast} highlights that extreme cold conditions will degrade the performance of SF$_6$ circuit breakers, reference \cite{an2022study} concludes that extreme heat and temperature rise of SF$_6$ has little impact on the SF$_6$ gas performance and its application in  SF$_6$ insulated power equipment. There are other studies in literature that have studied some gas mixtures like SF$_6$--CO$_2$ as insulating media in circuit breakers but do not discuss the performance under extreme temperatures \cite{wang2014fundamental}.

\textbf{Impact of Extreme Temperature Operations:}
Medium-voltage and low-voltage circuit breakers that are mainly for industrial and residential protection are significantly affected by extreme heat and cold. Reference \cite{sens_tech} provides details of de-rating and re-rating curves for low-voltage residential circuit breakers up to 63 A ratings. These clearly indicate that extreme heat and hot ambient air conditions will negatively impact the performance of the circuit breakers and can lead to tripping of the breakers much earlier than intended, leading to loss of load and often nuisance tripping. In aggregate, such behavior can cause significant impact on the load in the system leading, to unwanted cascaded actions on the distribution system level. Reference \cite{SE_CB} discusses the reliability concerns of other low-voltage, high-power circuit breakers in extreme weather conditions. Evaluating the temperatures experienced during transportation and storage is an important consideration for circuit breakers, and needs to be documented to ensure the grid operators and asset owners understand the expected life and reliability vulnerability of the equipment that is being commissioned.

 Reference \cite{NERC_lessons_learned} is a report by NERC that discusses examples of circuit breaker failures due to extreme cold. These occurrences are important to consider in various regions of the country	as the unstable polar vortex phenomenon is happening with increasing frequency and reaching regions of the country that normally did not expect extreme cold. 

 \textbf{Existing Methods to Measure the Impacts:}
Reference \cite{DYR_CB} Provides details of some methods for monitoring SF$_6$ gas conditions in SF$_6$ circuit breakers. Reference \cite{FEGS} provides a circuit breaker monitoring system that asset owners can use to monitor various operating conditions of the circuit breakers and can itself operate in temperatures ranging from -$40$\degree C to +65\degree C, which means that it might need additional thermal insulation to operate in extreme temperature exposures like during polar vortices or extreme hot days inside an enclosure. Other details specifications of this monitoring system can be found in reference \cite{FEGS2}.

 \textbf{Existing Methods to Mitigate the Impacts:}
While extreme heat-related failures can be managed to some extent by derating that compensates for the temperature rise caused due to loading of the equipment, extreme cold conditions are difficult to managed for outdoor equipment. Reference [7] provides links to other learning and reported material associated with winter preparedness for circuit breakers and other critical grid equipment. Winterizing the equipment that depend on fluids is important, and often high-voltage circuit breakers have some media around the contacts to extinguish the arc and break the circuit \cite{NERC_lessons_learned}.

%\subsection{Bushings}
%\textcolor{red}{Alok}

% \subsection{Capacitor Bank}

\subsection{Circuit Switcher}
% \textcolor{red}{done}

% \textcolor{blue}{Kishan self note: this section has reference numbers from word. Need to change them. Reason, need to find better references.}

Circuit switchers help to safely and reliably control the flow of power flow in the electric grid. They are also used to disrupt fault-induced current during faulty conditions. This isolates the faulty equipment, protecting against short circuits and overloading conditions.

\textbf{Impact of Extreme Temperature Operations:} Circuit switcher's function can be impacted by both extreme cold and heat.

In case of extreme heat, the issue faced by the circuit switcher is derating. The rated continuous current is the maximum current that the switch can carry continuously under normal operating conditions without exceeding the specified temperature limits. The specified temperature limits for non-enclosed are higher than enclosed switchers because the switchers that are enclosed have lower heat dissipation behavior \cite{cw_2}. % The switchers with copper or copper alloy have lower specified temperature limit than that of the silver (\textcolor{red}{Insert table but table is too big, need to figure out someway here}) \cite{cw_2}.

\textbf{Existing Methods to Measure the Impacts:}
Status of switch can be monitored from control room, and thermal image based equipment monitoring is also another option for monitoring. 

\textbf{Existing Methods to Mitigate the Impacts:} In case of extreme cold, having the circuit switcher in an enclosure with controlled climate can mitigate the breakdown of the equipment. In case of extreme heat, considering the derating behavior of the equipment can help safely operate the switch.

\subsection{Voltage Regulator}
% \textcolor{red}{done}

% \textcolor{blue}{Kishan self note: this section has reference numbers from word. Need to change them. Reason, need to find better references.}

Voltage regulation assists with reliable electric grid operation by maintaining the voltage variations within acceptable values. Reactive power compensation, tap-changing transformers, and voltage regulators help improve the reliability and control the power transfer in the grid \cite{Acceptable_Voltage_SN}\cite{Controlling_Voltage_SN}  %\href{https://www.enerdynamics.com/Energy-Currents_Blog/How-Electric-Operators-Maintain-Acceptable-Voltage.aspx}{Link}, %\href{https://eshop.se.com/in/blog/post/techniques-for-controlling-voltage-in-power-systems.html#:~:text=Therefore%2C%20techniques%20such%20as%20reactive,the%20distribution%20of%20electrical%20load.}{Link}.

Voltage fluctuations in distribution systems can occur for various reasons, including 1)~daily or seasonal variations in power demand, 2)~low demand to high demand, 3)~load density along the feeder, 4)~load balancing activities, and 5)~unbalanced loads and feeders \cite{monti2020converter_SN} %\href{https://www.sciencedirect.com/topics/engineering/voltage-regulation#:~:text=Voltage%20regulation%20of%20distribution%20system,unbalanced%20loads%20and%20feeders%2C%20etc.}{Link}. 

% Voltage regulation is a key factor in reliable electric grid operations. Voltage varies across the grid, and must be maintained within acceptable values at each key interface. Techniques such as reactive-power compensation, tap-changing transformers, and voltage regulators can help keep the system's voltage levels at ideal place, improve overall system reliability, and better control the distribution of electrical load [7, 8].

% Some causes of mechanical failure include [10]:
% \begin{enumerate}
%     \item Loose joints
%     \item Voids and contamination in insulation
%     \item Friction or tearing
%     \item Excessive electrical stress
%     \item Excessive temperature
%     \item Temperature cycling
%     \item Chemical and physical reaction with other materials
% \end{enumerate}

% Some causes of electrical failure of insulation include: Breakdown of insulating materials, Failure of the insulation of the magnetic circuit, and Overloading and "through" faults [10]. 

\textbf{Impact of Extreme Temperature Operations:} Voltage regulators' standard design ratings like kVA, voltage, and current (air cooled system) are defined based on an ambient temperature of 40\degree C. Further, the average temperature for an air-cooled system cannot exceed 30\degree C over a 24-hour period \cite{vg_1}. Also, the altitude consideration for rating design assume no more than 3,300 feet. This is because as altitude increases, the reduction is air density results in an adverse effect on temperature rise and dielectric strength (Tab.~\ref{tab:dielectric_table}) \cite{vg_1}. Tab.~\ref{tab:vg_table_2} shows the derating behavior of voltage regulator at different temperatures and altitudes. 

%Regulators should comply with ANSI Std. C57.15, “Requirements, Terminology, and Test Code for Step-Voltage and Induction-Voltage Regulators,” and ANSI Std. C57.95, “Guide for Loading Oil-Immersed Step-Voltage and Induction-Voltage Regulators.” (\textcolor{red}{add citations for the standards}).

%----------------------------------------------------

% Standard ratings of kVA, voltage, and current for air-cooled voltage regulators are based on ambient air temperature not exceeding 40C and on the average temperature of the cooling air for any 24-hour period not exceeding 30C \cite{vg_1}.

% Performance of a voltage regulator may be affected at unusual temperatures and altitude service conditions \cite{vg_1}. It is also shown that the voltage regulator derates at unusual temperatures and altitudes. For example, see the tables below \cite{vg_1}.

New York state reports regulator concerns either due to failure or maintenance issues. The most common failure mode is the failure of movable or stationary contacts \cite{Voltage_Regulator_SN}. %\href{https://documents.dps.ny.gov/public/Common/ViewDoc.aspx?DocRefId=%7B3F9379CD-46A3-415F-90E8-1EF3E510843B%7D}{Link}. 
The survey results showed that 90\% of regulators are less than 40 years old. The average age of a spare regulator is 31 years. There are a total of 2,980 operating units and 345 spares, out of which 1,742 operating units and 199 spare units have associated age data. This is because manufacturers in earlier years did not provide manufacture dates on their nameplates. There were 95 regulator-related events system wide over the last ten years (2000--2010). 56 out of 95 events were due to voltage regulator failure, leading to 52,797 customer interruptions. This is a failure rate of 0.19\% per year and is below the average yearly failure rate of 0.29\% based on survey involving ten North American utilities \cite{Voltage_Regulator_SN}% \href{https://documents.dps.ny.gov/public/Common/ViewDoc.aspx?DocRefId=%7B3F9379CD-46A3-415F-90E8-1EF3E510843B%7D}{Link}
. The failure rate of a voltage regulator increases when the ambient temperature increases, which also causes indirect impact on energy demand. This increase in energy demand shows a strong correlation with reactive power demand rise during summer peaks or heat waves resulting in malfunction \cite{vg_1}. Even though ambient temperature and altitude impacts the voltage regulator, their failures are often associated with a combination of maintenance and climate related factors.

\begin{figure}
    \centering
    \includegraphics[width=\linewidth]{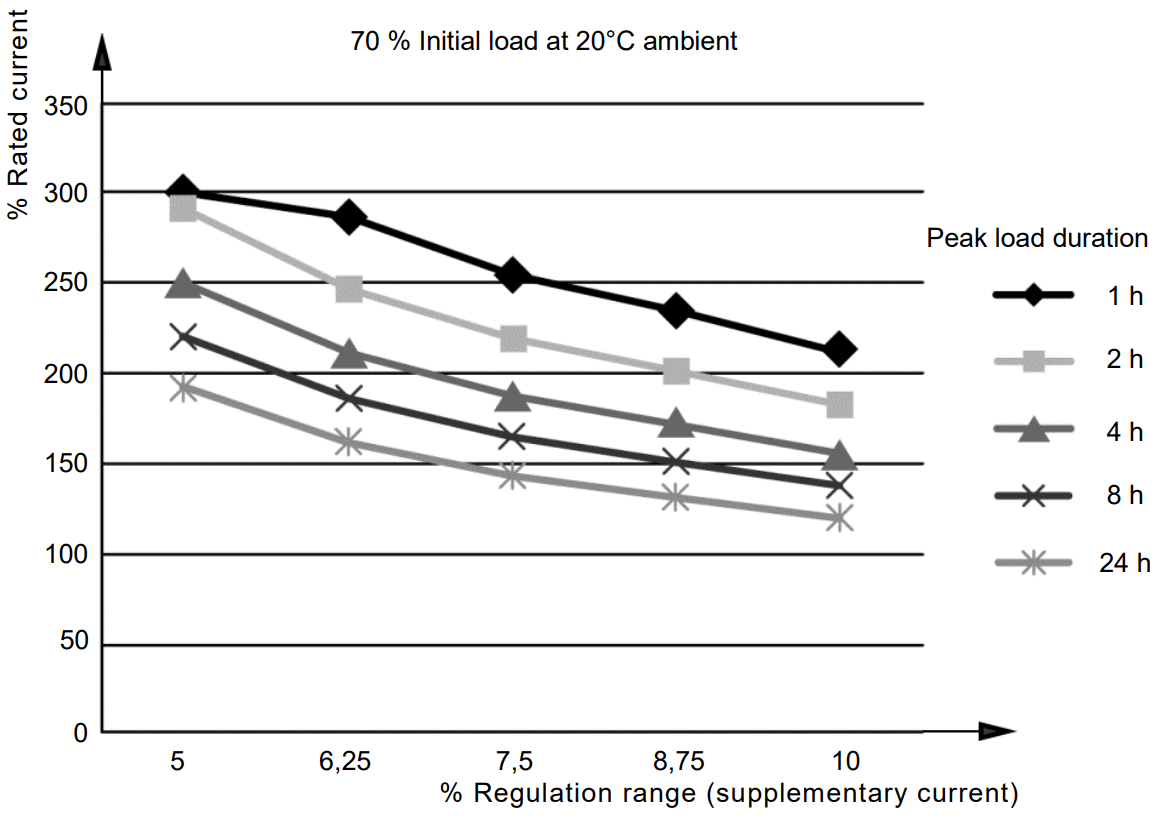}
    \caption{Impact on rated current of voltage regulator at a specific initial loading and ambient temperature for various percent regulation and peak load duration. It is observed that under different ambient temperatures, the rated current of regulator is affected from \cite{vg_1}.}
    \label{fig:volt_reg_5}
\end{figure}

%(section 9 from \href{https://documents.dps.ny.gov/public/Common/ViewDoc.aspx?DocRefId=%7B3F9379CD-46A3-415F-90E8-1EF3E510843B%7D}{Link}) Insert the age data of regulators in New York, Massachusetts, Rhode Island, New Hampshire, Vermont. 
The factors affecting voltage regulators overloading are the following \cite{vg_1}: 1) peak load; 2) tap position at peak load; 3) load prior to peak load; 4) tap position prior to peak load; 5) duration of peak load; 6) ambient temperature; 7) on-load tap-changer maximum through-current rating. All these factors are considered when defining the ratings of a voltage regulator. For example, Fig.~\ref{fig:volt_reg_5} shows how the rated current changes for various combination of the above factors. The effects of the ambient temperature, peak load, duration and initial load are extrapolated from calculations found in the loading guides IEEE Std C57.91 and IEC 60076-7 \cite{vg_1}. IEEE Std C57.91 and IEC 60076-7 provide the best known
general information for the loading of voltage regulators under various conditions \cite{vg_1}.

% \begin{figure}[h]
%     \centering
%     \includegraphics[width=\linewidth]{images2/volt_reg_1.png}
%     \caption{Source [1]}
%     \label{fig:volt_reg_1}
% \end{figure}

% \begin{figure}[h]
%     \centering
%     \includegraphics[width=\linewidth]{images2/volt_reg_2.png}
%     \caption{Source \cite{vg_1}}
%     \label{fig:volt_reg_2}
% \end{figure}

% Standard ratings of voltage regulators are based on an altitude not exceeding 3,300 feet. At higher altitudes, the decreased air density has an adverse effect on the temperature rise and the dielectric strength of voltage regulators.

% \begin{figure}[h]
%     \centering
%     \includegraphics[width=\linewidth]{images2/volt_reg_3.png}
%     \caption{Source \cite{Substations_SN}} %\href{https://pdhonline.com/courses/e473/e473content.pdf}{Link}}
%     \label{fig:volt_reg_3}
% \end{figure}

\begin{table}
\centering
\caption{Impact of altitude on dielectric strength of voltage regulators. Correction factor reduces the dielectric strength with altitude from \cite{vg_1}.}
\label{tab:dielectric_table}
\begin{tabular}{|l |l|} \hline   
Altitude & Correction factor \\ \hline 
3,300 & 1 \\ \hline 
4,000 & 0.98 \\ \hline 
5,000 & 0.95 \\ \hline 
8,000 & 0.86 \\ \hline 
10,000 & 0.8 \\ \hline 
14,000 & 0.7 \\ \hline 
15,000 & 0.67 \\ \hline
\end{tabular}
\end{table}

\begin{table}
\centering
\caption{Maximum allowable average temperature of cooling air when operated at rated kVA of voltage regulator from \cite{vg_1}.}
\label{tab:vg_table_2}
\begin{tabular}{|>{\centering\arraybackslash}p{0.3\linewidth}|>{\centering\arraybackslash}p{0.1\linewidth}|>{\centering\arraybackslash}p{0.1\linewidth}|>{\centering\arraybackslash}p{0.1\linewidth}|>{\centering\arraybackslash}p{0.1\linewidth}|} \hline     
Method of cooling apparatus & 3,300 ft & 6,600 ft & 10,000 ft & 13,200 ft \\ \hline 
Liquid-immersed self-cooled & 30\degree C & 28\degree C & 25\degree C & 23\degree C \\ \hline 
Liquid-immersed forced-air-cooled & 30\degree C & 26\degree C & 23\degree C & 20\degree C \\ \hline
\end{tabular}

\end{table}

% Performance of a voltage regulator may be affected at unusual temperatures and altitude service conditions \cite{vg_1}. It is also shown that the voltage regulator derates at unusual temperatures and altitudes. For example, see the tables below \cite{vg_1}.

% \begin{figure}[h]
%     \centering
%     \includegraphics[width=\linewidth]{images2/volt_reg_4.png}
%     \caption{Source \cite{vg_1}}
%     \label{fig:volt_reg_4}
% \end{figure}

\textbf{Existing Methods to Measure the Impacts:} Monitoring ambient temperature in conjunction with peak demand can ensure that the regulators' derated design values are respected, i.e., the regulator is operating within its design capabilities.

\textbf{Existing Methods to Mitigate the Impacts:} The following measures can mitigate the impacts of extreme heat on voltage regulators: 1) Voltmeters and other instruments used with regulators should be as accurate as the regulator. 2) If the regulator cannot be verified in neutral (lossy operating condition generates heat \cite{vg_1}), report the situation to the system operator. 3) When switching between substations, follow the procedure in the Substation Switching Procedures section. 4) Use a slightly larger bandwidth to reduce the number of operations. 5) Where the above points directly or indirectly increase the temperature in the regulator, consider the derating behavior of the equipment due to ambient temperature.

%\subsection{Distribution Busbar}
%\textcolor{red}{Alok}

\subsection{Frequency Changer}
% \textcolor{red}{done}

% \textcolor{blue}{Kishan self note: this section has reference numbers from word. Need to change them. Reason, need to find better references.}

Traditionally, frequency changers are electromechanical machines composed of motor-generator sets. But with the advent of solid-state electronics, electronic frequency changers also became a possibility. Electronic frequency changers consist of a rectifier stage (AC to DC) that changes the frequency to a desired value through an inverter. For voltage conversion, a transformer can be used, and it also provides galvanic isolation between input and output AC circuits. Electronic frequency changers have similar vulnerabilities to extreme heat to other electronics-based equipment discussed in previous subsections. Therefore, this section focuses on electromechanical frequency changers.

% “Traditionally, these devices were electromechanical machines called a motor-generator set”. “With the advent of solid state electronics, it has become possible to build completely electronic frequency changers. These devices usually consist of a rectifier stage (producing direct current) which is then inverted to produce AC of the desired frequency. The inverter may use thyristors, IGCTs or IGBTs. If voltage conversion is desired, a transformer will usually be included in either the AC input or output circuitry and this transformer may also provide galvanic isolation between the input and output AC circuits”. Here we discuss about frequency changers that are electromechanical machines. These type of machines are relatively less common due to high adoption of solid state electronics. 

\textbf{Impact of Extreme Temperature Operations:}
The impact on frequency changers can be assessed directly by understanding the impact of ambient temperature on the motor and generator. The primary challenge is derating. Ambient temperature can decrease the efficiency of generator-transformer set (GT), which further affects the performance of the static transformer (ST). Manufacturer design ratings assume that operation is maintained at ambient temperatures below 40\degree C. However, the alternator derates by 3\% for every 5\degree C increase in ambient temperature \cite{Derating_SN}.%\href{https://genesalenergy.com/en/communication/articles/how-temperature-and-elevation-affect-generators/#:~:text=Generator%20derating%20ambient%20temperature&text=The%20majority%20of%20manufacturers%20guarantee%20the%20power%20of%20their%20alternators,for%20each%20additional%205%C2%BA%20C.}{Link}.

% “Ambient temperature can cause a decrease in the efficiency of a generator-transformer set (GT), which can indirectly affect the electric energy production of a static transformer (ST). This is because the temperature increase causes an indirect decrease in the efficiency of each GT, and ST” [3]. For example, “The majority of manufacturers guarantee the power of their alternators, as long as they operate at an ambient temperature of below 40°C. At higher values, the derating in an alternator is generally of 3\% for each additional 5º C.” \href{https://genesalenergy.com/en/communication/articles/how-temperature-and-elevation-affect-generators/#:~:text=Generator%20derating%20ambient%20temperature&text=The%20majority%20of%20manufacturers%20guarantee%20the%20power%20of%20their%20alternators,for%20each%20additional%205%C2%BA%20C.}{Link}. 

\textbf{Existing Methods to Measure the Impacts:}
One approach to measure the impacts of extreme temperatures on the frequency changer is by monitoring the ambient temperature immediately surrounding the equipment, then referring the observed ambient temperature in the derating curve to understand the derated maximum current rating \cite{Harwin_Derating_SN}.%\href{https://www.harwin.com/blog/derating-curves-power-ratings-maximum-current-ratings}{Link}. 
Additionally, it is important to note the altitude. As in voltage regulators, altitude variations can result unusual ambient temperature effects on alternators.

% One approach the impact of ambient temperature on alternator is by utilizing the derating curve. “A Derating curve is a graph that shows how the maximum current rating of a component decreases as the ambient temperature increases. It can also be called a power rating curve or a current carrying capacity curve or graph \href{https://www.harwin.com/blog/derating-curves-power-ratings-maximum-current-ratings}{Link}”. Furthermore, altitude can also worsen the effect of ambient temperature on alternator.

\textbf{Existing Methods to Mitigate the Impacts:} Generally, these issues are resolved by placing the frequency changer in a climate-controlled environment or enclosure. This is especially useful when ambient conditions can change drastically at a specific location.

\subsection{Substation Grounding Resistance}
% \textcolor{red}{done}

% \textcolor{blue}{Kishan self note: this section has reference numbers from word. Need to change them. Reason, need to find better references.}

Substation ground resistance is the resistance in ohms between substation neutral and earth resistance (zero-potential reference). The NEC 250 recommends the grounding of panel and equipment to maintain a resistance of 25~ohms or less. In substations the grounding should maintain a resistance of 1 to 5~ohms, depending on size and voltage levels \cite{national2007nfpa}.

\textbf{Impact of Extreme Temperature Operations:} The literature discusses different approaches to measure the ground resistance. \cite{IEEE_Safety_SN} %[\href{IEEE Standard 80-2013}{Link IEEE Standard 80-2013 ??}]
provides an equation to calculate the substation ground resistance when grid area, soil resistivity, and total length of buried ground rods are known. Substation ground resistance is given by
\begin{align}
    R_g &= \dfrac{\rho}{4}\cdot \sqrt{\dfrac{\pi}{A}}+\dfrac{\rho}{L_T} \label{eq:1}
\end{align}
where $R_g$ is the substation ground resistance in $\Omega$, $\rho$ is the soil resistivity in $\Omega$m, $A$ is the area occupied by the ground grid in m$^2$, and $L_T$ is the total buried conductor length in m. 

\cite{IEEE_Fusing_SN}
%\href{https://ieeexplore.ieee.org/document/4110448/similar#similar}{Link} 
provides another expression considering buried depth of the ground grid which is given by
\begin{align}
    R_g &= \rho \left[\dfrac{1}{L_T} + \dfrac{1}{\sqrt{20A}}\left(1 + \dfrac{1}{1+h\sqrt{\dfrac{20}{A}}}\right)\right] \label{eq:2}
\end{align}
where $h$ is the depth of the ground grid in m. The soil resistivity ($\rho$) for different types of earth like wet organic soil, moist, soil, dry soil, and bedrock are given by $10$, $100$, $1,000$, and $10,000$ $\Omega$m respectively \cite{Power_Engineering_SN}
%\href{https://voltage-disturbance.com/grounding-and-bonding/estimating-substation-ground-grid-resistance/}{Link}. 
It can be observed from equations \eqref{eq:1} and \eqref{eq:2} that the substation ground resistance is directly dependent on the soil resistivity. Furthermore, soil resistivity has a negative temperature coefficient, which means that as temperature decreases, soil resistivity increases \cite{Soil_Resistivity_SN}. %\href{https://eepower.com/technical-articles/an-introduction-to-soil-resistivity/#}{Link}. 
For example, Tab.~\ref{tab:resistivity_tab} \cite{Soil_Resistivity_SN} % (\href{https://eepower.com/technical-articles/an-introduction-to-soil-resistivity/#}{Link}) 
shows that when the temperatures are above freezing point, the change in resistivity for temperature variation is smaller when compared to the ambient temperature below the freezing point. Equation \eqref{eq:2} also shows that the substation ground resistance is directly proportional to the soil resistivity ($\rho$) and inversely proportional to the depth of ground grid ($h$). Therefore, in colder climate conditions, it is important to construct the ground resistance mesh deep inside the ground.
% \begin{figure}[h]
%     \centering
%     \includegraphics[width=\linewidth]{images2/soil_Resistivity.PNG}
%     \caption{Source}
%     \label{fig:soil_resistivity}
% \end{figure}

\begin{table}
\centering
\caption{Effect of temperature on soil resistivity from \cite{Soil_Resistivity_SN}.}
\label{tab:resistivity_tab}
\begin{tabular}{|c|c|} \hline   
Temperature & Resistivity \\ \hline 
20 & 7,200 \\ \hline 
10 & 9,900 \\ \hline 
0 (water) & 13,800 \\ \hline 
0 (ice) & 30,000 \\ \hline 
-5 & 79,000 \\ \hline 
-15 & 130,000 \\ \hline
\end{tabular}
\end{table}

\textbf{Existing Methods to Measure the Impacts:} 

Once construction of substation ground resistance is complete, it is challenging to measure the zero volts between ground resistance and earth. This is because of the sphere of influence of the grid. For large substations, measurements must be taken outside the sphere of influence, which can mean long measurement traverses. In most cases, it is not practical \cite{Power_Engineering_SN}.%\href{https://voltage-disturbance.com/grounding-and-bonding/estimating-substation-ground-grid-resistance/}{Link}.
These measurements can be taken by a current probe and ground resistance measurement device. For example, IEEE 81:2012 \cite{IEEE_Grounding_SN} %(\href{https://standards.ieee.org/ieee/81/4889/}{Link}) 
requires the current probes to be used at least 5 times the maximum diameter of the grounding system. Another approach is by using the temperature impact on soil resistivity combined with the equations \eqref{eq:1} and \eqref{eq:2} to estimate the effect of extreme temperatures on the ground resistance.

\textbf{Existing Methods to Mitigate the Impacts:} Ref. \cite{gressit_1} proposed a method to reduce the ground resistance values in three steps: (1) drilling deep holes in the ground; (2) developing cracks in the soil by means of explosions in the holes; (3) filling the holes with low resistivity materials (LRM) under pressure. Most of the cracks around the vertical conductors will be filled with LRM, and a complex network of low resistivity tree-like cracks linked to the substation grid is formed.

% \subsection{Power Fuses}

\subsection{Lightning Arrestors and Surge Arrestors}
% \textcolor{red}{done}

% \textcolor{blue}{Kishan self note: this section has reference numbers from word. Need to change them. Reason, need to find better references.}

Lightning arrestors protect equipment against damaging voltages when lightning strikes. These devices divert the current from lightning surge through the arrestor into the ground.

\textbf{Impact of Extreme Temperature Operations:} Regular operating conditions (from standards that govern performance) for lightning arrestors are 1) ambient temperatures between $-40$\degree C and 40\degree C, 2) altitudes of no more than 1000~m, and 3) wind speeds less than or equal to 34 m/sec \cite{Lightning_arrestor_SN, nfpa_arrestor_light}. However, these equipment can be deployed in mountainous regions above 1000~m as well, in which case their derating nature needs to be considered. %(\href{https://www.nfpa.org/codes-and-standards/nfpa-780-standard-development/780}{Link})%(\href{https://www.wiringo.com/lightning-arrester.html}{Link}).

\textbf{Existing Methods to Measure the Impacts:} Various monitoring methods are used to monitor the health of lightning arrestors. On-line measurement methods are preferred, because they do not require a substation to be taken out of service to perform the evaluations. Furthermore, unlike equipment, lightning arrestor performance and health can only be monitored by comparing with the baseline measurements. Some of the online methods are 1) infrared imaging, 2) monitoring resistive component of current, 3) third harmonic measurement, 4) absolute temperature and comparison, and 5) power loss at maximum continuous over voltage \cite{Surge_arrestor_SN}. %(\href{https://www.inmr.com/monitoring-condition-surge-arresters/}{Link)}. 
Monitoring of arrestors was dominated by third harmonic measurement (also known as leakage current test) at the base of the arrestor. However, this method requires the arrestor to be isolated from the earth for accurate data collection \cite{Arrestor_condition_SN}. %(\href{https://www.inmr.com/assessing-arresters-using-surface-temperature-profile/}{Link}). 
To account for this, infrared imaging method is employed for online monitoring of arrestors. However, estimation of arrestor health using infrared imaging is not easy because of the lack of information from suppliers on absolute temperature correlation with arrestors' health. For example, \cite{NEMA_Arrestors_SN}% (\href{https://www.nemaarresters.org/wp-content/uploads/2018/10/Thermal-Imaging-of-Arresters.pdf}{Link}) 
shows how infrared scanning can help identify failures in work in three arrestors when compared to the each other. According to \cite{Arrestor_condition_SN}%(\href{https://www.inmr.com/assessing-arresters-using-surface-temperature-profile/}{Link})
, a temperature of 5\degree C between two partner arrestors warrants concern and anything greater than 10\degree C means a failure. There are various failure modes for lightning arrestors; in particular, moisture ingress is known to be a major factor \cite{Surge_arrestor_SN}.% (\href{https://www.inmr.com/monitoring-condition-surge-arresters/}{Link}). 
Ref. \cite{Surge_arrestor_SN} %\href{https://www.inmr.com/monitoring-condition-surge-arresters/}{Link} 
presents some field tests showcasing the impact of different factors such as 1) time of day, 2) location of sensor on arrestor, 3) cardinal direction of sun, and 4) weather on the absolute surface temperature of arrestor. These tests are conducted to identify the correlation of absolute surface temperature as metric for health monitoring instead of comparing a group of arrestors for health monitoring that is currently used in the field.

\textbf{Existing Methods to Mitigate the Impacts:} Arrestor failure can only be mitigated by early detection. Current online monitoring methods that have been in use for decades, such as third harmonic measurement, require isolation from earth and reliable data collection techniques, while the infrared imaging method does not suffer from these issue. However, it suffers from a lack of standard methods to accurately correlate absolute temperature to the arrestors' health. 

\subsection{Relays}
Relays are devices that monitor electrical systems for abnormal conditions (overcurrent, overvoltage, faults, etc.) and respond by sending signals to open or close circuit breakers to prevent damage. They are essential to power systems because they protect the grid by isolating faults and preventing equipment failure. The main functions of a relay are fault detection, isolation, protection coordination, and monitoring. Some of the different types of relays used in power system based on their application are electromechanical relays, static relays, digital and numerical relays.

IEEE Std C37.91-2021 \cite{relays_transformers}, \textit{IEEE Guide for Protecting Power
Transformers}, %transformer relays reference
reports temperature standards about various other kinds of relays that are currently being used commercially in the industry, as shown below:

\begin{itemize}
\item Adaptive overcurrent relays: In this type of relay, the pickup current value is obtained using an inverse-time characteristic curve. This curve shows that the pickup current increases with a decrease in ambient temperatures and vice versa. However, an exact formulation to modify the pickup current as a continuous function of the changing ambient temperature has not been suggested in this standard. It was also reported in \cite{relays_transformers} that in some utilities, an alternate practice is being followed in which the pickup current is increased by 25\% in the winter in severe climate swing locations.
\item Gas accumulator relay: This type of relay, also known as a Buchholz relay, is typically only used in transformers with conservator tanks due to no gas space being available inside the transformer tank. It is reported that these relays undergo heating issues due to increase in ambient temperature.
\item Temperature relay: This type of relay, also known as the replica relay, typically measures and applies the phase current in the transformer to its heater units placed inside the relay. It was recommended that the installed relay should not be ambiently compensated and should experience same ambient temperature as the transformer it is protecting.        
\end{itemize}

\textbf{Impact of Extreme Temperature Operations:} Some of the typical challenges faced by relays due to changes in ambient temperature as follows. The relay coil in electromechanical relays is made of copper wire, and the resistance of the coil increases by 0.4\% per degree C. This means that as the temperature increases, the available current in the coil to operate the contacts in the relay is not sufficient, which results in failure of relays to operate when required \cite{relay_link_1}. This diminished capacity is an example of derating.  The impact on the temperature derating factor (which proportionally decreases the maximum ambient temperature conditions by the manufacturers) for varying levels of altitude has been reported in \cite{relays_general}. The contacts of the relay at their full rated load may cause ambient temperature to rise by 10\degree C. For example, if a relay first switches at 20\degree C, within the short duration, the ambient temperature can rise to 30\degree C  \cite{relay_link_1, relay_1_kg}. Therefore, the relays are affected directly and indirectly through changes in ambient temperature. Ref. \cite{relay_2_kg} also showed that the functional coordination of components is affected by the layout, causing overheating of the components. Ref. \cite{gutierrez2005relays} shows that the combination of high ambient temperatures, limited ventilation, and solar radiation can cause relay failures.

Electromechanical relays use coils that are typically located near the conductor and are affected by the ambient temperatures. However, most electromechanical relays have been replaced by digital relays, which use CTs and PTs as sensors. The failure modes of digital relays follow the failure modes of CTs and PTs. Digital relays are also housed or enclosed and not exposed to the ambient environment. Note that ambient temperature can cause  sensitivity and loss of coordination in sensing technologies.

\textbf{Existing Methods to Measure the Impacts:} The impacts due to extreme temperatures can be measured by monitoring the load current in the coil and change in coil resistance due to temperatures. This helps to identify the derating behaviors.

\textbf{Existing Methods to Mitigate the Impacts:} The mitigation strategy to avoid the impact of ambient temperature on sensors is to adjust the pickup current based on the derating behavior of the device. This will help ensure that the relays operate as designed. This also ensure coordination of protection system. 

For digital relays, there are new methods to determine the protection settings accurately based on data analytics. These methods use historical event datasets to update and develop better protection strategies aimed at extreme weather readiness \cite{yan2024strategy}. There is also active research in the field of settingless protection that uses dynamic state estimation to determine the health of the system that the replay is protecting \cite{settingless_protection}. This can be particularly helpful during the extreme weather operations where relay coordination may tend to fail; however, this method can theoretically identify an unhealthy downstream system.

Another solution is to use predictive methods. The approaches to predict the hottest spot temperatures and take preventive action have been discussed in \cite{relays_transformers}. It was suggested to use the equations reported in Appendix D in \cite{relays_transformers}, to make these predictions, provided that the load current and ambient temperature remains the same in the foreseeable future. Typically, these predictions can be assumed to hold true for the next 30 minutes and would need to be recalculated automatically every few minutes to account for the changes in ambient temperature after that short time window. Based on the predictions computed using this approach, many corrective or monitoring actions can be taken, such as an automatic activation of the full cooling system based on the predicted hotspot temperature rather than the actual hotspot temperature, an alarm being triggered, or displaying the overtemperature or the time to excessive loss of life.

% \textcolor{red}{Kishan}

\subsection{Shunt Reactors}
%\textcolor{red}{Kishan}

%\textcolor{blue}{Kishan self note: this section has reference numbers from word. Need to change them. Reason, need to find better references.}
Shunt reactors are important reactive management power devices in the power grid. They are crucial in ensuring that the voltages in the power network do not rise to dangerously high values, causing the insulation in the system to fail. There are both fixed-rating and variable shunt reactors; fixed rating reactors are either dry or oil immersed, and variable reactors are mostly oil immersed. Fixed shunt reactors cannot provide regulation, while variable shunt reactors provide regulation capability (which is necessary with variable renewable generation like wind and solar \cite{sr_1}). Variable shunt reactors have multiple components including a tap changer and cooling system that are vulnerable to extreme temperatures.

\textbf{Impact of Extreme Temperature Operations:}
Like oil-cooled transformers, shunt reactors also have temperature-vulnerable components like insulation, windings, tap changer, etc. Reference \cite{sr_2} provides details of the temperature limitations of medium-voltage to high-voltage oil-immersed transformers. Extreme heat (temperatures greater than 40 \degree C) and extreme cold (temperatures less than $-25$ \degree C) ambient conditions are defined as stressed ambient conditions for oil-immersed shunt reactors \cite{sr_2}. There are further details about average temperatures: The monthly average temperature should not exceed 30 \degree C and annual average should not exceed 20 \degree C \cite{sr_2}. If the temperatures exceed these ranges, appropriate measures are needed to prevent impacts on the reliability and life of the shunt reactor. These temperatures should be maintained during transport and storage \cite{sr_2}.
Reference \cite{sr_3} provides details of mechanical failure and microcracks that can develop in the dry type of shunt reactors. It shows the mechanism of this failure in extreme cold conditions. This can lead to structural failure and eventual fire hazard. This impact is not only the failure of the reactor, but can easily cascade into a fire risk during polar vortex conditions where the wind speeds are relatively high.

\textbf{Existing Methods to Measure the Impacts:}
There are not online measurement methods, however Refs. \cite{sr_4} and \cite{sr_5} provide some general guidelines on the performance of the shunt reactors that could be used to assess the impacts of extreme temperatures on the shunt reactors' performance or the impact of extreme temperatures.

\textbf{Existing Methods to Mitigate the Impacts:}
There are not clear insights on the different methods of mitigating the impacts of extreme temperatures on shunt reactors. Because the size and form factor for some of these high-voltage reactors are fairly large, it is a challenging problem to shield these from external extreme ambient temperature conditions.

\subsection{SVCs and FACTS Devices}
% \textcolor{red}{done}

% \textcolor{blue}{Kishan self note: this section has reference numbers from word. Need to change them. Reason, need to find better references.}

SVC (static VAR compensator) and STATCOM (static synchronous compensator) are shunt devices in the FACTS family. SVCs based on thyristor technology appeared in the field during 1970s. STATCOMs based on GTOs (gate turn-off thyristors) appeared in 1980s, but with the help of high-power IGBTs (insulated gate bipolar transistor) in the 1990s, became commercially available. Both SVC and STATCOM assist in increasing the maximum transfer capability of power lines. This in turn helps with delaying the the need to build new power lines in face of ever-growing power demand on the electric grid \cite{gandoman2018review}. When the space in substations is a constraint. Furthermore, STATCOMs do not have capacitors like SVCs, which comes with a few advantages in terms of vulnerability to extreme temperatures.

% The initial of the two, SVC, based on high power thyristor technology, appeared in the field in the 1970s. STATCOM, based on GTO (Gate Turn-off Thyristors) came into use in the 1980s, and subsequently, with high power IGBT (Insulated Gate Bipolar Transistor) becoming commercially available, STATCOM based on this technology platform came on line in the 1990s. With SVC and STATCOM, the power transmission capability of lines can be increased considerably. Squeezing more power out of existing lines can eliminate or at least postpone the need to build new lines, which all adds up to reduced environmental impact, and significant cost and time savings.

% “Developed and refined over the years in a number of ways, SVC is still ruling as the main controllable shunt compensation device. Simultaneously, STATCOM is gaining momentum and is to an increasing SVC and STATCOM: An overview degree catching the interest of utilities looking for options that can offer additional benefits to those traditionally available, in particular where space in sub-stations is scarce, the reactive power output needs to be controllable more or less independent of the AC system voltage, or a speed of response one order of magnitude greater than what is possible with thyristors is desired. \href{https://library.e.abb.com/public/27b269233c494c81adf24df52c9dc343/Shunt%20compensation_Aug2019.pdf}{Link}”

SVC and STATCOM offer several benefits: 1) voltage control during normal and contingency conditions, 2) reactive power support during contingencies, 3) voltage stability, 4) overvoltage reduction during lightly loaded conditions, 5) damping oscillations, 6) maximizing power transfer capability, and 7) power quality.

% SVC:
% SVC can be built using a variety of designs. However, the controllable elements used in most systems are similar. The commonly used controllable elements are: • Thyristor-controlled reactor (TCR) • Thyristor-switched capacitors (TSC) • Thyristor-switched reactor (TSR) \href{https://library.e.abb.com/public/27b269233c494c81adf24df52c9dc343/Shunt%20compensation_Aug2019.pdf}{Link}

% STATCOM:
% SVC Light® is a STATCOM device, based on a chainlink modular multilevel (MMC) voltage source converter (VSC) concept, particularly adapted for power system applications. It is capable of yielding high reactive power input to the grid more or less unimpeded by suppressed grid voltages, and with high dynamic response. This is useful, for instance, to support more or less weak grids loaded by a large percentage of air conditioners in hot and humid climates, and to improve the availability of large wind farms under varying grid conditions. Insulated Gate Bipolar Transistors (IGBT) and Insulated Gate Commutated Thyrustirs (IGCT) are key components in SVC Light \href{https://library.e.abb.com/public/27b269233c494c81adf24df52c9dc343/Shunt%20compensation_Aug2019.pdf}{Link}

\textbf{Impact of Extreme Temperature Operations:} SVCs have similar points of failures to capacitors and electronics-based devices because of the components within. STATCOMs, which do not have capacitors, have similar but fewer vulnerabilities. Some of the conditions that affect all outdoor equipment are as follows \cite{SVC_SN}% \href{https://www.linkedin.com/pulse/what-do-your-static-var-compensator-svc-performing-expected-esteban-3c/}{Link}
, 1) ambient temperature; 2) seismic activity; 3) wind; 4) climatic conditions including high humidity, heavy rainfall or snow, sand-storms, ice load, etc.; and 5) pollution, including dust, corrosive atmosphere, etc.

% The scope of preventive maintenance for SVCs of steel mills usually includes a site survey, condition evaluation of the SVC equipment and recommendations for further maintenance and upgrades. 

\textbf{Existing Methods to Measure the Impacts:}
Remote monitoring of ambient temperature of the immediate air surrounding the equipment is the main method for assessing potential impacts. \cite{SVC_SN}.% \href{https://www.linkedin.com/pulse/what-do-your-static-var-compensator-svc-performing-expected-esteban-3c/}{Link}.

\textbf{Existing Methods to Mitigate the Impacts:}
Similarly to frequency changers, these issues can be resolved by placing the equipment in a climate-controlled environment/enclosure. This is especially useful when climate conditions can change drastically at a specific location.

% \subsection{FACTS Devices (STATCOM, SSSC, GUPFC, IPFC)}
% Done

\subsection{Switchgear Panels and Control Panels}
% \textcolor{red}{Alok}
\label{subsec:switchgear_ss}
Switchgear is a class of equipment that houses many different critical components in a substation, switchyard, or other key locations where multiple paths of the power network interface. Indoor switchgear is shielded from the environmental conditions to some extent, but outdoor switchgear is prone to impacts of environmental conditions. In medium-voltage and low-voltage systems, the outdoor switchgear is often metal enclosed. These ``weather-proof'' housings, however, cannot eliminate the impacts of extreme heat and extreme cold caused by heat domes and polar vortices or winter storms. 

\begin{figure}[H]
    \centering
    \includegraphics[width=\linewidth]{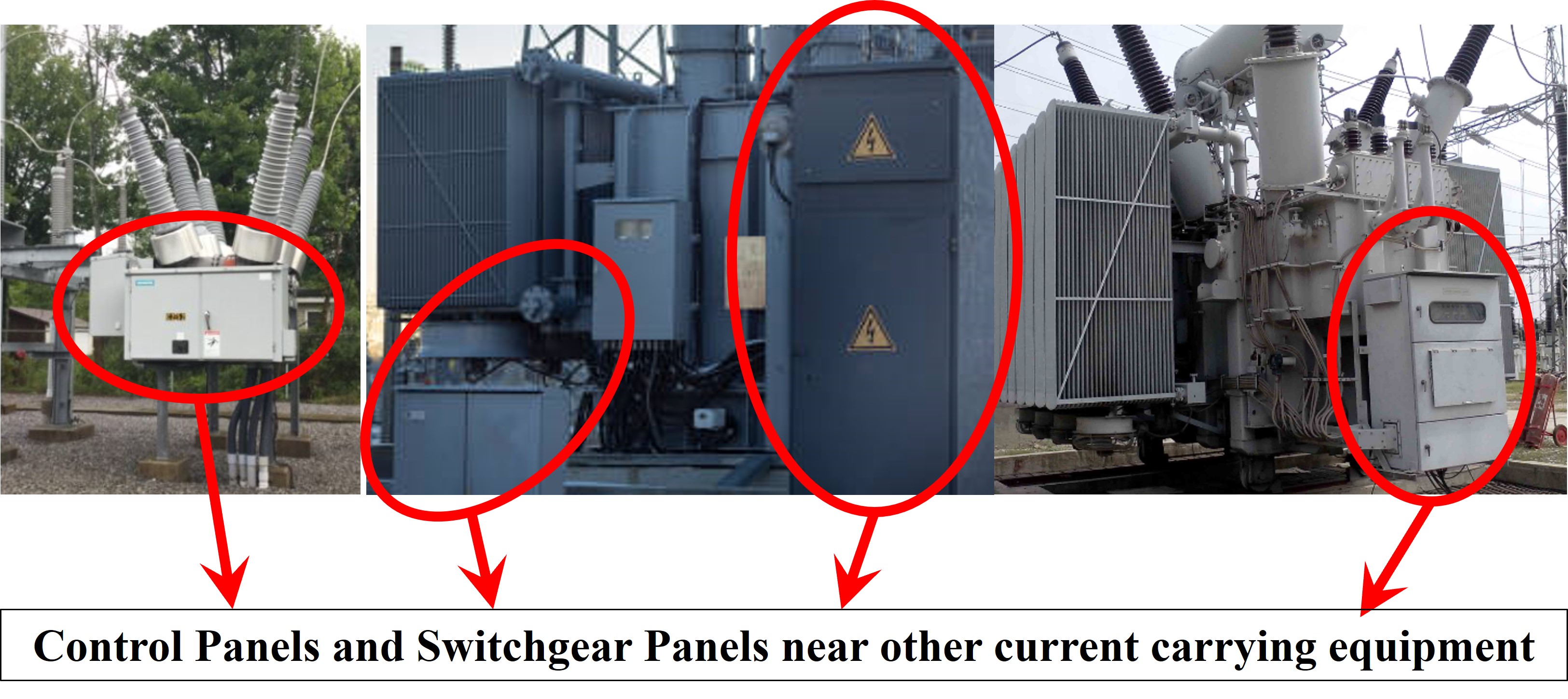}
    \caption{Panels close to other primary equipment}
    \label{fig:panelvicinity}
\end{figure}

There is also outdoor switchgear, typically located in substations, for protecting/managing key substation equipment like transformers and circuit breakers \cite{sg_1}. The radiators and oil pumps are all dedicated low-voltage auxiliary components like motors that are managed with their own switchgear. These are also called the control cabinets, and are mounted either on or near the transformer (\ref{fig:panelvicinity}). 

\textbf{Impact of Extreme Temperature Operations:}
Overheating and insulation failures are the key failure modes for switchgear \cite{sg_2}. The switchgear equipment located in the substation yard is not only subject to weather-related ambient conditions but also the heat generated by neighboring equipment. Especially during extreme heat conditions, the load/demand in the system is high (considering a significantly high air conditioning load). This condition tends to drive a high load current in all the current-carrying equipment, causing internal equipment temperature rise. This, coupled with the high ambient temperature, can lead to failure of multiple components like insulation, adhesives, etc., which in turn can lead to complete equipment failure.

Another aspect that is possible in the field but not reported in the literature is the heat exhaust from surrounding equipment increasing the ambient temperature, which can cause equipment like switchgear and switchgear panels to be exposed to extreme heat.

Fig. \ref{fig:panelvicinity} shows the vicinity of the switchgear and switchgear panels around large power transformers. The heat generated within the power transformer is radiated out to cool the transformer; however, this causes additional ambient temperature-rise around the transformer where the auxiliary equipment switchgear panels also located.

During some weather conditions the switchgear panels have a risk of ice deposition or condensation inside the panel enclosures \cite{sg_3}. This may result in arc flash risk, leading to major failures.

\textbf{Existing Methods to Measure the Impacts:}
There is some technology available in the market to monitor temperatures in low- and medium-voltage switchgear. This temperature monitoring system can be used for detecting irregularities in the following: 1)~the effects of constant thermal cycling on the joints; 2)~corrosion of the metal components damaging insulation; assets aging, mechanical damage, or environmental factors; and intrusion and human error because of maintenance processes.
 
The key failure modes of switchgear are failure of fastened components that can lead to arc flash during extreme heat and icing conditions, causing insulation failure.

\begin{figure}[H]
    \centering
    \includegraphics[scale=0.5]{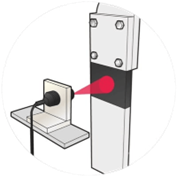}
    \caption{Thermal monitoring inside the switchgear panels and control panels from \cite{sg_4}}
    \label{fig:panelmonitor}
\end{figure}

\textbf{Existing Methods to Mitigate the Impacts:}
The key mitigation methods for outdoor switchgear for extreme heat are periodic inspections and maintenance; proper ventilation and (if needed) forced convection cooling using draft fans, etc.; thermal imaging and alarm systems to detect and manage early failures (\ref{fig:panelmonitor}); and load management, which is mostly derating the equipment \cite{sg_7}.

For extreme cold operating conditions, mitigation measure include understanding and revising the expected extreme cold temperatures and icing conditions followed by conducting the specific type-tests to handle such operation. Preparation for extreme cold and ice storms includes reinforcements of structures and use of composite insulation with hydrophobic properties. For cabinets, enclosures, and panels, the use of special gasket materials, anti-condensation heaters to prevent ice build up inside the panels, cabinets and enclosures is needed \cite{sg_5, sg_6}.

Further developments of technologies suggested in the literature include advanced sensors, data-driven and physics-aware monitoring, and mitigation methods \cite{sg_7}.

\subsection{HVDC Converter Stations}
\begin{figure}[H]
    \centering
    \includegraphics[width=\linewidth]{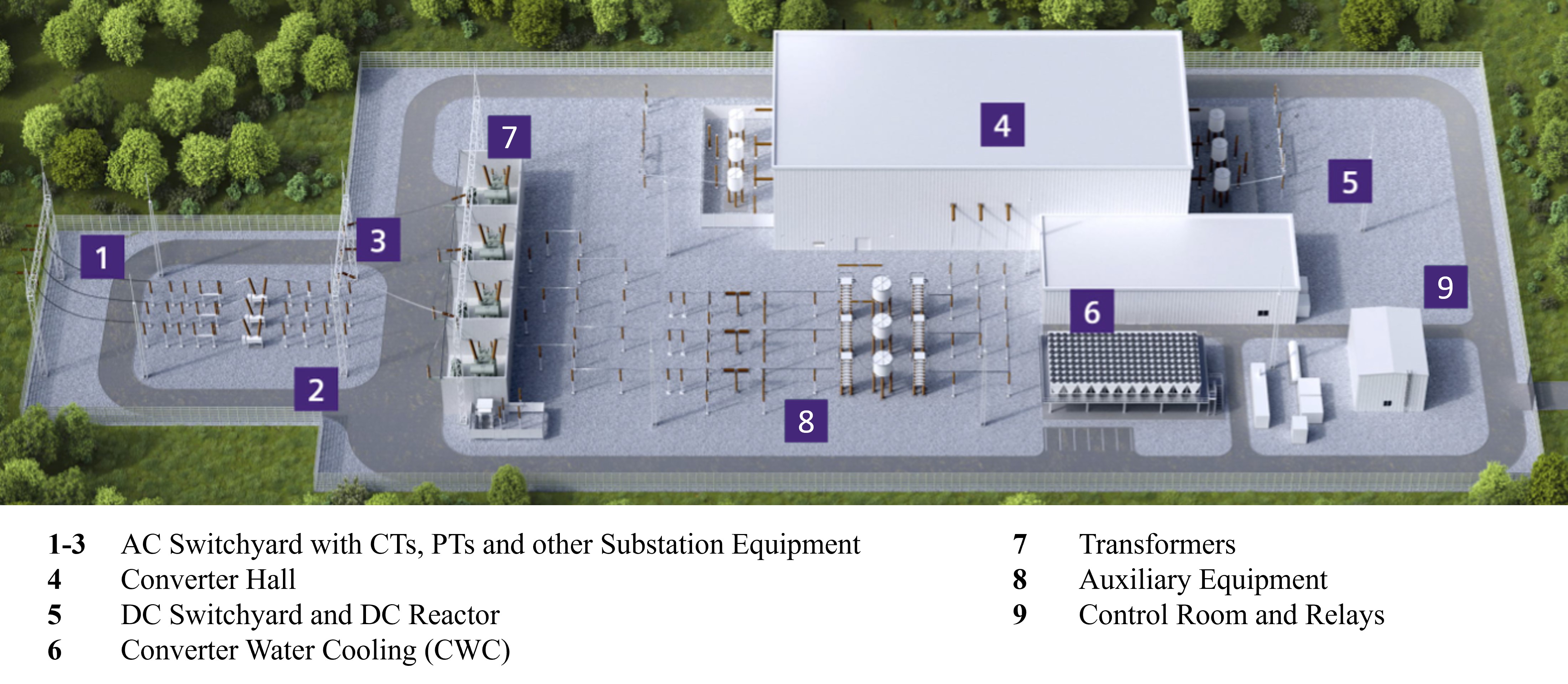}
    \caption{Key parts of HVDC converter station from \cite{hvdc_1}}
    \label{fig:hvdc}
\end{figure}

% \textcolor{red}{Alok}
The high voltage direct current (HVDC) converter stations and associated filter components are a set of infrastructure in themselves \cite{hvdc_6}. Stations are divided into indoor and outdoor components, as shown in Fig.~\ref{fig:hvdc}. 

Other references that showing the key components of a HVDC converter stations also illustrate their indoor-outdoor nature \cite{hvdc_1a,hvdc_2}. There are more than 15 HVDC converter stations in USA with a total capacity greater than 20 GW.  Fig.~\ref{fig:hvdc} shows that the main power electronic converters are housed inside a structure called converter hall that protects the converters from extreme ambient conditions; however, there could be indirect impact of ambient conditions.

%There are other references that show the key components of a HVDC converter station that also illustrate the indoor-outdoor nature of it \cite{hvdc_1a,hvdc_2}. There are more than 15 HVDC converter stations in North America with a total capacity greater than 20 GW \cite{hvdc_3}.  The Fig.~\ref{fig:hvdc} shows that the main power electronic converters are housed inside a structure called converter Hall that protects the converters from extreme ambient conditions, however, there could be indirect impact of ambient conditions.

\textbf{Impact of Extreme Temperature Operations:} Outdoor components of an HVDC converter station are directly affected by extreme heat, cold, and icing conditions. The outdoor components are standard AC equipment and are addressed in the previous sections. However, HVDC converter stations are cooled from external cooling agents like water. In extremely cold conditions and icing conditions, it is important to note that freezing is a common occurrence due to which power electronics can over heat. This failure mode is similar to the failure mode in batteries when the cooling system fails, as discussed in subsection \ref{subsec:batteries_ds}.
Another failure mode for HVDC systems is also the impact of extreme cold and icing on the HVDC conductors \cite{hvdc_7} and this is also similar to failure modes for overhead line conductors.
There are also indirect impacts of extreme temperatures, i.e., the overloading of the HVDC converter stations. During heat domes and heat waves, the efficiency of cooling systems also goes down, which can also lead to indirect failures and possible misoperations of the HVDC converter stations \cite{hvdc_8}. This can also be a result of extreme temperatures affecting the operations of the reactors and auxiliary equipment present in the HVDC converter stations \cite{hvdc_8}.

\textbf{Existing Methods to Measure the Impacts:} The literature does not report major methods for measuring the impact of extreme temperatures on HVDC converter stations. However, because the functioning of converter stations depends on the many types of equipment that compose them, assessing all auxiliary equipment in a converter station can help to measure its overall vulnerability \cite{hvdc_8}.

\textbf{Existing Methods to Mitigate the Impacts:} Apart from other substation equipment vulnerabilities, the vulnerability of cooling systems to extreme cold and icing conditions is important to address. NERC reports on lessons learned \cite{hvdc_4} and guidelines on winterizing and extreme cold weather preparedness and operations \cite{hvdc_5} are good references to consider for mitigation of impacts of extreme temperatures. One of the key aspects in the guidelines is recommendations on winterizing liquid-cooled systems, ensuring that the insulating media and coolants do not freeze during extreme cold conditions.

%\subsection{Harmonic Filters}
%\textcolor{red}{Alok}

\subsection{Power Line Carrier Communication Systems}
% \textcolor{red}{Alok}

\begin{figure}[H]
    \centering
    \includegraphics[width=\linewidth]{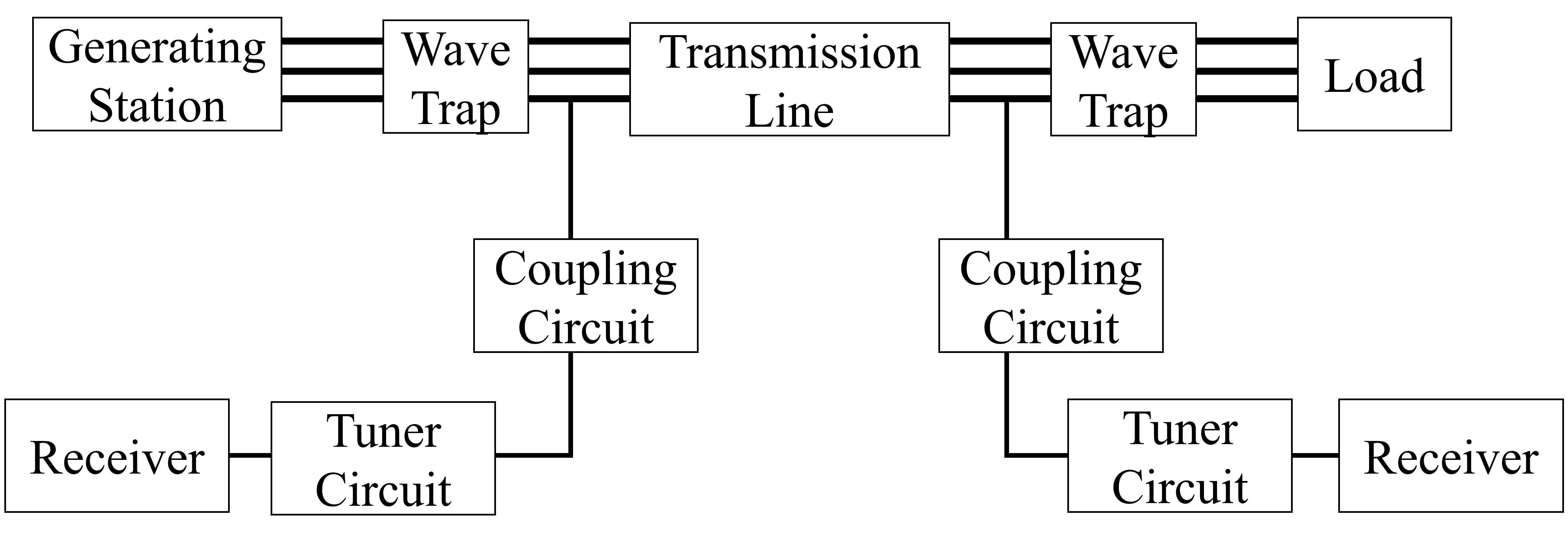}
    \caption{Key subsystems in a PLCC from \cite{plcc_3}.}
    \label{fig:plcc_system}
\end{figure}

Power line carrier communication (PLCC) equipment is typically used for communications over the power carrying conductors. The PLCC system has many different components: coupling capacitor; line trap unit; transmitters and receivers; hybrids and filters; line tuners; master oscillator and amplifiers; protection and earthing of coupling element\cite{plcc_1, plcc_2}. The PLC systems have several critical components that need tuning, and for accurate functioning, the communication systems have a specific range of operating temperatures. 
The usual setup of a PLCC system or a power line carrier (PLC) system is illustrated in Fig.~\ref{fig:plcc_system} \cite{plcc_3}.

Ref. \cite{plcc_3} indicates several applications of PLCC systems in the future power grid, and Ref. \cite{plcc_4} provides the summary of the PLCC market and indicates an increase in the PLCC system market. 

\textbf{Impact of Extreme Temperature Operations}:
Considering their increased applications \cite{plcc_3,plcc_4}, it is important to understand the limitations of the PLCC systems in the future power grid under a changing climate. While parts of the system are mounted in a control panel either indoors or outdoors, parts of the PLCC system are outdoors and house the coupling circuit and the line traps. Reference \cite{plcc_6} provides details of the PLCC system applications that highlights the indoor and outdoor PLCC systems and provides some details of the ambient operating conditions. These are mainly for the indoor part of the PLCC systems, which are usually housed in a panel or enclosure.  
The key components outdoors components of PLCC equipment are the line traps (also called wave traps); these are mainly parallel LC circuits (parallel inductor capacitor circuits) in parallel with a lightning arrestor \cite{plcc_7,plcc_8}. There is some work done on assessing and considering the impacts of temperatures on line traps \cite{plcc_9}. While reference \cite{plcc_9} provides many critical details, it was last updated more than 15 years ago. It may need to be reevaluated in light of recent climate changes.

\textbf{Existing Methods to Measure the Impacts:}
The key impact of extreme temperature on PLCC equipment is malfunction of the communication equipment. Generally, the impact is expected to be temporary and includes increase in the noise levels and attenuation. However, in some rare cases, the damage could be permanent and require replacement of equipment. Reference \cite{plcc_10} provides details of monitoring and measurement methods to identify malfunctions of various subsystems in a PLCC system.

\textbf{Existing Methods to Mitigate the Impacts:} The key failure mode of PLCC equipment is malfunction of equipment, and using the monitoring methods \cite{plcc_10} can identify the faulty equipment. The mitigation methods could include having back-up communication methods for essential control applications in the power grid.

\subsection{Disconnect Switches and Isolators}
% \textcolor{red}{done}

% \textcolor{blue}{Kishan self note: this section has reference numbers from word. Need to change them. Reason, need to find better references.}

Disconnect switches and isolators are electrical switches that control the flow of electricity and isolate circuits when needed for maintenance.

\textbf{Impact of Extreme Temperature Operations
 :} Disconnect switches and isolators are susceptible to the changes in operating temperature. This is because the ratings of the switches and isolators are dependent on the ambient temperature. According to IEEE standard C37.30-1997 \cite{voltage_switch_SN}% (\href{https://ieeexplore.ieee.org/document/663234}{Link})
, the usual service conditions are 1) ambient temperatures between -30\degree C and 40\degree C, 2) altitude does not exceed 1000 $m$, and 3) wind velocity not exceed 37 $m/sec$. The IEEE standard C37.30-1997 \cite{voltage_switch_SN} %(\href{https://ieeexplore.ieee.org/document/663234}{Link})
also suggests the unusual service conditions such as 1) higher altitudes will have lower dielectric strength resulting in higher temperature rise in cooling medium (derating/correction factor must be used appropriately), 2) switches within enclosure must use immediate air around the equipment rather than the ambient temperatures, etc.

PJM interconnection reported their derating calculations for switches and insulators for summer and winter \cite{PJM_Switches_SN}%(\href{https://www.pjm.com/-/media/planning/design-engineering/maac-standards/air-disconnect-switches.ashx}{Link})
. The temperature considered for summer and winter are 35\degree C and 10\degree C respectively. Furthermore, emergency allowable ratings are considered to be 20\degree C above the normal allowable maximum temperatures.

In case of extreme cold, the switches has to pass the ice breaking test which states that the rated ice-breaking ability are 9.5~mm and 19~mm \cite{circuit_switchers}. The rated ice-breaking ability is the maximum thickness of ice deposited on a device that will not interfere with its successful opening or closing. Under certain circumstances, an ice storm can cause a deposit of ice of such thickness that overhead lines fail and operation of specific switching equipment is impaired \cite{cw_2}. Ice is produced naturally in two ways, they are 1) clear ice appears from rain falling through air with a temperature between 0\degree C (+32F) and –10\degree C (+14F)., and 2) Rime ice, characterized by a white appearance from the air entrapped during ice formation, appears from rain falling through air with a temperature below –10\degree C (+14 °F) or from condensation of atmospheric moisture on cold surfaces.

\textbf{Existing Methods to Measure the Impacts:} The method to measure the impacts of temperatures on switches and isolators is by monitoring the ambient temperatures and the associated derated maximum rating value. For example, PJM interconnection reported that the derated maximum ratings for winter and summer are 131\% and 108\% respectively \cite{PJM_Switches_SN}% (\href{https://www.pjm.com/-/media/planning/design-engineering/maac-standards/air-disconnect-switches.ashx}{Link})
. It also reported that the maximum rating can change based on the duration of the different loading on the equipment. A control room based monitoring of the status of the switch can also be used.

\textbf{Existing Methods to Mitigate the Impacts:} A method to mitigate the impact on switches and isolators is by installing protection system (for motorized switches). However, the protection system's settings must be adjusted based on the derating behavior of the switches which is not typical.

\subsection{Backup Batteries}
\label{subsec:backup_battery_ss}
Most of the substation equipment like relays, motors for circuit breakers, re-closers, isolators, sectionalizers, etc., need electric power supply to operate. Most substations have a dedicated backup power supply in the form of batteries that are mainly used to support the substation operation including the relay and other critical equipment that is housed inside the substation control room.

\textbf{Impact of Extreme Temperature Operations:}
The battery systems in the substations are usually stored in a separate battery room and this room needs to be maintained at a temperature of 25$\degree$C. Else the battery life is significantly shortened. Extreme heat causes a shortening of the lifespan of the battery. Extreme cold causes the a reduced charge and increased Sulfation leading to corrosion and overall deterioration of the battery \cite{bupbatt_1}. Some specifications show that life of a Ni-Cad battery  is better compared to a lead-acid battery with increased temperatures \cite{bupbatt_1}. Extreme heat can cause irreversible damage to the battery systems \cite{bupbatt_2}.

\textbf{Existing Methods to Measure the Impacts:} There are some devices available to measure the health of a stationary battery system like the ones used for backup power in substations called battery testers \cite{bupbatt_3}. While these battery testers are well developed for lead-acid battery chemistry. There are other methods that are more suitable for other battery chemistry that are described in detail in reference \cite{bupbatt_4}, where there are methods and metrics that are evaluated to measure the health of a battery system. Since extreme cold and extreme heat operations of the battery impact the state of the charge, battery end-of-life (EOL), the methods described in reference \cite{bupbatt_4} can be directly applied.

\textbf{Existing Methods to Mitigate the Impacts:} The key methods for mitigating the adverse impacts of extreme temperature operation are to have a temperature controlled environment of the batteries. This is however a challenge in peak summer and peak winter durations, especially under events like heat domes and polar vortices as the external temperature often tends to limit the amount of cooling/heating a heating ventilating and air conditioning (HVAC) system could provide \cite{bupbatt_5}. ANSI/IEEE 484 provides more details of the battery room requirements \cite{bupbatt_6}. Reference \cite{bupbatt_7} provide details of operating temperatures and other considerations for designing and operating battery rooms in a substation.

% \section{CASCADING FAILURE EFFECTS}
% \label{sec:cascade}
% \input{tex_files/cascade}

\section{STANDARDS}
\label{sec:ST}
In this section, the current standards that are available in the literature for various substation, transmission and distribution components when subjected to extreme weather conditions have been discussed in detail.

\subsection{Lightning Arrestors}
\label{subsec:Lt_Ar}
\begin{figure}
    \centering
    \includegraphics[scale=0.5]{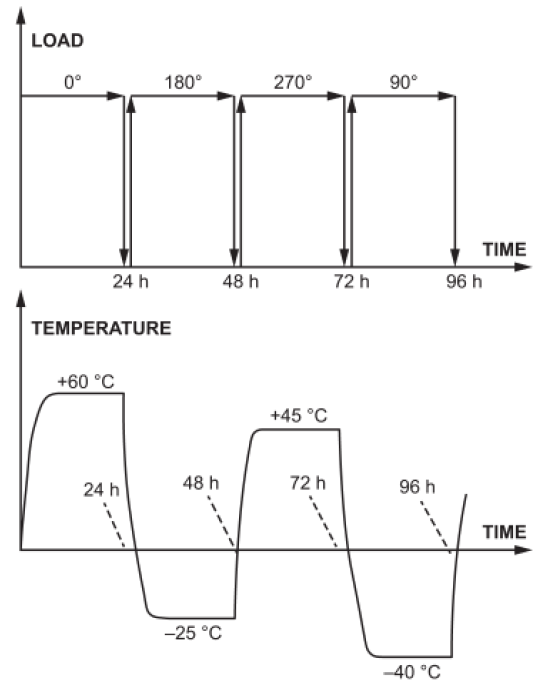}
    \caption{Thermal stress tests behavior of lightning arrestors from \cite{lightning_arrestors}.}
    \label{fig:thermal_test_LT_AR}
\end{figure}

IEEE Std C62.11-2020 \cite{lightning_arrestors}, \textit{IEEE Standard for Metal-Oxide Surge
Arresters for AC Power Circuits (>1 kV)}, states that before utilizing them in the field, the lightning arrestors are typically made to go through a process called thermomechanical conditioning by applying controlled thermal, mechanical, and electrical stress to them to improve their performance and reliability \cite{lightning_arrestors}. From Fig.~\ref{fig:thermal_test_LT_AR}, it can be observed that over a 24-hour period, these components would be undergoing thermal stress between $-40\degree C$ to $+60\degree C$. For example, an extreme cold event like the great Texas freeze in 2021 \cite{lightning_arrestors}, led to cold temperatures for prolonged period of time up to 8 days. Furthermore, this testing does not account for temperatures lower than $-40\degree C$ and also for longer durations.  This standard also provides the ambient air temperature and arrestor temperature limits as shown in Tab.~\ref{Temp limits for arrestors}   

\begin{table}[H]
\caption{Temperature limits for arrestors from \cite{lightning_arrestors}}
\begin{tabular}{lllll}
\cline{1-3}
\multicolumn{1}{|l|}{}                & \multicolumn{1}{l|}{\begin{tabular}[c]{@{}l@{}}\textbf{Minimum air} \\ \textbf{temperature} ($\degree C$)\end{tabular}} & \multicolumn{1}{l|}{\begin{tabular}[c]{@{}l@{}l@{}}\textbf{Maximum} \\ \textbf{air/arrestor} \\ \textbf{temperature} ($\degree C$)\end{tabular}} &  &  \\ \cline{1-3}
\multicolumn{1}{|l|}{Deadfront}       & \multicolumn{1}{l|}{-40}                                                                          & \multicolumn{1}{l|}{65/85}                                                                                 &  &  \\ \cline{1-3}
\multicolumn{1}{|l|}{liquid-immersed} & \multicolumn{1}{l|}{-40}                                                                          & \multicolumn{1}{l|}{95/120}                                                                                &  &  \\ \cline{1-3}
                                      &                                                                                                   &                                                                                                            &  & 
\end{tabular}
\label{Temp limits for arrestors}
\end{table}

 Equation \ref{accl_aging} \cite{lightning_arrestors}, provides the relationship between the accelerated aging of the arrestors as a function of continuous elevated levels of temperature exposed to the arrestor. However, this formula does not talk about what happens to the aging factor of the arrestor at extremely cold temperatures such as -40 C and below. The impact of variance in temperatures away from the nominal ambient temperature has been presented in Tab.~B.2 in \cite{lightning_arrestors}, which shows that the equivalent time of usage of dead front type arrestors and the remaining type arrestors accelerates 15 times and 142 times faster compared to liquid-immersed type arrestors assuming a 40C  average temperature use. 

\begin{equation}
\label{accl_aging}
AF = 2.5 ^{\frac{\Delta t}{10}}
\end{equation}

where $\Delta t$ is the difference between test temperature and weighted average use temperature.

In \cite{lightning_arrestors}, it was also reported, that, 70\% of all distribution arrestor failures were caused by moisture ingress due to atmospheric pressure and temperature excursions which cause the water vapor to get seeped into a defective seal and get trapped inside the arrestor leading to total arrestor failure over time after sufficient moisture penetration in this manner.

\subsection{Grounding Resistors}
\label{subsec:Lt_Ar}
  IEEE Std C57.32-2015 \cite{groundresistors}, \textit{IEEE Standard for Requirements,
Terminology, and Test Procedures for
Neutral Grounding Devices}, reports that, under usual service conditions, the ambient temperature limits for the grounding resistors should be between $-20\degree C$ and $40\degree C$. The maximum average ambient temperature limit over 24 hours is recommended to not exceed $30\degree C$.

Equation \ref{Eq_Temp_coeff_of_R_in_terms_of_R2} in \cite{groundresistors}, provides the relationship between the variance in the grounding resistance as a function of changes in its temperature and its temperature coefficient. It was also recommended that the temperature coefficient of resistance should not exceed 0.000263 per degree C to ensure that the resistance does not change drastically between $30\degree C$ and $790\degree C$ to avoid the scenario of the very low fault currents which could render the protection devices to be ineffective. However, this standard does not talk about what happens to the impact on the variance of grounding resistance at extremely low temperatures (less than $-20\degree C$) which is critical because if the resistance value becomes smaller at lower temperatures then it might defeat the purpose of using the grounding resistors to limit the fault currents to a safe level.

\begin{equation}
\label{Eq_Temp_coeff_of_R_in_terms_of_R2}
R_{new} = R_{old} (1+ \alpha  (\theta_{new} - \theta_{old}))
\end{equation}

$R_{old}$ and $R_{new}$ are resistances in ohms at temperatures $\theta_{old}$ and $\theta_{new}$ in degree celsius respectively, and $\alpha$ is the temperature coefficient of resistance. 

 The limits for maximum temperature (hottest spot temperature) rises for the rated time and steady-state time of thermal current ratings ($760\degree C$) and continuous currents ratings ($385\degree C$)  respectively have also been provided in Tab.~25 of \cite{groundresistors}. However, the cold spot temperature limits for the current carrying parts of this component have not been provided in the literature.

\subsection{Capacitor Banks}
\label{subsec:Lt_Ar}
IEEE Std 1036-2020, \cite{capacitor_banks}, \textit{IEEE Guide for the Application of Shunt Power Capacitors}, recommends the maximum average ambient temperature of the capacitors (indoor or outdoor) is to be $46\degree C$ with a peak limit of $55\degree C$ in \cite{capacitor_banks}. Additionally, the capacitors should be capable of operation at a minimum ambient temperature of $-40\degree C$ and it is suggested that some users in more harshly colder climates can specify a minimum ambient temperature of $-50\degree C$. It is also recommended to consider different average maximum ambient temperatures depending on the type of mounting arrangement for the capacitors as shown below in Tab.~\ref {capacitor_mounting}.

\begin{table}[H]
\centering
\caption{Maximum average temperature limits of the capacitors based on its mounting arrangement}
\begin{tabular}{|c|c|}
\hline
\textbf{Mounting   Arrangement}       & \multicolumn{1}{l|}{\begin{tabular}[c]{@{}l@{}} \textbf{Maximum average} \\ \textbf{ambient temperature ($\degree C$)}\end{tabular}} \\ \hline
Isolated capacitor                    & 46   \\ \hline
Single row of capacitors              & 46    \\ \hline
\multicolumn{1}{|c|}{\begin{tabular}[c]{@{}l@{}}Multiple rows \\ and tiers of capacitors\end{tabular}} & 40                   \\ \hline
\multicolumn{1}{|c|}{\begin{tabular}[c]{@{}l@{}}Metal-enclosed \\ or housed equipment\end{tabular}}    & 40                 \\ \hline
\end{tabular}
\label{capacitor_mounting}
\end{table}

% \begin{tabular}{|c|c|}
% \Xhline{2\arrayrulewidth}
% \makecell[c]{\textbf{Mounting Arrangement}} & \makecell[c]{\textbf{Maximum Average Ambient Temperature (°C)}} \\
% \Xhline{2\arrayrulewidth}
% Isolated capacitor & 46 \\
% \hline
% Single row of capacitors & 46 \\
% \hline
% Multiple rows and tiers of capacitors & 40 \\
% \hline
% Metal-enclosed or housed equipment & 40 \\
% \Xhline{2\arrayrulewidth}
% \end{tabular}

\subsection{Harmonic Filters}
\label{subsec:Lt_Ar}
From IEEE Std 1531‐2020, \cite{harmonic_filters}, \textit{IEEE Guide for the Application and Specification of Harmonic Filters}, it can be seen that the ambient temperature experienced by the harmonic filter reactor and capacitors plays an important role in the design of the harmonic filters. Typically, the harmonic filters are designed to operate at a maximum ambient temperature of $40\degree C$. It is also stressed that during the harmonic filter design, the placement location of the harmonic filter reactor is critical, especially that of an iron core reactor.    

It was suggested to have auxiliary cooling or fans to ensure the filter components operate below their rated operating temperatures for hot environments. Importantly, thermostatically controlled heaters are recommended to maintain optimum temperature within the filter enclosure for colder climates to prevent the formation of condensation inside the filter. The absence of the heaters may lead to the failure of the filter.

%\subsection{HVDC Converters}
%\label{subsec:Lt_Ar}
%\input{tex_files/HVDC_converters_st}

\subsection{Power Fuses}
\label{subsec:Lt_Ar}
IEEE Std C37.41-2016 \cite{power_fuses}, \textit{IEEE Standard Design Tests for High-Voltage (> 1000 V) Fuses and Accessories}, talks about the standards used for the design tests of the high voltage fuses (> 1 kV) devices rated upto 170 kV that are typically used on the distribution systems. This standard typically talks about two types of fuses - expulsion fuses and current-limiting fuses and mentions that this standard can also be applied to other types of fuse devices such as non-expulsion type fuses. 

The rated current of a power fuse is defined as the constant RMS current in amperes, aa t constant frequency, at a maximum ambient temperature of $40\degree C$ and for a specified maximum component temperature rises \cite{power_fuses}. By agreement between the user and the manufacturer, the rated current can be designed to a fuse device for ambient temperatures above $40\degree C$ which has been termed as the maximum permissible continuous current. Additionally, the manufacturer provides another rating  - rated maximum application temperature (RMAT) - which represents the maximum ambient temperature at which a device can be used without causing any deterioration that would inhibit its ability to interrupt the
circuit \cite{power_fuses}.

In this standard, under normal service conditions, the ambient temperature limits of the surrounding medium for the fuses are defined to be not above $40\degree C$ and not below $-30\degree C$ if the RMAT of the considered fuses has a maximum limit of $40\degree C$. In special circumstances, for fuses whose RMAT value is greater than $40\degree C$, it should be ensured that its surrounding medium ambient temperature does not exceed its corresponding RMAT value and should be greater than $-30\degree C$. Although this standard does not specifically account for extreme cold weather conditions below $-30\degree C$, it has been reported here that some designs have been considered for the fuses in the past for ambient temperatures below $-30\degree C$. 

The limits of temperature and the temperature rise seen in the fuses components, whose RMAT is less than $40\degree C$, have been provided in Tab.~2 of \cite{power_fuses}. These limits have been provided for various types of contacts in air, other insulating gases, and liquid insulating material (spring-loaded contacts, bare, tin-coated, silver, nickel-coated coated, and other coatings). The maximum temperature rise limit has been observed to be for insulation class type F with a temperature rise of $115\degree K$ and its corresponding maximum temperature limit to be $155\degree C$.

\subsection{Relays}
\label{subsec:Lt_Ar}
In IEEE Std C37.90-2005 \cite{relays_general}, \textit{IEEE Standard for Relays and
Relay Systems Associated with
Electric Power Apparatus}, the operational temperature overall limits range of the relays has been defined to be between $-40\degree C$ and $70\degree C$. Specifically, for a specific relay, any of the following temperature ranges can be chosen before being deployed in the field for corresponding operation/non-operational purposes (transportation, storage, and installation)

\begin{itemize}
\item $-40\degree C/\hspace{-0.05in}-\hspace{-0.05in}50\degree C $ to $70\degree C/85\degree C$
\item $-30\degree C/\hspace{-0.05in}-\hspace{-0.05in}40\degree C$ to $65\degree C/75\degree C$
\item $-20\degree C/\hspace{-0.05in}-\hspace{-0.05in}30\degree C$ to $55\degree C/65\degree C$
\item Custom range can be defined by the manufacturer but must be within $-20\degree C/\hspace{-0.05in}-\hspace{-0.05in}30\degree C$ to $55\degree C/65\degree C$ limits
\end{itemize}

The insulation and the temperature ratings would be derated for the relays at higher altitudes due to the impact of the relative decrease in air density on the dielectric strength and the cooling effect of air insulation of the relays. The Impact on the temperature derating factor (which proportionally decreases the maximum ambient temperature conditions by the manufacturers) for varying levels of altitude has been reported in \cite{relays_general}. In this table, the temperature derating factor range has been provided from 0.6-1 (for any altitude above 1500 m this factor gradually decreases until 0.6 for a maximum of 6000 m.  

For extreme weather events such as temperature excursions, it is recommended that the manufacturer consider it but didn’t specify how exactly they need to account for it in their calculations.

Typically, electromechanical direct current (DC) auxiliary relays, in hot conditions, pick up at higher proportions of the rated voltage (80\% or less) compared to cold conditions (72\% or less of rated voltage). For alternate current (AC) relays, they should be able to continuously withstand 110\% of the rated voltage with maximum temperature rises as presented in Tab.~5 \cite{relays_general}. In Tab.~5 from \cite{relays_general}, limits of temperature rise for coils for various insulation classes (class 105 – class 220) for varying ambient temperatures of both DC and alternate current (AC) type auxiliary relays are given. 

Additionally, IEEE Std C37.90.3-2023 \cite{relays_discharge_tests}, \textit{IEEE Standard for Electrostatic Discharge Tests for Protective Relays}, recommends conducting electrostatic discharge tests of the relays within the ambient temperature limits of  $15\degree C$ to $35\degree C$.

\subsection{Shunt Reactors}
\label{subsec:Lt_Ar}
IEEE Std C57.21-2021 \cite{shunt_reactors}, \textit{IEEE Standard for Requirements, Terminology, and Test Code for Shunt Reactors Rated Over 500 kVA}, points out that the ambient temperature plays a significant role in determining the life of insulating materials used in the shunt reactors. 

Under usual service conditions, it is recommended that the ambient temperature of the cooling air surrounding the reactor at any point in time and an average calculated over 24 hrs should not exceed $40\degree C$ and $30\degree C$ respectively. It was also suggested that special consideration should be given during the design stage by the manufacturers if the ambient temperature is out of the specified range but did not specify how exactly the design specifications would change for large changes in ambient temperature.

When operated at the maximum operating voltage, it was recommended that the maximum \textit{rise} in the hottest spot-conductor temperature above the ambient temperature should not exceed the limits provided in Tab.~3 from \cite{shunt_reactors}. This table contains the temperature rise limits for both liquid-immersed and dry-type reactors for various insulation temperature classes (105, 130, 155, 180, and 220) for the latter. It was also mentioned that the allowable hottest spot temperature rise can be increased in cases where the average ambient temperature is observed to be lower than $30\degree C$, however, if the average ambient temperature is more than the recommended $30\degree C$, it was not specified if any modifications can be made for the allowable hottest spot temperature rise. 

The ambient temperature limits for the temperature rise tests of the two types of shunt reactors - liquid-immersed iron-core (Class ONAN) type and dry-type air-core (Class AN) reactors should be within the $10\degree C$ and $40\degree C$. However, for the latter reactor type, the correction factor should be used as a function of ambient temperature in this range and is calculated as shown below in \ref{Ambient_Temp_Correction_reactors}.

\begin{equation}
\label{Ambient_Temp_Correction_reactors}
%C = \frac{T_m  + 30\degree C}{T_m + T_a}
C = \frac{T_m  + 30^{\circ} C}{T_m + T_a}
\end{equation}

Where, $T_m$ = 234.5\degree C and 225\degree C for copper and aluminium respectively and $T_a$ corresponds to the ambient temperature

It is also recommended that any correction factor should not be utilized for the Class ONAN-type reactors within the specified ambient temperature limits. However, for ambient temperatures outside this range, if a suitable correction factor is applicable then it can be applied to both the reactor types.

% \begin{itemize}
% \item Class ONAN type - The ambient temperature of the surrounding air should be within the $10\degree C$ and $40\degree C$. Any correction factor would not be applied outside of this temperature range during these tests.
% \item Class AN type - The ambient temperature of the surrounding air should be within the $10\degree C$ and $40\degree C$. The temperature range 
% \end{itemize}

\subsection{Composite Insulators}
\label{subsec:Lt_Ar}
IEEE Std 987-2001 \cite{composite_insulators}, \textit{IEEE Guide for Application of Composite Insulators}, reports that typically there would not be any unusual effects on the composite insulators in extreme hot or cold conditions, and the manufacturers have agreed to design these insulators in a general ambient temperature range between $-50\degree C$ and $50\degree C$. However, it is also discussed that external changes on an insulator such as surface cracking and oozing can be observed due to its polymer aging under electrical stress because of temperature excursions.   

\subsection{Circuit Switchers}
\label{subsec:Lt_Ar}
IEEE Std C37.30.1-2022 \cite{circuit_switchers}, \textit{IEEE Standard Requirements for AC High-Voltage Air Switches Rated Above 1000 V}, under usual service conditions, considers the ambient temperature limits of the circuit switchers to be between $-30\degree C$ and $40\degree C$ and the wind velocity must not exceed $36 m/s $. It was also recommended that under unusual service conditions, such as at very high or low ambient temperatures, manufacturers responsible for the design of these components need to be notified about this to enable them to factor these conditions into their calculations. 

Tab.~1 in \cite{circuit_switchers}, provides the altitude correction factors used to derate the current carrying capability of the circuit switchers assuming a maximum ambient temperature of $40\degree C$. However, this standard does not mention what correction factor needs to be considered for extremely hot conditions (more than $40\degree C$).

The maximum continuous current that a switch can carry as a function of the ambient temperature, has been presented in \ref{Eq_allowable_cont_curr} \cite{circuit_switchers}, without exceeding the maximum allowable temperature of any of its parts.

\begin{equation}
\label{Eq_allowable_cont_curr}
I_a = I_r \left( \frac{\theta_{max} - \theta_a}{\theta_{ri}}\right)^{1/2}
\end{equation}

where
\begin{equation*}
\begin{aligned}
 \theta_a -& \text{ambient temperature in ($^{\circ} C$)} \\
I_a -& \text{allowable continuous current at ambient temperature, $\theta_a$} \\
I_r -& \text{rated continuous current}\\
\theta_{max} -& \text{allowable maximum temperature in ($^{\circ} C$) of switch part}\\ %&\text{ from Table 3} \\ 
\theta_{ri} -& \text{limit of observable temperature rise in ($^{\circ} C$) at rated }\\ &\text{ current of switch part} \\ 
\end{aligned}
\end{equation*}

The limits of the temperature rise in the circuit switchers has been defined based on its type - indoor (enclosed switch) and outdoor (non-enclosed switch), as shown below in \ref{Eq_theta_ri} and \ref{Eq_theta_ri_2} respectively. 

\begin{equation}
\label{Eq_theta_ri}
\theta_{ri} = \left( \frac{\theta_{max} - \theta_c}{1.5}\right)
\end{equation}

\begin{equation}
\label{Eq_theta_ri_2}
\theta_{ri} =  min\left(\frac{\theta_{max} - \theta_{am_1}}{1.5},\theta_{max} - \theta_{am_2}\right)
\end{equation}

Where, $\theta_c$, $\theta_{am_1}$ and $\theta_{am_2}$ are assumed to be $25\degree C$, $40\degree C$ and $55\degree C$ respectively at a loadability factor to be 1.22. It should be noted that loadability factor ($LF$) of a switch, at $25\degree C$ ambient temperature, is defined as follows

\begin{equation}
LF =   \frac{I_{a}}{I_{r}}
\end{equation}

The values of  $\theta_{ri}$, $\theta_{max}$ for various individual types of parts of the circuit switchers such as insulator caps, pins, bushing caps, contacts considering different types of material types (copper, copper alloy, silver, silver alloy) are provided in Tab.~3 in \cite{circuit_switchers}.

The ambient temperature limits considered for different types of testing conducted for the circuit switchers before deploying them into the field are provided in Tab.~\ref{Amb temp limits switchers}. Although, the considered tests do not consider extreme weather conditions (less than -$40\degree C$ and more than $40\degree C$) for long durations. 

\begin{table}[H]
\caption{Ambient temperature limits for various tests from \cite{circuit_switchers}}
\begin{tabular}{|l|c|c|ll}
\cline{1-3}
\textbf{Test type}                                                               & \multicolumn{1}{l|}{\begin{tabular}[c]{@{}l@{}}\textbf{Min ambient} \\ \textbf{temperature} (\degree C)\end{tabular}} & \multicolumn{1}{l|}{\begin{tabular}[c]{@{}l@{}}\textbf{Max ambient} \\ \textbf{temperature} (\degree C)\end{tabular}} &  &  \\ \cline{1-3}
\begin{tabular}[c]{@{}l@{}}Continuous \\ current\end{tabular}           & 10                                                                                           & 40                                                                                           &  &  \\ \cline{1-3}
Switch                                                                  & 25                                                                                           & 30                                                                                           &  &  \\ \cline{1-3}
\begin{tabular}[c]{@{}l@{}}Ice loading \\ (Indoor/Outdoor)\end{tabular} & -6/-15                                                                                       & 3/-3                                                                                         &  &  \\ \cline{1-3}
\end{tabular}
\label{Amb temp limits switchers}
\end{table}

\subsection{Circuit Breakers}
\label{subsec:Lt_Ar}
IEEE Std C37.04-2018 \cite{circuit_breakers}, \textit{IEEE Standard for Ratings and
Requirements for AC High-Voltage
Circuit Breakers with Rated Maximum
Voltage Above 1000 V}, recommends that, under usual service conditions, the ambient temperature at any point of time for both the indoor and outdoor type circuit breakers should not exceed $40\degree C$ and on average should not exceed $35\degree C$. Additionally, the ambient temperature of the indoor circuit breakers should not drop below $-5\degree C$, whereas for the outdoor circuit breakers, it should not drop below $-30\degree C$, and the ice coating should be below $20mm$.

For unusual service conditions, at extreme temperatures, the following ambient temperature limits are recommended for the circuit breakers

\begin{itemize}
\item $-50\degree C $ to $40\degree C$ for extremely cold climates
\item $-40\degree C $ to $40\degree C$ for very cold climates
\item $-30\degree C $ to $40\degree C$ for cold climates (indoor circuit breakers only)
\item $-15\degree C $ to $40\degree C$ for moderate climates (indoor circuit breakers only)
\item $-5\degree C $ to $55\degree C$ for hot climates
\item $-15\degree C $ to $55\degree C$ for hot and dry climates
\end{itemize}

Tab.~1 in \cite{circuit_breakers}, provides the maximum temperature rise limits of the individual components within a circuit breaker such as insulation material (types O, A, B, F, H, C and Oil) connections (bare-copper, bare aluminium, bare-copper alloy, bare-aluminium alloy, and silver/Nickel/Tin coated) and surfaces (depending on whether the external surfaces are handled, accessible or not accessible for an operator) assuming a nominal ambient temperature of  $40\degree C $. It should be noted that the temperature rise limits provided in Tab.~1 in \cite{circuit_breakers} are provided for various types of circuit breakers depending on the usage of either non-reactive gases (SF6, N2, CO2, and CF4) or reactive gases (ambient/dry air).

It was also reported in IEEE Std C37.119-2016 \cite{circuit_breaker_generator}, \textit{IEEE Guide for Breaker Failure
Protection of Power Circuit Breakers},
that exposure to prolonged low ambient temperatures, leads to the failure in the operation of the circuit breakers to act in time, typically used for the protection of generators, especially when the generators are being reconnected to the system. This could lead to severe damage to the generator and turbine, due to the former being out of phase with the system when the circuit breakers fail to close within the expected time. However, in this standard, it has not been mentioned for how long and how low the ambient temperatures need to be for the circuit breakers to fail in this manner which could lead to loss of costly and crucial equipment such as the generators.

\subsection{Overhead Lines/ Underground Cables}
\label{subsec:Lt_Ar}
The existing standards do not report any recommended ambient temperature limits nor the impact of extreme weather conditions on the overhead lines and the underground cables. 

However, IEEE Std 835-1994 \cite{cables_1994}, \textit{IEEE Standard Power Cable Ampacity Tables}, and IEEE Std 835a-2012 (the latest revised version of the former standard) \cite{cables_2012}, reported the ampacity of the cables as a function of changing ambient temperature, assuming a nominal ambient temperature of $25\degree C$ and $40\degree C$ for underground cables and overhead lines respectively as shown in \ref{ampacity_lines}. 

\begin{equation}
I_{amp} = \sqrt{\frac{T_{mc} - T_{am}'}{T_{mc} - T_{am}}}I_{0}
\label{ampacity_lines}
\end{equation}

where
\begin{equation*}
\begin{aligned}
T_{mc} &- \text{is the lookup maximum conductor temperature} \\
T_{am} &- \text{is the lookup ambient temperature} \\
T_{am}' &- \text{is the ambient temperature at any particular instant} \\
I_{0} &- \text{is the lookup ampacity of the cable} \\
I_{amp} &- \text{is the adjusted ampacity of the cable }
\end{aligned}
\end{equation*}

It should be noted that $T_{mc}$, $T_{am}$ and $I_{0}$ can be obtained from lookup tables provided in \cite{cables_1994}.

IEEE Std 738-2023 \cite{cables_2023}, \textit{IEEE Standard for Calculating the
Current-Temperature Relationship
of Bare Overhead Conductors}, also do not give any recommendations for ambient temperature limits. It mainly considers the conductor temperature and its variance impact on the line ratings using a thermal model. The users are recommended to use their own assumptions while considering the ambient temperature and wind speeds.

\cite{cables_cigre2006guide}, %CIGRE document
reports certain methods to use weather data (considering both ambient temperature and wind speed correlation) to adjust the line ratings. For example, it is suggested to reduce the assumed wind speed based on its correlation to the ambient temperature changes while adjusting the baseline rating. It is also suggested to use different assumptions for the ambient temperature and the wind speeds depending on the time of the day - daytime/nighttime to account for varying radiation heat losses and precipitation cooling encountered by the line throughout the day.       

\subsection{Instrument Transformers}
\label{subsec:Lt_Ar}
IEEE Std C57.13.5-2019, \cite{instrument_tranformers}, \textit{IEEE Standard for Performance and
Test Requirements for Instrument
Transformers of a Nominal System
Voltage of 115 kV and Above}, reports that the dissipation factor should not be more than  $0.5\% $ and $0.15\% $ for oil-immersed and gas-filled instrument transformers respectively assuming an ambient temperature of $20\degree C$. However, the consideration of the dissipation factors at extreme hot/cold ambient temperatures has not been reported. 

For gas-filled instrument transformers, the accepted leakage rates of the SF$_{6}$, SF$_{6}$-N$_{2}$ gas mixtures, is reported to be 0.5\% per year assuming a nominal ambient temperature range between $10\degree C$ and $40\degree C$.  It has also been recommended that increased leakage rates at extreme ambient temperatures are acceptable as long as the leakage rates revert to permissible levels at nominal ambient temperatures. The acceptable leakage rates at extremely low ambient temperatures up to $-40\degree C$ and $-50\degree C$ are $1.5\%$ and $3\%$ respectively. However, the leakage rates at extremely high ambient temperatures above $40\degree C$ have not been provided. An ambient temperature range between $10\degree C$ and $40\degree C$ with $20\degree C$ as the reference temperature has been recommended to conduct leakage test, short-circuit, and internal arc tests to evaluate the performance of the instrument transformers before deploying them in the field.

\subsection{Battery}
\label{subsec:Lt_Ar}
IEEE Std 1635-2022 \cite{battery_ashrae}, \textit{IEEE/ASHRAE Guide for the Ventilation and
Thermal Management of Batteries for
Stationary Applications}, recommends the ideal ambient temperature limits for various types of batteries typically used in substations such as vented lead-acid (VLA), valve-regulated lead acid (VRLA), or Nickel-Cadnium (NiCd) type batteries to be between $19\degree C$ and $24\degree C$. This is because the internal battery temperature is approximately $1\degree C$ higher than its surrounding ambient temperature and the battery internal optimum electrolyte temperature is typically recommended to be between $20\degree C$ and $25\degree C$.

For closed cabinet battery installations that use VRLA-type batteries, it is recommended that the maximum rise in temperature within the cabinet should be $2\degree C$ by having either forced ventilation or sufficient vent openings. It was reported that the thermal runaway issues seen in the VRLA batteries seem to be exacerbated by elevated ambient temperatures. Therefore, an increase in ambient temperature leads to reduction in battery life but the available capacity of the battery decreases with a decrease in ambient temperature.  The cell-size correction factors used for the VLA and VRLA batteries capacity design as a function of the ambient temperature is provided in Tab.~1 in IEEE Std 485-2020 \cite{battery_VRLA_Sizing}, \textit{IEEE Recommended Practice
for Sizing Lead-Acid Batteries
for Stationary Applications}.

\begin{figure}
    \centering
    \includegraphics[width=\linewidth]{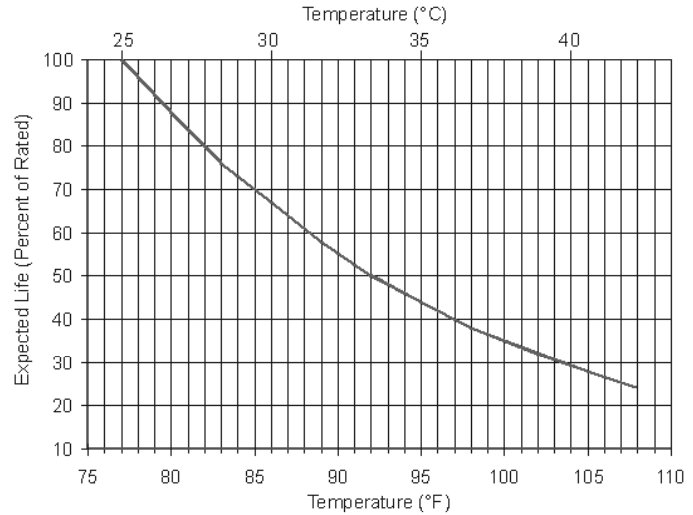}
    \caption{Impact of electrolyte temperature on the battery life from \cite{battery_ashrae}}
    \label{fig:Battery_Life}
\end{figure}

The expected lifespan of the lead-acid type batteries as a function of their electrolyte temperature is presented in Fig.~\ref{fig:Battery_Life}. NiCd battery life is typically less impacted by extreme temperatures than lead-acid batteries. For example, IEEE Std 484-2019 \cite{battery_VRLA_installation}, \textit{IEEE Recommended Practice for
Installation Design and Installation
of Vented Lead-Acid Batteries
for Stationary Applications}, reports that operating the lead-acid type batteries continuously at elevated temperatures would reduce its life span by 50\% for every $8\degree C$ increase above $25\degree C$ whereas $30\degree C$ above the $25\degree C$ is required to reduce the life span of NiCd type batteries by 50\%.  

It should be noted that the recommended ambient temperature limits for the more popularly used lithium-ion type batteries in various grid applications are not currently available in the literature. 

\subsection{Transformers}
\label{subsec:Lt_Ar}
Under usual service conditions, IEEE Std C57.12.00-2021 \cite{transformers_liquid_immersed}, \textit{IEEE Standard for General
Requirements for Liquid-Immersed
Distribution, Power, and Regulating
Transformers}, and IEEE Std C57.12.01-2020 \cite{transformers_dry_type},\textit{IEEE Standard for General
Requirements for Dry-Type
Distribution and Power Transformers},  reports that for both dry-type and liquid-immersed type transformers (distribution, power, and autotransformers), the ambient temperature limits when air-cooled would be $-30\degree C$ and $40\degree C$ and the average temperature of the cooling air (over 24 hrs) should not exceed $30\degree C$. Additionally, when water-cooled, the recommended ambient temperature limits are  $-20\degree C$ and $30\degree C$, and the average temperature of the cooling water (over 24 hrs) should not exceed $25\degree C$. 

However, in IEEE Std C57.91-2011 \cite{transformers_mineral_oil}, \textit{IEEE Guide for Loading Mineral-
Oil-Immersed Transformers and
Step-Voltage Regulators}, it has been recommended that the above-mentioned ambient temperature maximum and average limits need to be increased by $5\degree C$ due to the following reasons -

\begin{itemize}
\item For air-cooled types: The increase in aging of these transformers at higher average temperatures would not be compensated by the decrease in aging at lower average temperatures. 
\item For water-cooled type transformers: To account for the possible loss of cooling efficiency due to deposits formed on the cooling coil surfaces for these transformers.
\end{itemize}

\section{DISCUSSION}
\label{sec:discussion}
\begin{figure*}
    \centering
    \includegraphics[width=\linewidth]{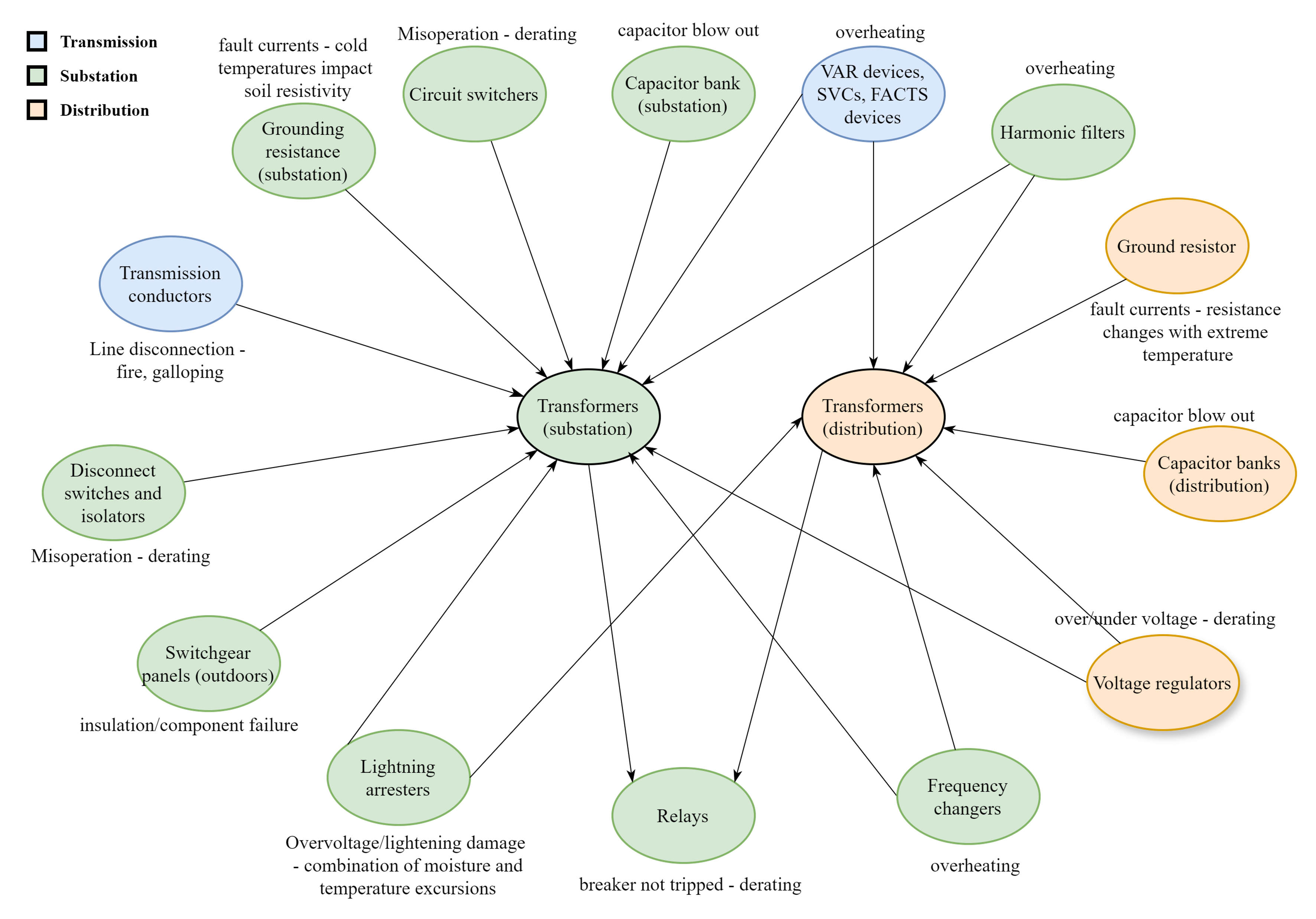}
    \caption{Example of inter-dependency of different equipment in power grids: transformers.}
    \label{fig:inter_1}
\end{figure*}

\begin{figure*}
    \centering
    \includegraphics[width=\linewidth]{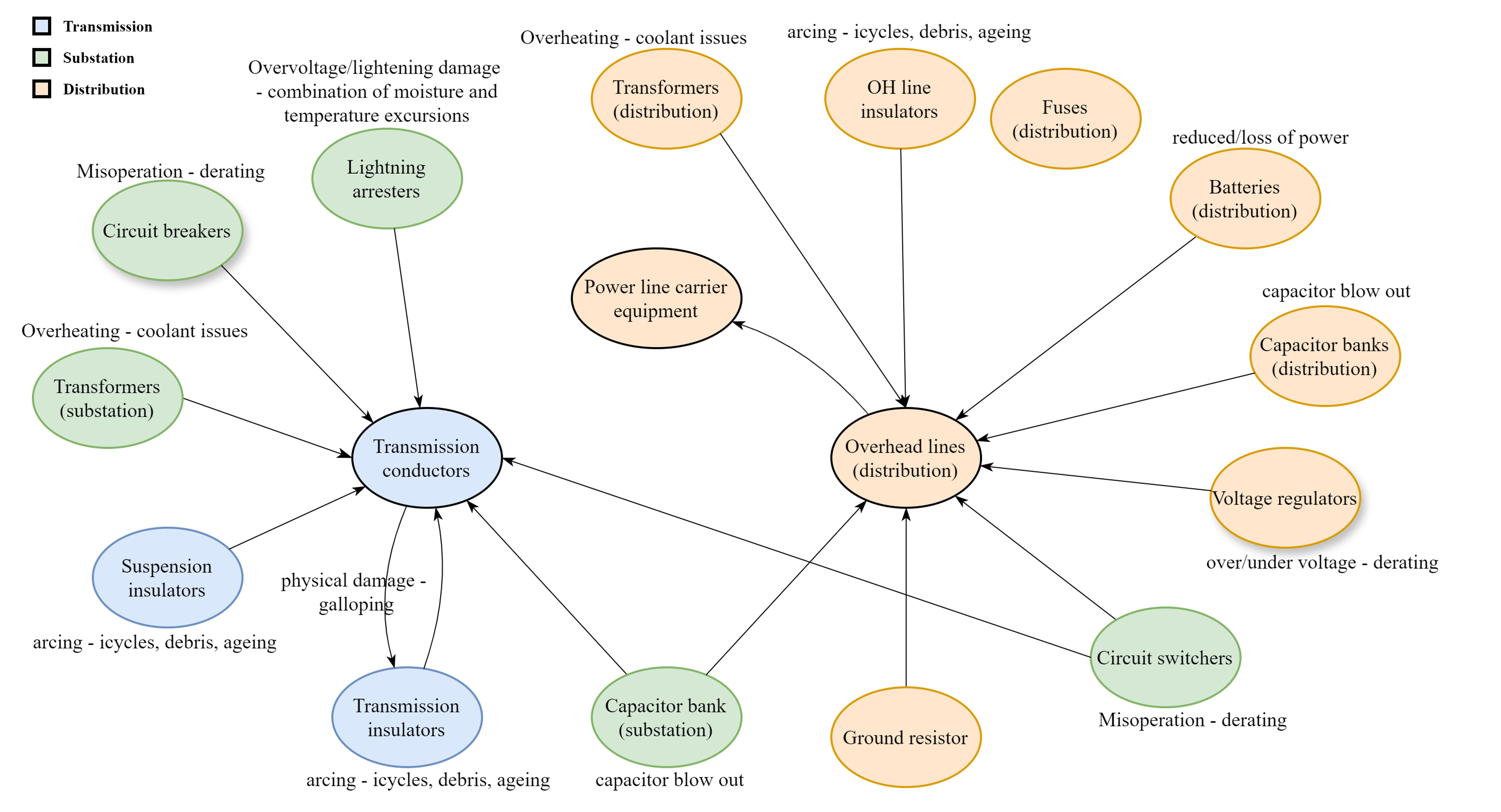}
    \caption{Example of inter-dependency of different equipment in power grids: power lines and power line carrier equipment.}
    \label{fig:inter_2}
\end{figure*}

\begin{figure*}
    \centering
    \includegraphics[width=\linewidth]{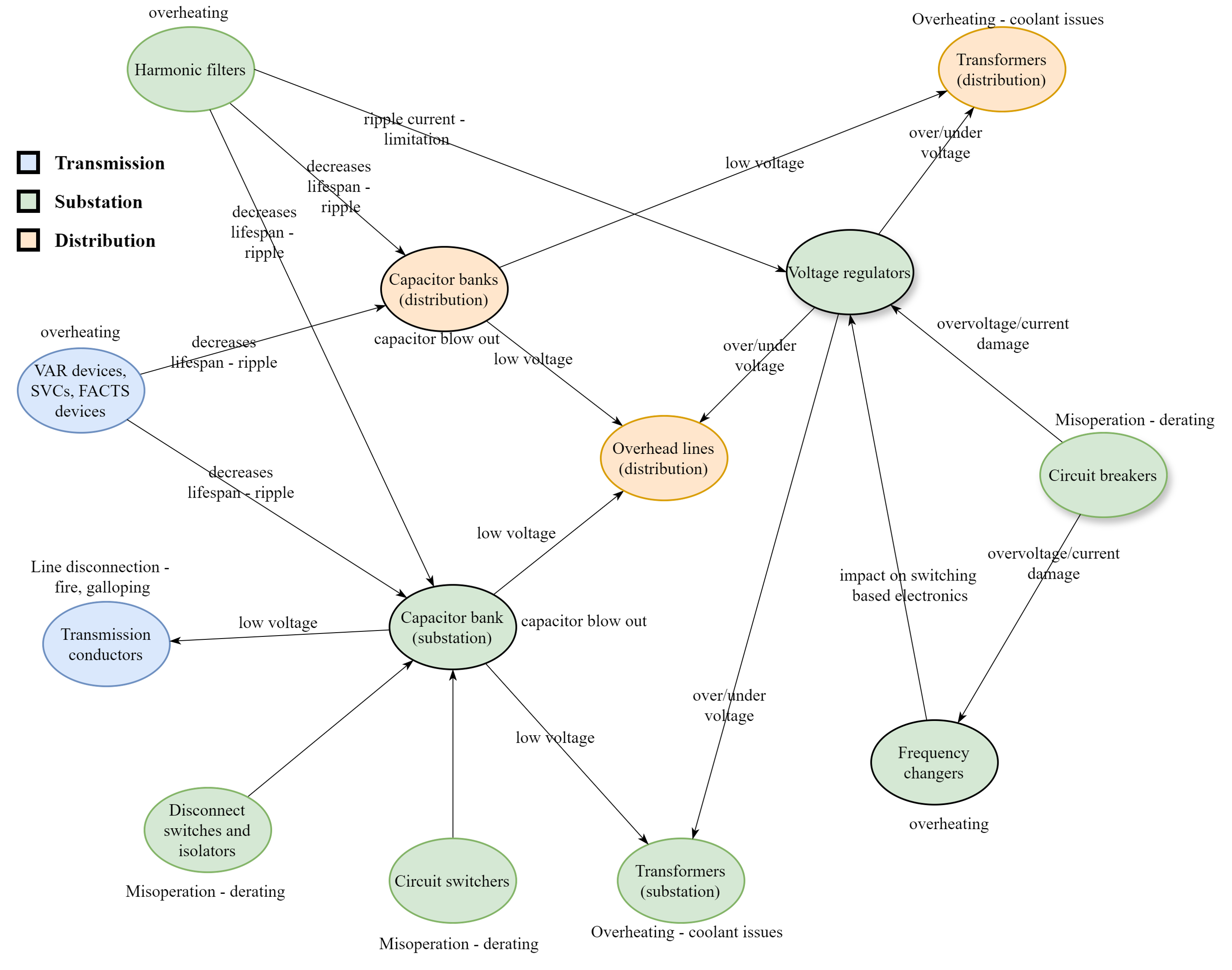}
    \caption{Example of inter-dependency on capacitor banks, voltage regulators, and frequency changer.}
    \label{fig:inter_3}
\end{figure*}

In this section, we consolidate and discuss the information on different topics presented in this paper so far. These topics include the following: 1) A summary of the direct and indirect impact of extreme heat and cold on different equipment (categorized as primary and secondary), 2) A summary of various equipment standards and violation of their recommended operational temperature limits across U.S. geography are presented to show the need for revision of these standards in light of heat waves and polar vortex, 3) Inter-dependency of equipment failures, and 4) Possible cascade failure due to equipment outage.

\subsection{Summary of equipment impacts due to extreme heat and cold temperatures} In this section, Tab.~\ref{tab:data_stats} and Tab.~\ref{tab:data_stats2} present the equipment impacts/failure modes due to extreme temperatures on primary and secondary grid equipment, respectively. In these tables, the impacts of extreme temperatures (heat and cold) are separated as direct and indirect impacts. A direct impact is a failure mode for equipment directly caused by the extreme temperature. In contrast, an indirect impact is a failure mode for equipment that is indirectly caused by other factors in conjunction with extreme temperatures. For Example, the conductor in Tab.~\ref{tab:data_stats} has derating and thermal loading as direct and indirect impacts during extreme heat scenarios. In extreme cold scenarios, its failure modes are ice and snow accumulation, causing reduced right-of-way and galloping (combination of winds and heavy icing on conductors) for direct and indirect impacts respectively.
% \begin{table*}[ht]
% % \vspace{-2em}
%     \caption{{Stability and Operational Assessment}}
% 	\centering
% 	\renewcommand{\arraystretch}{1.3}
% 	% \begin{tabularx}{\columnwidth}{|X|X|X|X
%  % |X|}
%  \begin{tabularx}{\textwidth}{|s|s|b|b|b|}
% 		\hline \hline
% 		{Reference} & {ML method} & {Data} & {Strength} & {Shortcoming} \\ \hline%& \% of solar and wind\\ \hline
%           kjsafh Synchrophasor applications  & GAN and GRL (combination of GCN and LSTM)  &  the New England 10-machine system,  Nordic system,  Iceland network system & 
%                Learn from both the graphical (network) and temporal characteristics of the power system dynamics. & large graph size computation.
%           \\ \hline
% 	\end{tabularx}
% 	\label{tab:comparison}
% \end{table*}

%secondory equip: insulator, coronaa ring, power line carrier communication system, lightening arrestor, switchgear panels and control panels,

% %primary equip: conductor, fuses, capacitor bank, batteries, transformer, ciruit breaker, circuit switcher, voltage regulator, frequency changer, grounding resistance, svc and facts devices, disconnect switches and isolators 
% \begin{table*}[ht]
% % \vspace{-2em}
%     \caption{Primary equipment impacts}
% 	\centering
% 	\renewcommand{\arraystretch}{1.3}
% 	% \begin{tabularx}{\columnwidth}{|X|X|X|X
%  % |X|}
%  \begin{tabularx}{\textwidth}{|b|b|b|b|b|}
% 		\hline \hline
% 		{Equipment} & {Failure mode} &{Direct/indirect impact} & {Polar vertex} \\ \hline%& \% of solar and wind\\ \hline
%           ABC & GHY & XYZ &LMNOP \\ \hline
% 	\end{tabularx}
% 	\label{tab:comparison}
% \end{table*}

\begin{table*}[h]
\caption{Primary equipment impacts.}
\centering
\renewcommand{\arraystretch}{1}
\begin{tabular}{|>{\centering} p{2cm}|>{\centering} p{3.5cm}|>{\centering} p{3.5cm}|>{\centering} p{3.5cm}|p{3.6cm}|}
\hline \hline
\multirow{3}{*}{} & \multicolumn{2}{c|}{Extreme heat} & %
    \multicolumn{2}{c|}{Extreme cold} \\ %& \multirow{3}{*}{D}\\
\cline{2-5}
Equipment & Direct impact & Indirect impact & Direct impact & Indirect impact \\ \hline
Insulators and Bushings & - & Structural damage due to cycling of extreme temperatures & Flashover due to icing and dust deposition & Puncture due to flashover because of polymeric material \\ \hline
 Circuit breaker & Derating & - & Derating but due to change in insulating medium properties & - \\ \hline
 Conductor & Derating & Thermal loading & Ice and snow accumulation reduces right-of-way, Metal contraction which increases line tension causing damage to structure & Galloping \\ \hline
 Fuses & Derating & - & Structural damage due to contraction & - \\ \hline
 Capacitor bank & Continuous temperature rating & Large loading changes & - & - \\ \hline
 Corona rings & Reduced corona onset voltage & - & Higher corona loss & - \\ \hline
 Battery & Efficiency reduction & Failure due to combination of heat and cooling system failure & Reduced capacity, frozen electrolyte & - \\ \hline
 Transformer & Derating due to impact on coolant temperature & Increased demand on the windings, cooling system failure & Oil viscosity, waxing, condensation, etc. & - \\ \hline
 Circuit switcher & Derating & - & Ice-breaking capacity& - \\ \hline
 Voltage regulator & Derating and continuous temperature rating & Large loading changes & - & Condensation and freezing in moving parts \\ \hline
 Frequency changer & Derating and continuous temperature rating & - & - & - \\ \hline
 FACTS devices & Derating and continuous temperatre rating when not in climate controlled environment & - & - & - \\
\hline \hline
\end{tabular}
\label{tab:data_stats}
% \vspace{-1.5em}
\end{table*}

\begin{table*}[h]
\caption{Secondary equipment impacts.}
\centering
\renewcommand{\arraystretch}{1}
\begin{tabular}{|>{\centering} p{2cm}|>{\centering} p{3cm}|>{\centering} p{3cm}|>{\centering} p{3cm}|p{2.6cm}|}
\hline \hline
\multirow{3}{*}{} & \multicolumn{2}{c|}{Extreme heat} & %
    \multicolumn{2}{c|}{Extreme cold} \\ %& \multirow{3}{*}{D}\\
\cline{2-5}
Equipment & Direct impact & Indirect impact & Direct impact & Indirect impact \\ \hline
 Substation grounding & Soil resistivity changes & - & Soil resistivity changes & - \\ \hline
 Lightning arrestor & Manufacturer operational temperature & - & Manufacturer operational temperature & - \\ \hline
 Relay & Loss of sensitivity and coordination for EM relays & Loss of coordination due to CT/PT/sensor inaccuracy & Loss of sensitivity and coordination for EM relays & Loss of coordination due to CT/PT/sensor inaccuracy \\ \hline
 Switchgear and control Panels & Derating & Arc flash due to insulation failure & - & Arc flash due to condensation and freezing \\ \hline
 Converters & Derating & Auxilary equipment failure & - & Freezing of water cooling system \\ \hline
 % Harmonic filters & 0.0017 & 0.0018& 0.5770& 0.3937\\ \hline
 PLCC & - & Latency, and noise due to component's temperature sensitivity & - & Latency, and noise due to component's temperature sensitivity \\ \hline
 Disconnect switches and Isolators & Derating & - & Ice-breaking capacity & - \\
\hline \hline
\end{tabular}
\label{tab:data_stats2}
% \vspace{-1.5em}
\end{table*}

\subsection{Summary of standards and equipment operational temperature violations due to climate change}
A summary of all the standards provided in this paper and potential modeling gaps present in the literature for each equipment has been presented in Tab.~\ref{fig:Equipment_Standards_Summary}. For Example, in all the standards except for Circuit switchers and Circuit breakers, the permitted level of ice loading on the equipment has not been discussed, which would be critical information to know for the manufacturers or the equipment designers to account for ice loading due to extreme cold events such as the Polar vortex. Additionally, although the recommended maximum leakage rates for the instrument transformers at extremely cold conditions (up to $-50\degree C$) have been provided in \cite{instrument_tranformers}, the corresponding recommended leakage rates at extremely high ambient temperatures (at higher than $40\degree C$) are not currently available in the literature.  

% \begin{figure*}[h]
%     \centering
%     \includegraphics[width=\linewidth]{images2/Final_Table_Standards_Summary.PNG}
%     \caption{Equipment standards summary}
%     \label{fig:Equipment Standards Summary}
% \end{figure*}

% \begin{table*}[h]
%   \centering
%   \caption{{Equipment standards summary}}
%   \includegraphics[width=\linewidth]{images2/Final_Table_Standards_Summary.PNG}
%   \label{fig:Equipment_Standards_Summary}
%   % \vspace{-2em}
% \end{table*}

\begin{table*}[h]
\centering
\caption{Equipment standards summary}
\renewcommand{\arraystretch}{1}
% \resizebox{\textwidth}{!}{
\begin{tabular}{|>{\centering} p{2cm}|>{\centering} p{4cm}|>{\centering} p{2cm}| p{8cm}|}
\hline \hline
\textbf{Equipment}                          & \textbf{Standards}                                       & \textbf{Year}    & \textbf{Modeling Gaps}                                                                               \\ \hline
Lightning   arrestors                       & IEEE   Std C62.11                                        & 2020             & Thermal   stress testing for longer than 24 hrs, Ice loading                                        \\ \hline
Grounding   resistors                       & IEEE   C57.32                                            & 2015             & Cold   spot temperature limits, resistance variance at extreme cold temperatures and   Ice loading   \\ \hline
Capacitor   Banks                           & IEEE Std 1036                                            & 2020             & Ice loading                                                                                         \\ \hline
Harmonic   filters                          & IEEE   Std 1531                                          & 2020             & Ambient   temperature limits at extreme cold conditions, Ice loading                                \\ \hline
Power fuses                                 & IEEE   Std C37.41                                        & 2016             & Ice loading                                                                                          \\ \hline
Relays                                      & IEEE   Std C37.90, IEEE Std C37.91, IEEE Std C37.90.3    & 2005, 2021, 2023 & Ice loading                                                                                         \\ \hline
Shunt reactors                              & IEEE   Std C57.21                                        & 2021             & Ambient   temperature limits at extreme cold conditions, Ice loading                                 \\ \hline
Composite   insulators                      & IEEE   Std 987                                           & 2001             & Ice loading                                                                                          \\ \hline
Circuit   switchers                         & IEEE   Std C37.30.1                                      & 2022             & Current   carrying capability temperature correction factor at extreme hot ambient   temperatures   \\ \hline
Circuit   breakers                          & IEEE   Std C37.119, IEEE Std C37.04                      & 2016, 2018       & Adjustment   needed for recommended ambient temperature limits for unusual service   conditions      \\ \hline
Overhead lines/ Underground cables          & IEEE   Std 835, IEEE Std 835a, IEEE Std 738              & 1994, 2012, 2023 & Recommended ambient temperature limits, Ice   loading                                                \\ \hline
Instrument transformers                     & IEEE   Std C57.13.5                                      & 2019             & Leakage   rates recommendations at extremely high ambient temperatures, Ice loading                 \\ \hline
Batteries                                   & IEEE Std 484, IEEE Std 485, IEEE Std   1635              & 2019, 2020, 2022 & Recommendations   for extreme hot/cold ambient temperatures, Ice loading                            \\ \hline
Distribution,   Power and Auto Transformers & IEEE   Std C57.91, EEE Std C57.12.01, IEEE Std C57.12.00 & 2011,2020, 2021  & Ice loading                                              
\\ \hline \hline

\end{tabular}
% }
\label{fig:Equipment_Standards_Summary}  
\end{table*}

Typically, before deploying them in the field, the ambient temperature limits of the equipment, based on the available standards, are considered so that the limits reflect the usual type of extreme temperatures experienced in its corresponding region. For Example, in Arizona, where extremely hot weather conditions are observed, emphasis would be placed on having ambient temperature limits higher than the recommended limits during the equipment design process. However, due to the recent climate changes, many unusual events (anomalies) such as the heat haves occurring in usually cold climates (extreme heat wave in Chicago in 1995) and freezing events occurring in usually warm locations (the great Texas freeze in 2021). Therefore, it is vital to analyze if the standards of the considered equipment ambient temperature limits recommendations consider these anomalies.

For this reason, two categories have been considered in this work, as shown below, to analyze the ambient temperature violations that can be observed for various types of regions based on the current standards recommended ambient temperature (under usual service conditions) limits and has been presented in Tab.~\ref{fig:Equipment_Violations_Standards} -
\begin{itemize}
\item \textit{Category 1:} Potential future scenarios where ambient temperatures might become very extreme. These scenarios are defined as "Extreme heat in cold region" and "Extreme cold in hot region" in Tab.~\ref{fig:Equipment_Violations_Standards}. 
\item \textit{Category 2:} The historical extreme ambient conditions already recorded across the continental United States in the last 100 years \cite{wiki_temp_records}. These conditions are referred to as "Historical data in the US" in Tab.~ \ref{fig:Equipment_Violations_Standards}.
\end{itemize}

 The temperature data in \cite{wiki_temp_records} has been obtained from the National Oceanic and Atmospheric Administration (NOAA) \cite{NOAA_First_SN}. For these two categories, two types of extreme conditions have been considered - extremely cold conditions in a "hot" state and extremely hot conditions in a "cold" state. The classification considered whether a state is "cold" or "hot" is done based on the ranking of their average recorded temperatures over the last five years \cite{NOAA_SN}. The top five "hot"/"cold" states considered here are as follows -

\begin{itemize}
\item Florida/North Dakota 
\item Louisiana/Minnesota
\item Texas/Wyoming
\item Georgia/Montana
\item Missisipi/Maine
\end{itemize}

As mentioned earlier, the main reasoning behind considering cold conditions for a "hot" state and vice-versa is because typically the manufacturers do not account for equipment's operating conditions in hotter regions do not consider extreme cold conditions as they are an anomaly. In Tab, ~\ref{fig:Equipment_Violations_Standards},for example, for the considered five hottest regions, the historical extreme cold temperatures are binned into (-$20\degree C$, $30\degree C$) and < $30\degree C$. We considered cold temperatures since the equipment's operating conditions in hotter regions do not consider extreme cold conditions as they are an anomaly. However, the Texas great freeze is a good example of hot climatic region that experienced unexpected cold conditions. In the Tab, ~\ref{fig:Equipment_Violations_Standards}, the standard for equipment such as power fuses says that their recommended standard ambient temperature ranges are between -$30\degree C$ and $40\degree C$ which aligns with observed cold temperatures between -$20\degree C$ and $30\degree C$ but does not align with extreme observed cold temperatures of < -$30\degree C$. 
Additionally, from Tab.~\ref{fig:Equipment_Violations_Standards}, it can be observed that except for capacitor banks and the relays, the ambient temperature limits from standards for all the other equipment undergo violations in at least one of the considered two anomaly categories in this work. Especially for battery and grounding resistor equipment, their recommended ambient temperature limits would be violated for all the considered extreme scenario anomaly conditions.
This shows the need for revision of standards due to increased frequency, intensity, and duration of heat waves (Fig. \ref{fig:heatwave_charac}) and polar vortex type events. This clearly shows a critical need to revisit and revise the current standards to ensure that the power system equipment performance is considered robust under all possible extreme weather conditions.

%[!htb]

% \begin{figure*}[h]
%     \centering
%     \includegraphics[width=\linewidth]{images2/Final_Table_Standards_Violations_Latest.PNG}
%     \caption{Ambient temperature violations possibilities for various equipment based on existing standards}
%     \label{fig:Equipment_Violations_Standards}
% \end{figure*}

% \begin{table*}[h]
%   \centering
%   \caption{{Ambient temperature violations possibilities for various equipment based on existing standards}}
%   \includegraphics[width=\linewidth]{images2/Final_Table_Standards_Violations_Latest.PNG}
%   \label{fig:Equipment_Violations_Standards}
%   % \vspace{-2em}
% \end{table*}

\begin{table*}[h]
  \centering
  \caption{{Ambient temperature violation possibilities for various equipment based on existing standards. The values in the table represents if a particular equipment's recommended ambient temperature range from standards is violated when compared with historical extreme cold/hot temperatures in hotter/colder regions in the U.S, respectively. Yes, No, and N/A in the table represents if there is a violation, no violation, and no available information respectively.}}
  \includegraphics[width=\linewidth]{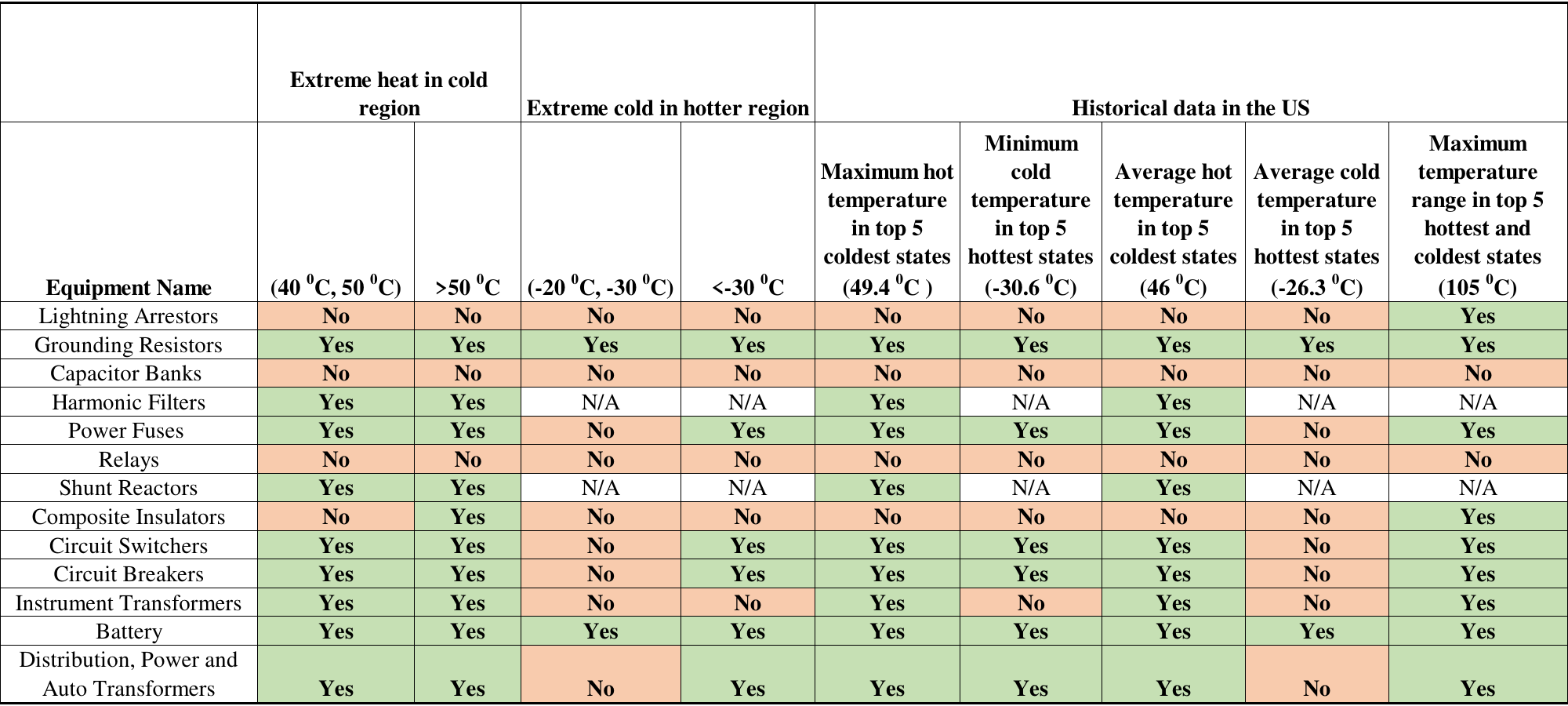}
  \label{fig:Equipment_Violations_Standards}
  % \vspace{-2em}
\end{table*}

\subsection{Inter-dependency of equipment}

Beyond the direct effects of heat domes and polar vortices on equipment, component failures can damage other parts of the power grid. This can result in cascading failures or damage that impacts grid reliability. In this section, some possible impacts of equipment on one another has been presented. However, these examples are not exhaustive and are only an indication of the interdependency of equipment in power grids.

Figs.~\ref{fig:inter_1}, \ref{fig:inter_2}, and \ref{fig:inter_3} demonstrate the impact of different equipment failures on two or three components each. These figures highlight each equipment and its corresponding failure mode due to extreme temperatures. The arrows indicate the impact of the equipment on the equipment of focus. For Example, in Fig.~\ref{fig:inter_1}, the equipment of focus are the Substation and distribution transformer, which are highlighted by a black outline.

As shown in Fig.~\ref{fig:inter_1}, many different equipment failures can affect some grid components. For Example, transformers are one of the most vulnerable components with inter-dependency; seven types of equipment failures can cause conditions that could lead to damage or failure of distribution transformers, and twelve types can do the same to substation transformers. Many of these dependencies occur in equipment that is either part of operation or protection, such as lightning arresters, grounding resistors, and relays that allow overcurrent or overvoltage to flow through the transformer. It is to be noted that the equipment considered to show interdependency is not exhaustive; it only serves as a demonstration of the interdependency of grid equipment and indication.

Similarly, Fig.~\ref{fig:inter_2} demonstrates the vulnerabilities of power lines (transmission and distribution conductors) and power line carrying equipment in transmission and distribution systems. Power lines can be impacted by numerous equipment throughout the grid, such as failures in circuit breakers, lightning arrestors, insulators, and capacitors. Most events that affect power lines cause or fail to protect the lines from overcurrent or overheating.

While Fig.~\ref{fig:inter_1} and Fig.~\ref{fig:inter_2} display the effects of some components on a few others, Fig.~\ref{fig:inter_3} shows a more interwoven chain of influence. Fig.~\ref{fig:inter_3} focuses on equipment with relationships to voltage regulators, frequency changers, and capacitor banks. Failures in components such as harmonic filters, volt-ampere reactive (VAR) devices, static VAR compensators (SVCs), and flexible alternating current transmission system (FACTS) devices cause the lifespans of capacitor banks to decrease. It is to be noted that not every grid infrastructure may have all the components mentioned above working together. However, it could be a combination of equipment for utility planning and operation requirements. Additionally, events in capacitor banks and voltage regulators can influence the distribution transformers. 

These examples are not exhaustive and are only an indication of interdependency in power grids. The objective here is for these influence maps to be useful in determining equipment of priority for research and development in the future to prepare grid equipment for specific failures elsewhere in the system.

% \begin{figure*}
%     \centering
%     \begin{subfigure}[t]{0.5\textwidth}
%         \centering
%         \includegraphics[width=\linewidth]{images2/inter_1.png}
%         \caption{Inter-dependency on transformers.}
%         \label{fig:inter_1}
%     \end{subfigure}%
%     ~ 
%     \begin{subfigure}[t]{0.5\textwidth}
%         \centering
%         \includegraphics[width=\linewidth]{images2/inter_2.png}
%         \caption{Inter-dependency on power lines and power line carrier equipment.}
%         \label{fig:inter_2}
%     \end{subfigure}
%     \caption{Example of inter-dependency of different equipment in power grids.}
%     % \vspace{-2em}
% \end{figure*}

\subsection{Cascade Connections}
\begin{figure*}
    \centering
    \includegraphics[width=0.8\linewidth]{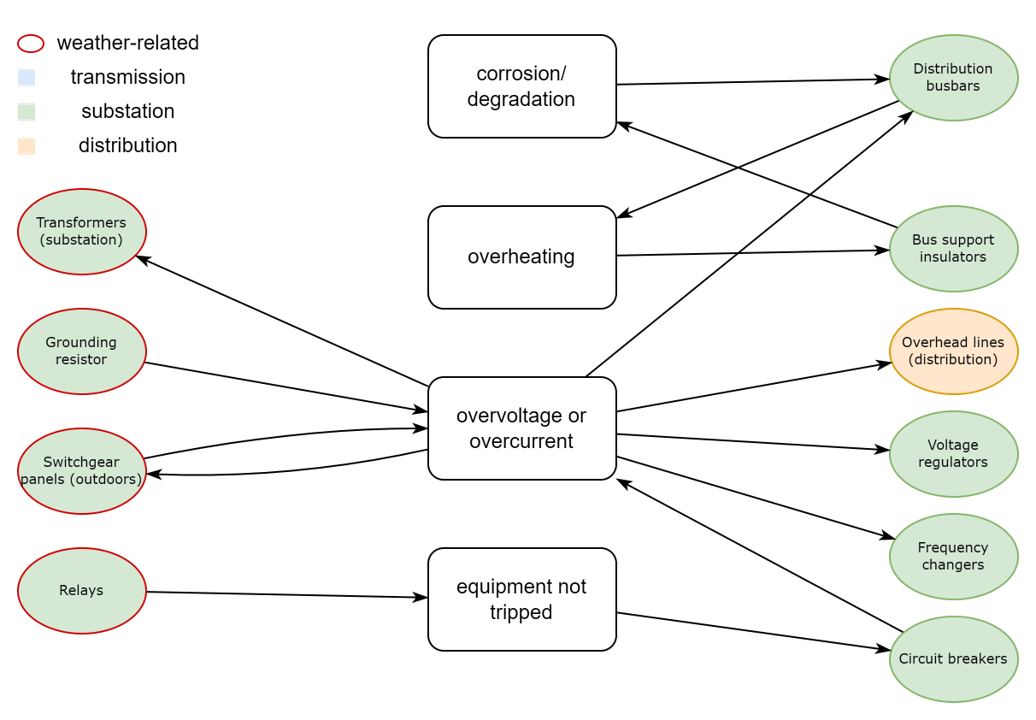}
    \caption{Example of cascades in power grids during extreme weather conditions: substation equipment during extreme heat.}
    \label{fig:cas_1}
\end{figure*}

\begin{figure*}
    \centering
    \includegraphics[width=0.8\linewidth]{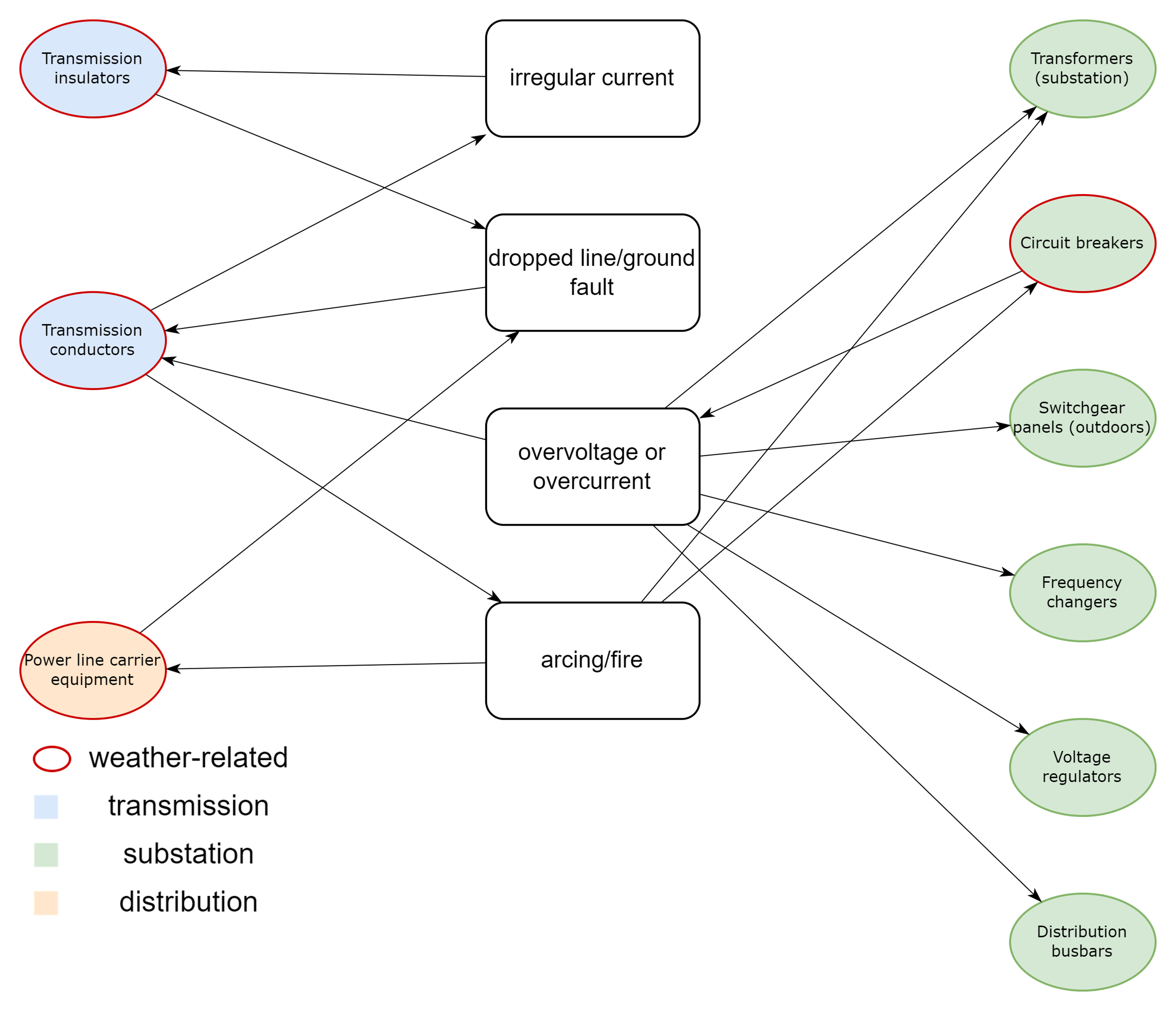}
    \caption{Example of cascades in power grids during extreme weather conditions: transmission during extreme cold and icing.}
    \label{fig:cas_2}
\end{figure*}

While Figs.~\ref{fig:inter_1}, \ref{fig:inter_2}, and \ref{fig:inter_3} characterize many specific inter-dependency relationships, they do not capture the potential for cascading effects in power grids. Figs.~\ref{fig:cas_1} and \ref{fig:cas_2} demonstrate a few possibilities for chains of equipment failures to occur under specific extreme weather conditions, although there are many possibilities that are not exhaustively presented in this work. 


Fig.~\ref{fig:cas_1} identifies potential cascades in substations under extremely high temperatures. For Example, a heat-related failure in a relay can prevent a circuit breaker from tripping when a fault is present in the substation. This can cause overcurrent or overvoltage to flow through the substation, potentially damaging many other Substation components such as transformers, voltage regulators, and distribution busbars. These conditions could cause distribution busbars to overheat, damaging bus support insulators. Similar cascading effects are observable if the chain begins with the other weather-related equipment failure events in Fig.~\ref{fig:cas_1}.

The contingency chain for transmission lines and carrier equipment in extreme cold and icing conditions is shown in Fig.~\ref{fig:cas_2}. Cold ambient temperatures and icing can affect transmission equipment, leading to substation issues. For Example, towers that carry transmission lines can be brought down by the weight of ice accumulation. This will likely cause an open circuit, ground fault, or short circuit in the transmission conductors, which can lead to arcing in a circuit breaker or transformer. A transformer or circuit breaker failure could lead to power outages and damage to substation equipment.

% \begin{figure*}
%     \centering
%     \begin{subfigure}[t]{0.49\textwidth}
%         \centering
%         \includegraphics[width=\linewidth]{images2/cas_1.png}
%         \caption{Example of cascades in power grids during extreme weather conditions: substation equipment during extreme heat.}
%         \label{fig:cas_1}
%     \end{subfigure}
%     \begin{subfigure}[t]{0.49\textwidth}
%         \centering
%         \includegraphics[width=\linewidth]{images2/cas_2.png}
%         \caption{Example of cascades on transmission during extreme cold and icing.}
%         \label{fig:cas_2}
%     \end{subfigure}
%     \caption{Example of cascades in power grids during extreme weather conditions.}
%     % \vspace{-2em}
% \end{figure*}
% \input{tex_files/discussion}

\section{CONCLUSION}
\label{sec:conclusion}

\subsection{Background Summary} The criticality of the power grid infrastructure is very high as it is a backbone and an essential system for other critical infrastructures to operate successfully. The power grid infrastructure is undergoing many changes as it evolves to account for the inclusion of technologies like renewable generation, battery storage, microgrids, etc. This calls for paradigm shifts in the way the power grid is planned and operated, therefore there are many changes required in the grid codes and standards for addressing these issues. 

In addition to the changes in the power grid described above, there is also a significant change in global climate that is impacting the overall average temperatures and extreme temperatures that the power grid needs to operate under. The past few decades have seen an increase in the frequency of occurrences and intensity of heat waves consistently indicating that the trend may not be changing in the near future. Not only the frequency, and intensity of the heat waves are increasing but also a more concerning aspect is that the duration of heat waves is also increasing. The overall hot conditions are leading to uncertain and relatively short duration of extreme cold scenarios owing to the phenomenon of disrupted polar vortex. This causes the cold winds to blow further south away from the poles bringing unexpected snow storms and severe ice conditions in normally hotter regions. The increased heat and increased cold in unexpected regions of the country have increased the vulnerability of the power grid equipment to heat domes, heat waves, polar vortices, and severe ice conditions. 

However, the power grid infrastructure has a life of several decades and therefore needs to be planned foreseeing operating conditions for several decades. The current power grid infrastructure was not planned expecting consistent extreme cold temperatures in regions like Texas, Florida, and Arizona and extreme heat and heat waves in the Pacific Northwest, and New England regions.

While the extreme conditions for heat and cold are increasing it is resulting in extreme stress on power grid infrastructure. While the naive way to assess the equipment vulnerability by comparing specifications can give some understanding if the equipment can withstand extreme temperatures, however it is not sufficient to determine if the grid infrastructure on the whole can withstand these conditions.

\subsection{Technical Conclusions} The present paper discusses the climate change phenomenon in a certain detail in the context of how the grid infrastructure is vulnerable to the changing climate. Therefore, this paper has discussed the most vulnerable set of T\&D grid equipment including the various components in a substation providing the details of the individual equipment failure modes, its immediate impact on the other equipment, and is extended to cascaded failures of equipment. The failure modes and failure relation to other equipment are derived in the form of failure influence diagrams and cascaded failure influence diagrams. These diagrams give a better understanding of the overall vulnerability of the grid infrastructure. For each equipment, the paper also discusses existing and potential approaches that the grid planners and operators can take to monitor, measure, and mitigate the impacts of climate change (particularly, heat domes, polar vortices, and icing). The paper also discusses the standards and their readiness in terms of the extreme temperatures experienced in the various regions of the different states.

\subsection{Future Work}
While this work establishes the fundamental understanding of infrastructure impacts due to extreme temperatures from the literature review, industry articles, etc., it is important to translate this information into actionable steps regarding equipment availability, system contingency analysis, and overall grid stability analysis. There are other important impacts of extreme cold and extreme heat related  events on the power grid equipment like corrosion or structural integrity failures that lead to loss of equipment. There are other impacts that accelerate the end of life of equipment that may suddenly lead to acute failures eventually. We will expand on providing these insights along with gap analysis and research priorities in the future.

% % \vspace{-1em}
% \section{Future Grid's Framework for Dynamic Simulation Studies}
% \label{sec:framework_overall}
% \input{input_files/sim_framework}

% % \section{Problem Description }
% % \label{sec:problem_description}
% % \input{input_files/problem_Description}

% % % \vspace{-1em}
% % \section{QV Curve Approach for Future Large Interconnected Grids}
% % \label{sec:pv_qv}
% % \input{input_files/pv_method}

% % \vspace{-1em}
% \section{Voltage Stability Critical Location Identification}
% \label{sec:dvsi}
% \input{input_files/dvsi_method}

% % \section{Proposed Framework to Identify Voltage Stability Critical Locations in Future Large Interconnected Grids}
% % \input{input_files/algo_overall}

% % \section{Analytical Comparison Between the QV Curve and Proposed D-VSI Methods}
% % \label{sec:comparison_section}
% % \input{input_files/comparison_section}

% % \vspace{-1em}
% \section{Simulation Results}
% \label{sec:simulation}
% \input{input_files/simulation_vsi}

% % \vspace{-1em}
% \section{Conclusion}
% \label{sec:conclusion}
% \input{input_files/conclusion}

% % % % \vspace{-1em}
% \section{Future Work}
% \label{sec:future_scope}
% \input{input_files/future_scope}

% \vspace{-1em}
\bibliographystyle{IEEEtran}
\bibliography{kishan.bib}

% \vspace{-4em}
\begin{IEEEbiography}
	[
	{\includegraphics[width=1in,height=1.25in,clip,keepaspectratio]{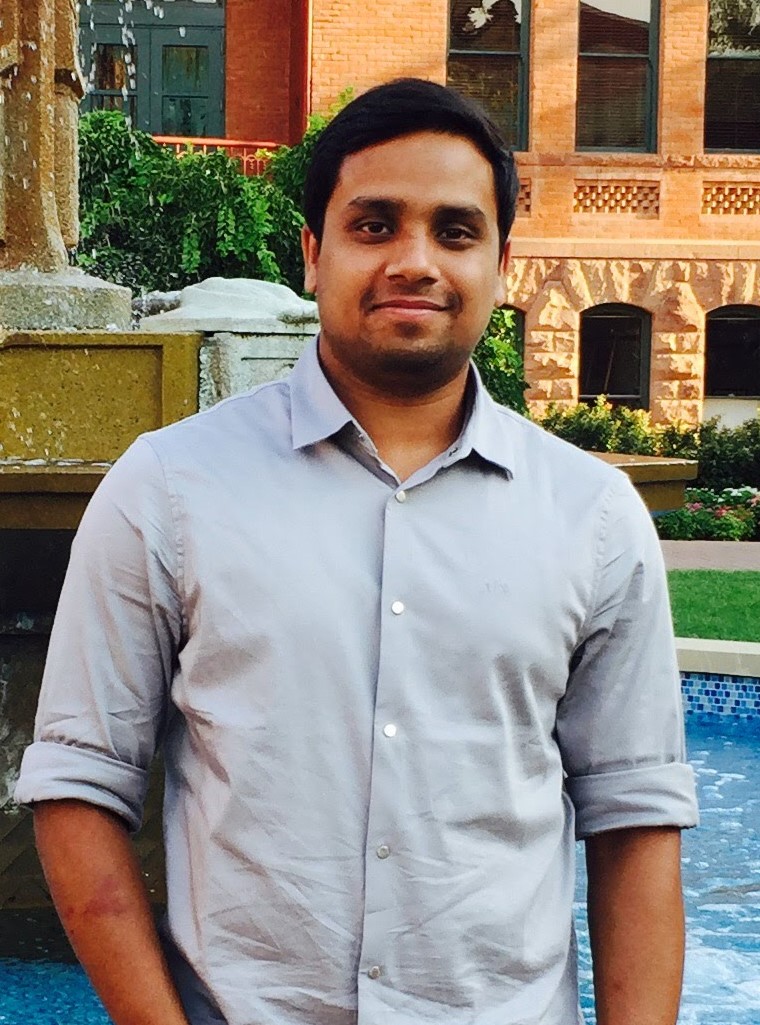}}
	] {Kishan Prudhvi Guddanti} (Member, IEEE) received his M.S and Ph.D. degrees in electrical engineering from Arizona State University, Tempe, AZ, USA in 2019 and 2021 respectively. His past work experience includes Hitachi America Ltd., RTE France. He is currently working as Power Systems Research Engineer with the Pacific Northwest National Laboratory. %, USA.  
 %His current research interests are in the interdisciplinary area of AI applications in power systems in addition to voltage stability, and data-driven techniques for power system risk assessment and control.
\end{IEEEbiography}

\begin{IEEEbiography}[
{\includegraphics[width=1in,height=1.25in,clip,keepaspectratio]{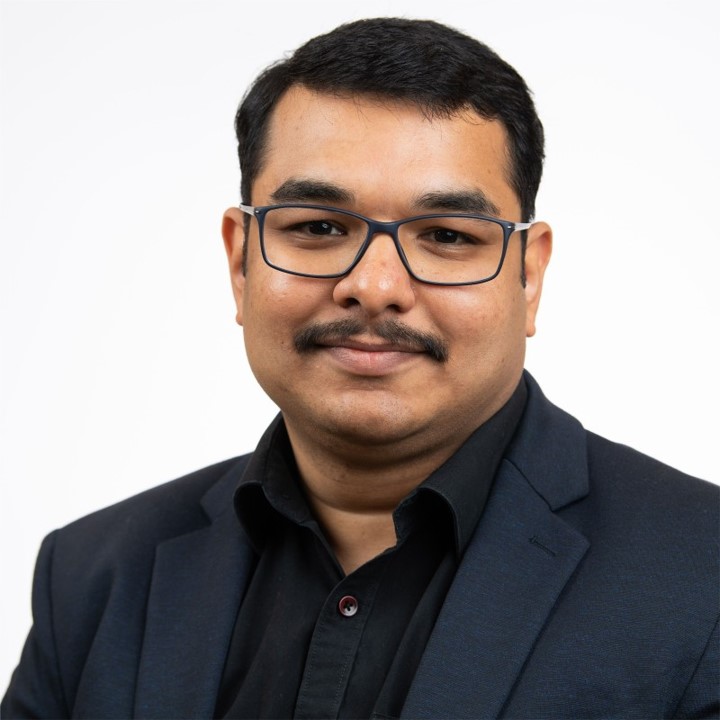}}
	] {Alok Kumar Bharati} (Member, IEEE) received the master’s degree in power electronics and power systems from the Indian Institute of Technology Hyderabad in 2011, the Ph.D. degree in electrical engineering from Iowa State University in 2021. He worked with GE from 2011 to 2017 in research and development of low-voltage switchgear and is an inventor on four inventions that resulted in multiple granted patents and patent applications. He is currently a Senior Power Systems Engineer with Pacific Northwest National Laboratory. His research interests include T\&D interactions and its impact on power system stability, renewable integration, EV integration to grid, T\&D co-simulation, multidomain co-simulation, and microgrids.
\end{IEEEbiography}

\begin{IEEEbiography}[
{\includegraphics[width=1in,height=1.25in,clip,keepaspectratio]{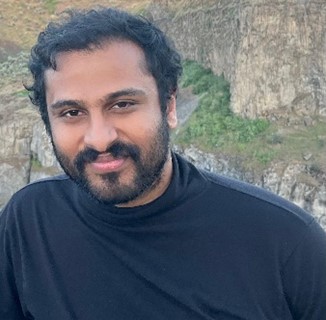}}
	] {Sameer Nekkalapu} (Member, IEEE) received the B.Tech. degree in electronics and electrical engineering from IIT Guwahati, India, in 2015, and the M.S. and Ph.D. degrees in electrical engineering from Arizona State University, Tempe, AZ, USA, in 2018, and 2022 respectively. Currently, he works as a power system research engineer at Pacific Northwest National Laboratory.
\end{IEEEbiography}

\begin{IEEEbiography}[
{\includegraphics[width=1in,height=1.25in,clip,keepaspectratio]{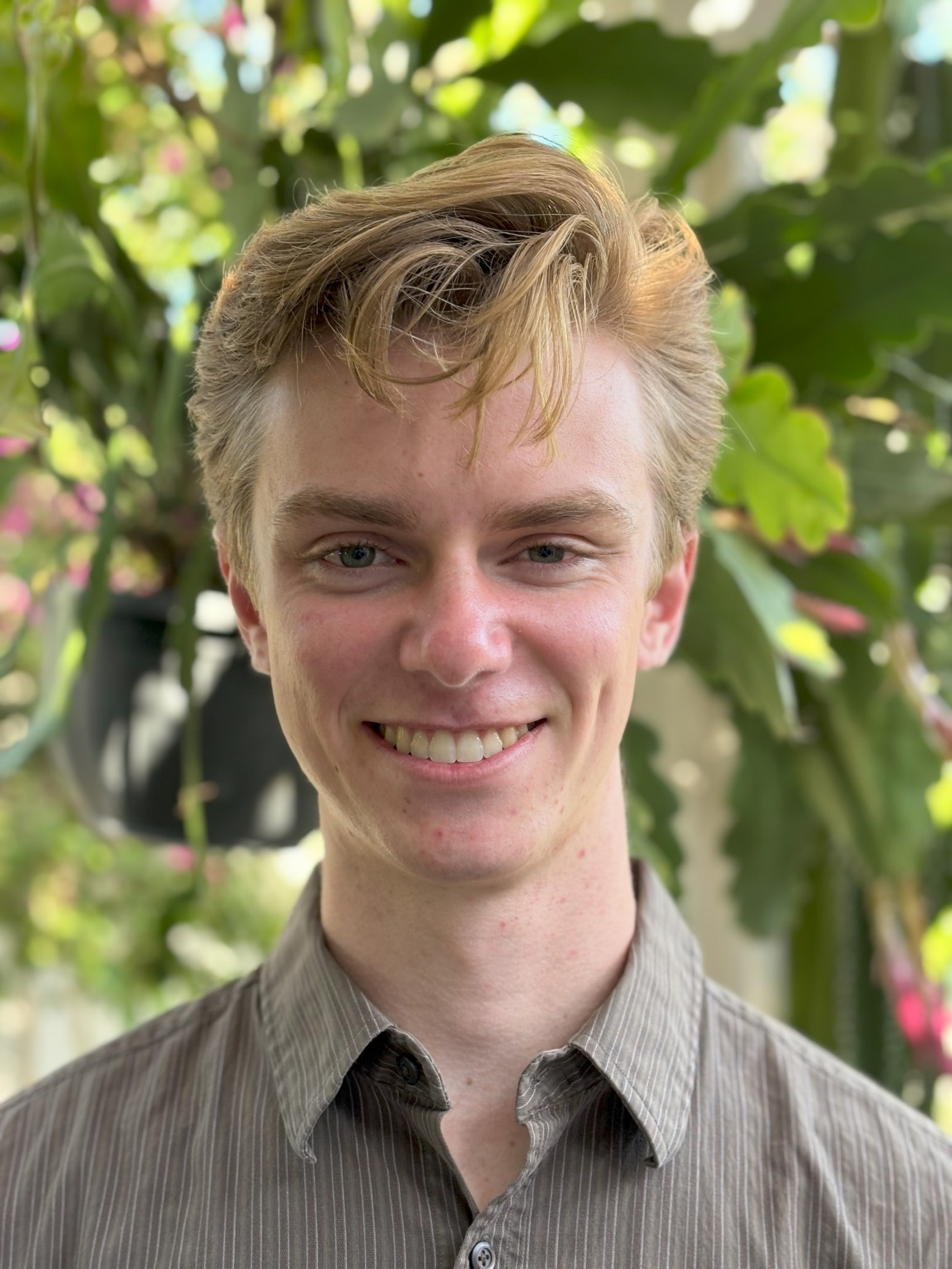}}
	] {Joseph McWherter} (Student Member, IEEE) is a junior pursuing his undergraduate degree in electrical engineering with a minor in physics at Chapman University. His research interests are power engineering, specifically renewable energy integration, grid resilience, and EV infrastructure.
\end{IEEEbiography}

\begin{IEEEbiography}[
{\includegraphics[width=1in,height=1.25in,clip,keepaspectratio]{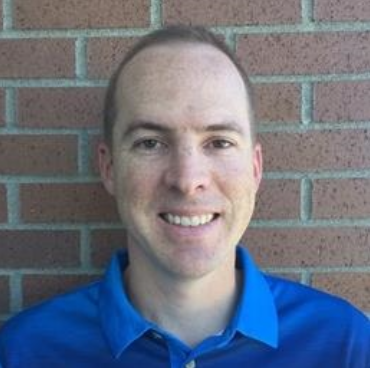}}
	] {Scott L Morris} received his master’s degree in statistics from Brigham Young University in 2012. He is currently a project manager and a team lead for a Grid Edge team with Pacific Northwest National Laboratory. He has been involved with cybersecurity research supporting the US Army's efforts to increase energy resilience, research supporting the US Department of State's technical assistance for southeast Asia and Central America regarding renewable energy generation integration efforts, power system resiliency, and electricity markets. His research interests include electrical power system resiliency, data analytics, and statistical modeling.
\end{IEEEbiography}

\EOD

\end{document}